\documentclass[pdflatex,sn-nature]{sn-jnl}

\usepackage{graphicx}%
\usepackage{multirow}%
\usepackage{amsmath,amssymb,amsfonts}%
\usepackage{amsthm}%
\usepackage{mathrsfs}%
\usepackage[title]{appendix}%
\usepackage{xcolor}%
\usepackage{textcomp}%
\usepackage{manyfoot}%
\usepackage{booktabs}%
\usepackage{algorithm}%
\usepackage{algorithmicx}%
\usepackage{algpseudocode}%
\usepackage{listings}%
\usepackage{longtable}
\usepackage{chngpage}
\usepackage{lineno}
\usepackage{caption}
\usepackage{subcaption}
\usepackage{float}

\theoremstyle{thmstyleone}%
\theoremstyle{thmstyletwo}%

\theoremstyle{thmstylethree}%

\begin{document}
\title[Article Title]{A spectacular multi-wavelength transient associated with an off-axis relativistic jet}

\author*[1]{\fnm{Delina} \sur{Levine}}\email{dlevine2@caltech.edu}

\author[1]{\fnm{Gregg} \sur{Hallinan}}

\author[2,3]{\fnm{Jean J.} \sur{Somalwar}}

\author[4]{\fnm{Dillon Z.} \sur{Dong}}

\author[5]{\fnm{Ehud} \sur{Nakar}}

\author[6]{\fnm{Kenta} \sur{Hotokezaka}}

\author[1]{\fnm{Vikram} \sur{Ravi}}

\author[7]{\fnm{Assaf} \sur{Horesh}}

\author[1]{\fnm{Jessie M.} \sur{Miller}}

\author[1]{\fnm{Casey} \sur{Law}}

\author[3]{\fnm{Steven T.} \sur{Myers}}

\author[1,8]{\fnm{Stella K.} \sur{Ocker}}

\author[8]{\fnm{Daniel D.} \sur{Kelson}}


\affil*[1]{\orgdiv{Cahill Center for Astronomy and Astrophysics}, \orgname{California Institute of Technology}, \orgaddress{\city{Pasadena}, \state{CA}, \country{USA}}}
\affil[2]{\orgdiv{Kavli Institute for Particle Astrophysics and Cosmology}, \orgaddress{\city{Stanford}, \state{CA}, \country{USA}}}

\affil[3]{\orgdiv{Department of Astronomy}, \orgname{University of California Berkeley}, \orgaddress{\city{Berkeley}, \state{CA}, \country{USA}}}

\affil[4]{\orgname{National Radio Astronomy Observatory (NRAO)}, \orgaddress{\city{Socorro}, \state{NM}, \country{USA}}}

\affil[5]{\orgdiv{School of Physics and Astronomy}, \orgname{Tel Aviv University}, \orgaddress{\city{Tel Aviv}, \country{Israel}}}

\affil[6]{\orgdiv{Research Center for the Early Universe}, \orgname{University of Tokyo}, \orgaddress{\city{Tokyo}, \country{Japan}}}

\affil[7]{\orgdiv{Racah Institute of Physics}, \orgname{The Hebrew University of Jerusalem}, \orgaddress{\city{Jerusalem}, \country{Israel}}}

\affil[8]{\orgname{Observatories of the Carnegie Institution for Science}, \orgaddress{\city{Pasadena}, \state{CA}, \country{USA}}}




\abstract{\textbf{Many of the most luminous extragalactic transients originate from the accretion of material onto a black hole (BH) via core-collapse \cite{gal-yam2012}, stellar mergers \cite{tsuna+lu25}, or the tidal disruption of a star \cite{rees+88}. Some produce energetic multi-wavelength emission, displaying short, blue optical flares \cite{perley+19a, yao+23, ho+23} and bright, long-lived radio afterglows \citep{ho+19, Zauderer+11, andreoni+22}. In rare cases, these have also launched powerful relativistic jets \citep{andreoni+22, Burrows+11} almost exclusively detected on-axis via high-energy emission. Here we report AT 2019ijn, a radio transient discovered in the Very Large Array Sky Survey, associated with a powerful relativistic jet viewed off-axis and accompanied by a luminous ($\nu L_{\nu, \rm{opt}} \sim 10^{44}$~erg/s) optical transient. Originating from a star-forming dwarf galaxy at $z = 0.273$, AT 2019ijn’s optical flare exhibited a fast rise ($t_{\rm{rise, rest}} \sim 7~\rm {d}$) and shallow decline ($t_{\rm{dec, rest}} > 38~\rm {d}$), followed a year later by an energetic ($E \sim 2\times10^{52}$~erg) radio counterpart persisting for $>6$ years. These combined properties are unprecedented and preclude known classes of stellar explosion powered by core-collapse. Instead, the implied accretion onto a BH and associated off-axis jet invokes an exotic BH-stellar merger \citep{tsuna+lu25, klencki+metzger25_axv} or a jetted tidal disruption event (TDE) \citep{teboul+23} by a $10^4 - 10^6~\rm{M_\odot}$ black hole. Favoring a TDE, this work determines the jet-launching occurrence, with the low redshift allowing constraints on the late-time evolution in contrast to prior events \citep{andreoni+22, Burrows+11}. Upcoming radio sky surveys will offer an unprecedented new window for further discovery of AT 2019ijn-like off-axis relativistic transients. 
}}

\maketitle
\section{Main}\label{sec:main}

We conducted a search for radio transients in the Very Large Array Sky Survey (VLASS) \citep{lacy+20}, with no detected counterpart in the PanSTARRS survey, to identify luminous and potentially jetted extragalactic transients with a large disparity ($> 1000\times$) between the radio spectral luminosity of the transient and the optical luminosity of its host. During this search, we discovered VLASS transient J131240+211339 (abbreviated as VT 1312+2113), which was undetected at $\sim 3~\rm{GHz}$ at a $3\sigma$ upper limit of 0.4 mJy in VLASS Epoch 1.1 on September 25, 2017, then detected in VLASS Epoch 2.1 on July 21, 2020 at a flux density of 6.2 mJy. It then faded to a 3 GHz flux density of 3.8 mJy in VLASS Epoch 3.1 on January 23, 2023 and 2.3 mJy in VLASS Epoch 4.1 on October 17, 2025. The source had no detected counterpart in the PanSTARRS-1 survey down to a $5\sigma$ limiting magnitude of $m_g \sim 23.3$. A cross-match of this source with major transient surveys revealed an optical transient, AT 2019ijn, co-located with the radio source ($0.2'' \pm 0.2''$ offset between the optical and radio positions) and announced by the Zwicky Transient Facility (ZTF) 
on June 24, 2019 \citep{Nordin+19}, approximately one year prior to the radio detection. The probability of chance association between these sources is $\sim 10^{-8}$ (see Methods) and we therefore determine these to be the same event and hereafter refer to VT 1312+2113 as AT 2019ijn. 

\begin{figure}
\centering
\includegraphics[width=1.1\textwidth]{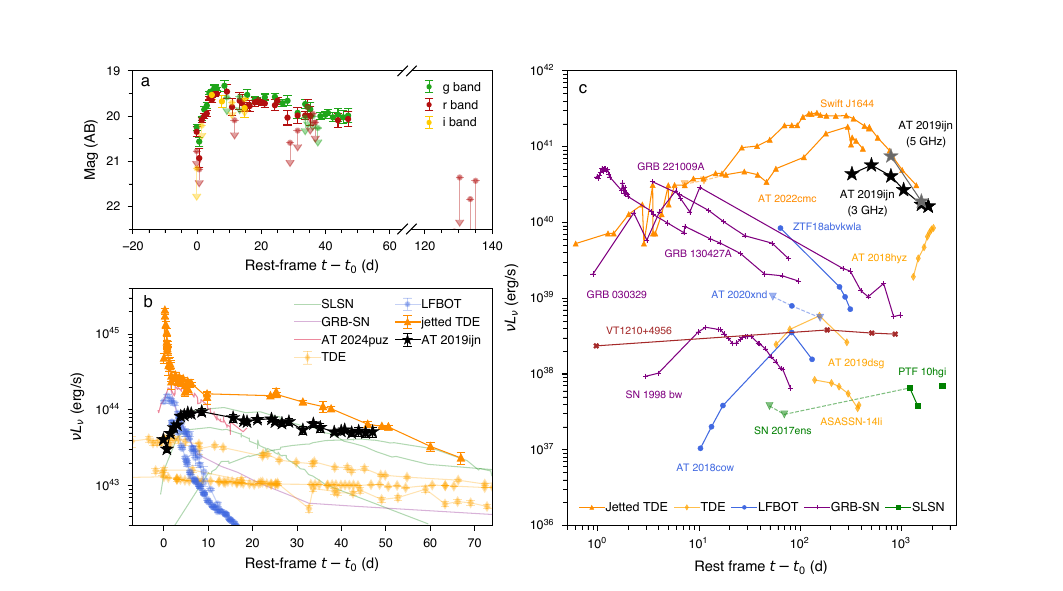}
\caption{\small \textbf{AT 2019ijn is a luminous optical and radio transient. a,} g-, r-, and i-band magnitudes obtained from ZTF forced photometry from May 11, 2019 to October 18, 2019. Data binned by day, upper limits marked with arrows. Optical emission for AT 2019ijn is detected in the ZTF photometry from May 31, 2019, to July 30, 2019. Break in time axis corresponds to when ZTF photometric data becomes unavailable, though extrapolating from the rate of decline, it is likely that the optical transient remains detectable beyond this end date. Data are corrected for Galactic extinction. \textbf{b,} comparison of g-band light curve to other luminous transients in the literature (see Section \ref{sec:figs}). Data are corrected for Galactic extinction with k-corrections applied. \textbf{c,} comparison of AT 2019ijn to radio light curves of other energetic transients at $< 10~\rm{GHz}$, adapted from \cite{ho+19}. 3 GHz (black) and 5 GHz (gray) measurements shown for AT 2019ijn for best comparison with jetted TDEs. Other light curves taken from the literature; see Section \ref{sec:figs}. }\label{fig:LC_panel}
\end{figure}

The optical source, as seen by ZTF, was initially detected on May 31, 2019 ($t_0$), approximately one year before the first VLASS detection (Figure \ref{fig:LC_panel}). The event had an initial magnitude of $m_g = 20.4 \pm 0.1$ and rose to a peak magnitude of $m_g = 19.4 \pm 0.1$ from half-maximum within 8.8 days in the observer frame. This sharp rise is followed by a shallower decline that can be fit by an exponential decay with a timescale of $\tau_{\rm{dec}} = 73 \pm 3$ days, though the light curve only remains visible for a further $\sim 50$ days after peak before further data is unavailable due to observing constraints. This gives a total visible optical transient duration of $\sim 59$ days in the observer frame. The g-r color of the transient at peak light is $g-r = -0.12 \pm 0.07$, indicating a blue transient that becomes slightly redder ($\frac{d(g-r)}{dt} = 0.001 \pm 0.005$ mag/d) throughout the decline of the optical emission (see Methods). 

\begin{figure}
\includegraphics[width=1\textwidth]{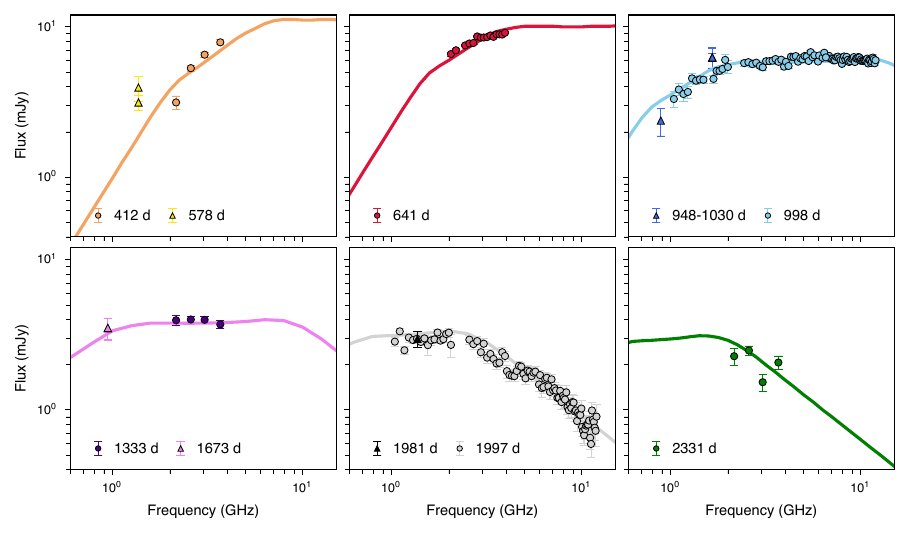}
\caption{\small \textbf{AT 2019ijn produced a relativistic jet viewed off-axis. a,} Observed VLA data shown, including VLASS in-band (2-4 GHz) spectra (412 days post-$t_0$, gold; 1333 d, indigo; 2331 d, green), VLA single-band (2-4 GHz) follow-up (641 d, red) and VLA multi-band (1-12 GHz) spectra (998 d, blue and 1997 d, gray). Spectra are divided into sub-bands for greater spectral resolution and off-axis jet models overlaid (see Methods section \ref{sec:radenergy} for derivation and section \ref{sec:figs} for specific model parameters shown). The jet is assumed to launch simultaneously with the optical transient. We stress that this is not the only possible set of parameters that can replicate the observed data (e.g., the jet may be launched at later times), nor is the underlying physics fully captured in this model (e.g., the actual jet spreading); rather, we demonstrate that an off-axis jet model can reproduce the observed radio spectra for all epochs with reasonable parameters. 
}\label{fig:radiojet}
\end{figure}

We obtained radio follow-up with the Karl G. Jansky Very Large Array (VLA) \citep{Perley+11} over three epochs, spanning 2021 - 2024, i.e., 2-5 years after first detection by ZTF (Figure \ref{fig:radiojet}). The multi-band data, ranging from 1-12 GHz, show significant evolution over this three-year period. Most notably, the radio spectrum in 2022 (998 d post-$t_0$ in the observer frame) was observed to be fairly flat ($F_{\nu} \propto \nu^{-0.2\pm 0.1}$; Extended Data Figure \ref{fig:radspec}) in the optically-thin regime across a significant fraction of the observed frequencies, evolving at later times to an optically-thin spectral index of $F_{\nu} \propto \nu^{-1\pm 1}$ in 2024 (1997 d).

We conducted a search for archival data in radio and X-ray bands. Additional radio observations were taken from 2019 to 2024 by the Australian Square Kilometer Array Pathfinder (ASKAP) as part of the Rapid ASKAP Continuum Survey (RACS). In X-ray, a $3\sigma$ upper limit of $F_X < 1.4\times10^{-13}~\rm{erg/s/cm^2}$ in the 0.2-5 keV band was produced from eROSITA observations of the source location approximately 6 months after the ZTF detection of the optical transient and $\sim 8$ months before the initial radio detection with VLASS.

Although undetected in PanSTARRS, a $5 \sigma$ detection of an optical source at the location of the transient is present in the Dark Energy Spectroscopic Instrument (DESI) Legacy survey in g, r, and i bands at de-reddened magnitudes of $m_g = 24.5 \pm 0.2$, $m_r = 23.8 \pm 0.2$, and $m_i = 23.3 \pm 0.4$. The counterpart is additionally detected in z band at $2.6\sigma$ at a magnitude of $m_z = 23.7 \pm 0.4$. The g, r, and z detections were made pre-discovery of the optical transient, with the most recent detections in 2018 in g and r bands and 2017 in z band. The i-band detections are from 2021, 2 years post-transient, and are not considered in our characterization of the Legacy source due to possible transient contamination. The presence of a faint optical source preceding the transient indicates a potential host galaxy for AT 2019ijn. 

\begin{figure}
\centering
\begin{subfigure}[b]{1\textwidth}
\includegraphics[width=1\textwidth]{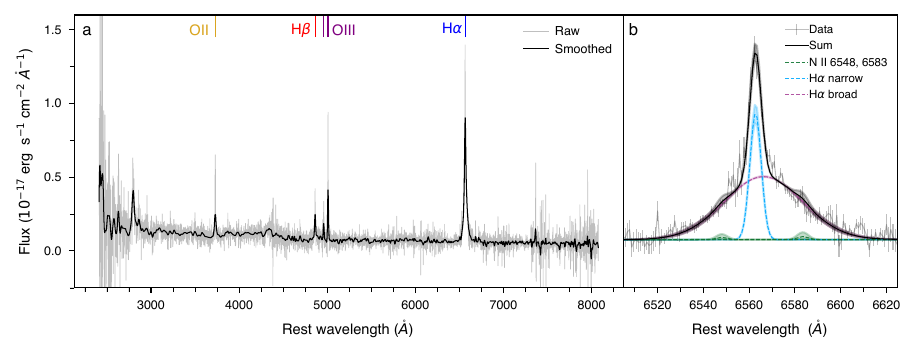}
\end{subfigure}
\hfill
\begin{subfigure}[b]{1\textwidth}
\includegraphics[width=1\textwidth]{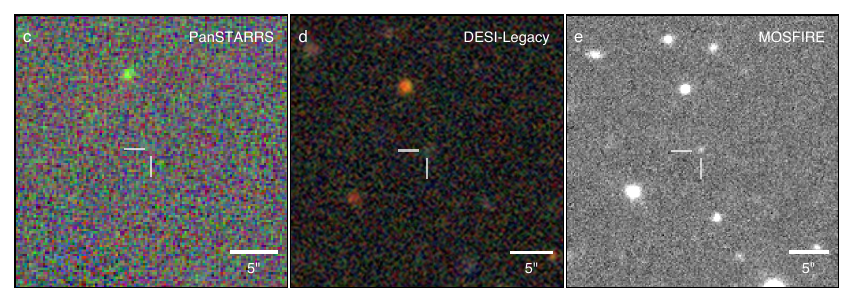}
\end{subfigure}
\caption{\small \textbf{AT 2019ijn is located in a star-forming host galaxy at $z = 0.273$. a,} optical spectrum of host galaxy from $\sim 3000-8000 \mathring{A}$. The smoothed spectrum, produced by convolving the spectrum with a Gaussian kernel with a standard deviation of $5~\mathring{A}$, is overlaid on the raw spectrum for clarity of spectral lines. Visible lines are labeled, and are consistent with a star-forming galaxy at a redshift of $z = 0.273$. \textbf{b,} H$\alpha$ complex, including broad and narrow H$\alpha$ components and NII 6548, 6583 lines. The gaussian best-fit to each line and sum of individual line contributions are also plotted (see Methods). \textbf{c,} PanSTARRS image of the location of the radio transient. \textbf{d,} DESI-Legacy image of the location of the transient. The host galaxy is faintly visible in the image, unresolved. \textbf{e,} MOSFIRE image of the host galaxy.}\label{fig:optspec}
\end{figure}

We conducted follow-up observations, including optical spectra and near-infrared (NIR) Ks-band images, of the DESI-Legacy counterpart. A number of spectral lines were detected in the optical spectra of this counterpart (Figure \ref{fig:optspec}), including Balmer lines H$\alpha$ $6563~\mathring{A}$ and H$\beta$ $4861~\mathring{A}$, and [OIII] $4959, 5007~\mathring{A}$.  The [OII] doublet at $3727, 3729~\mathring{A}$ and the [MgII] doublet at $2796, 2803~\mathring{A}$ were clearly detected in the latest spectrum at 2160 d. These lines are consistent with a galaxy at redshift $z = 0.2729 \pm 0.0004$, corresponding to a luminosity distance of $D_L \sim 1400~\rm{Mpc}$. 



The H$\alpha$ line displays both a broad and a narrow component, with a narrow-component luminosity of $L_{H\alpha} = (1.6 \pm 0.2)\times10^{40}$ erg/s, consistent with host-galaxy star formation at a rate of $\rm{SFR} \leq 0.12 \pm 0.01 ~\rm{M_\odot/yr}$, assuming no transient or active galactic nucleus (AGN) contamination. The broad component has a luminosity of $L_{H\alpha} = (4.6 \pm 0.2)\times10^{40}$ erg/s and a full width at half-maximum of $1900 \pm 100$ km/s.  All other detected emission lines are single-component narrow lines consistent with star formation. Using strong-line ratios, we determine this host to be of sub-solar metallicity at $12 + \rm{log}(O/H) \sim 8.572 \pm 0.001$. 

In the NIR, the counterpart was detected in a 3150s MOSFIRE image at a magnitude of $m_{Ks} = 21.2 \pm 0.2$. The combination of optical color and Ks-band magnitude, chosen for its relative insensitivity to extinction, implies a host galaxy mass of $M_* \leq (9 \pm 5) \times10^8~M_\odot$ and specific SFR of $\geq 1.4 \times10^{-10}~\rm{yr^{-1}}$ (limits due to possibility of transient contamination; see Methods).

\begin{figure}
\centering
\includegraphics[width=0.5\textwidth]{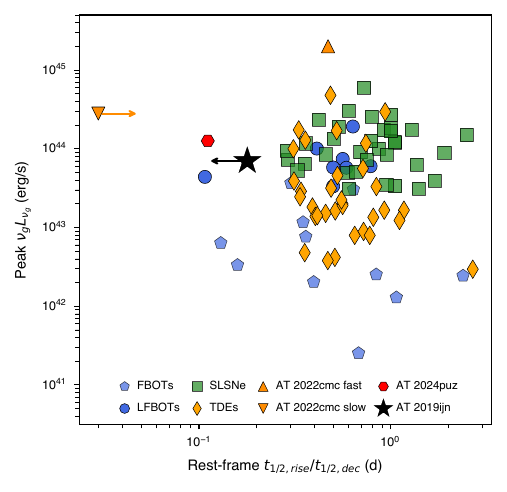}
\caption{\small \textbf{The optical flare of AT 2019ijn is unusually asymmetric compared to other luminous optical transients.} Comparison of peak g-band optical luminosity vs. ratio of rise time to decline time, defined as time from half-max to peak, for energetic extragalactic transients in the literature (see Methods section \ref{sec:figs}), partially adapted from \cite{ho+22b, andreoni+22}. AT 2019ijn (black star) shown with an arrow to account for lower limit on decline time. SLSNe shown in green, TDEs shown as orange diamonds, FBOTs/LFBOTs shown in blue. Jetted TDE AT 2022cmc shown as orange triangles, with upward triangle for the fast initial flare from the jet and a downward triangle for the slower, thermal emission \citep[see][]{andreoni+22, hammerstein+26}. Rise time for the thermal component is shown as a lower limit. AT 2024puz, an energetic, TDE- or FBOT-like transient hosted by a dwarf galaxy, shown in red. Luminosities are k-corrected according to Section \ref{sec:kcorr}.}\label{fig:phase}
\end{figure}

The redshift allows us to constrain the peak, k-corrected (see Methods) radio luminosity to $\nu L_{\nu_{\rm{peak, rad}}} \sim 1.8\times10^{41}$ erg/s, indicating an uncommonly luminous radio transient. The relatively flat synchrotron spectra over multiple epochs may indicate broad spectral peaks, which have been seen in some supernovae \citep[e.g. SN 2019oys;][]{sfaradi+24} and attributed to inhomogeneities in the circum-explosion medium, or in AGN, where the flatness has been attributed to the superposition of multiple synchrotron-emitting regions \citep[e.g.][]{falcke+00, nagar+02}. As seen in AGN and X-ray binaries, flat synchrotron spectra may also be indicative of a relativistic jet \citep[e.g.][]{fender+01, fender06, kaiser06}.

We attempt to determine the physical origin of the radio emission via detailed modeling of the radio spectra. The radio emission of many energetic transients can be modeled by a spherical blastwave \citep[e.g.][]{chevalier98}, from which the radio is produced by synchrotron emission from the forward shock as the front of the blastwave interacts with the surrounding medium. In the case of AT 2019ijn, the flat spectra seen at 998 d and 1997 d sets more observational constraints than the free parameters of the spherical blastwave model, and there is no spherical solution that can easily fit the late-time observations (see Methods). 
However, the observed spectra in all epochs are well-reproduced at the observed timescales by a single narrow, off-axis jet with an energy of $E \sim 2\times 10^{52}~\rm{erg}$ propagating in a density profile that drops slightly faster than $r^{-2}$ (Figure \ref{fig:radiojet}). In the absence of early-time data, we cannot constrain the exact launch time and velocity of the jet, but this model allows for an outflow launch at any time between the optical flare and the first radio observation at 412 d. At the observation at 1997 d, the jet would be expected to be mildly relativistic with $\Gamma \sim 3$ and have reached a radius of $R\sim 10^{19}~\rm{cm}$.

This therefore determines AT 2019ijn as a transient co-located with a star-forming dwarf galaxy and characterized by a short, luminous optical flare and extremely energetic, long-lasting radio emission due to an off-axis jet. The combination of these properties is unique relative to any transient discovered to date and restrict consideration to the phase-space occupied by the most extreme transient classes. 

While perhaps intuitive to invoke a long gamma-ray burst (LGRB) as an example of a jetted multi-wavelength transient, we quickly disfavor this interpretation for AT 2019ijn, as it is absent confirmation of high-energy emission and the optical supernova (SN) signature expected of this class, while also displaying a radio afterglow that is years longer and $10-100\times$ more luminous than on-axis LGRB radio afterglows observed at late times (e.g. GRB 030329, see Figure \ref{fig:LC_panel}c). The radio emission is also difficult to reconcile with superluminous supernovae (SLSNe), though thought to be able to produce jets in some cases \citep{gottlieb+metzger24, desai+26_axv}. Long-lasting radio emission has been detected for a small number of SLSNe, but these have produced emission at a luminosity $>10^3\times$ fainter than AT 2019ijn \citep{eftekhari+19, margutti+23}. 

The properties of the optical flare present an additional challenge, as they cannot be easily aligned with those observed from known energetic transients to date. The high peak absolute magnitude ($M_{\rm{g,rest}} = -21.1 \pm 0.1$) and luminosity ($\nu L_{\nu_{\rm{peak, opt}}} \sim 7.8\times10^{43}$ erg/s), along with the blue color at peak light, have been seen in prior SLSNe \citep{perley+20}, luminous fast blue optical transients (LFBOTs) \citep{ho+23}, and some tidal disruption events (TDEs) \citep{yao+23}. However, the short optical rise time (6.9 d in the rest frame) is highly atypical for SLSNe and classical TDEs (Figure \ref{fig:phase}). By contrast, short rise times are common for LFBOTs and events suggested as extensions of this class (e.g. AT 2024puz \citep{somalwar+25}, Dougie \citep{Vinko2015}), but these fade faster than the shallow decline seen in AT 2019ijn (Figures \ref{fig:LC_panel}b, \ref{fig:phase}), and may cool at late times \citep{Vinko2015}, which is not seen in AT 2019ijn. The broad H$\alpha$ line, similar to those commonly seen in SNe and LFBOTs \citep{yan+15, inserra+18, perley+19a} and also seen in some TDEs \citep{somalwar+25b}, implies a rich surrounding medium, and likely indicates outflowing material that is still interacting with this medium at late times.

The presence of the jet implies a central engine for AT 2019ijn. Some energetic transients are thought to be powered by the spin-down of a millisecond magnetar \citep{metzger+15, margalit+18} following core-collapse of a massive star, which may produce a jet, though current models predict shorter and less luminous emission than that seen in AT 2019ijn \citep[][see Methods]{Gottlieb2022, gottlieb+metzger24, desai+26_axv}. To date, jets of this power have only been seen in transients thought to originate from massive (nominally $10^6 <M_{\rm{BH}} (\rm{M_\odot}) < 10^8$) black holes (BHs) \citep[e.g.][]{Burrows+11, andreoni+22}. For AT 2019ijn, we obtain a constraint on a possible BH mass from $M_* - M_{\rm{BH}}$ relations of galaxies in the mass range $10^8 < M_* (\rm{M_\odot}) < 10^{12}$ \citep{Greene+20}, which suggest that the host could support a BH of mass $1.1\times10^4~\rm{M_\odot} < M_{\rm{BH}}< 7.3\times10^5~\rm{M_\odot}$ (for a more detailed discussion, see Methods). 

A BH progenitor further restricts the landscape of possible transient scenarios for AT 2019ijn. Common-envelope events involving the inspiral, merger, and subsequent tidal disruption of a stellar companion by a neutron star or BH have been suggested for energetic transient classes such as LFBOTs \citep{metzger+22a, tsuna+lu25, klencki+metzger25_axv, nugent+26_axv}, with key components being their energetic and quickly-evolving optical emission and a dense, extended circumstellar medium (CSM). These models can reproduce the optical color, luminosity, and quick rise seen in AT 2019ijn, and may produce sufficiently energetic jets in the case of a BH progenitor. Current models are tuned to the observational characteristics of LFBOTs, resulting in predictions of quickly-declining optical emission and shorter, less energetic radio emission. However, the broad H$\alpha$ line, implied dense interior CSM, and star-forming host galaxy are well-aligned with predictions of these models. An extension of this class could therefore be a possible explanation for AT 2019ijn. 

Perhaps a more natural explanation associates AT 2019ijn with a TDE accompanied by a relativistic jet. Observations of TDEs, currently detected from BHs of masses $10^5 \lesssim M_{\rm{BH}} (\rm{M_\odot}) \lesssim 10^8$, imply that only $\sim 1\%$ launch relativistic jets \citep{teboul+23}, with very few candidates presently observed \citep{Burrows+11, cenko+12, brown+15, andreoni+22}. In the lower-mass BH regime, however, expectations for super-Eddington accretion as a function of mass ($\frac{\dot{M}_{\rm{peak}}}{\dot{M}_{\rm{Edd}}} \sim M_{\rm{BH}}^{-3/2}$) favor multiple types of luminous outflows, including these jets \citep{strubbe+quataert09, stone+metzger16, Dai+18, teboul+23}. 

AT 2019ijn does bear remarkable similarities to prior jetted TDE candidates, the most well-known being Swift J1644 \citep{Burrows+11, Zauderer+11}, and most notably, the more recent, optically-selected candidate AT 2022cmc \citep{andreoni+22}. Both events have produced long-lasting, energetic radio emission (Figure \ref{fig:LC_panel}c) with a low-density medium similar to those inferred from our modeling at comparable radii \citep{eftekhari+18, rhodes+25}. The delayed onset of the radio emission in AT 2019ijn as compared to AT 2022cmc and Swift J1644 would be explained by the off-axis viewing angle, as the jet only becomes visible after decelerating at late times. Delayed radio flares have also been seen in non-relativistic TDEs \citep{horesh+21a, horesh+21b}, and may indicate the presence of off-axis jets in prior events (e.g. AT 2018hyz, a TDE now thought to have an off-axis jet; see Methods). 

AT 2022cmc's optical counterpart, the most well-studied of a jetted TDE to date, featured a fast, red optical flare, attributed to emission from the on-axis jet, followed by a slower, blue component attributed to thermal emission \citep{andreoni+22}. Though AT 2019ijn does not display this initial flare, consistent with an off-axis viewing angle, the optical decline of AT 2019ijn resembles the thermal component of AT 2022cmc, as well as that of some non-jetted TDEs (Figure \ref{fig:LC_panel}b). Though the short optical rise preceding this shallow decline has not been seen in a TDE scenario, jetted or otherwise, the similarities in peak luminosity and color may indicate that the optical emission can be explained as an extension of current TDE predictions (see Methods). We also note that the rise of the thermal phase in AT 2022cmc's optical light curve is not well-constrained. Broad H$\alpha$ features have not yet been identified in prior jetted TDE candidates, though differences in observing angle, lack of late-epoch observations, and lower sensitivity in observations of prior candidates may explain this absence.

The detected population of TDEs at present has been dominated by supermassive black holes (SMBHs) with hosts more massive and less star-forming than for AT 2019ijn, though the implied host galaxy mass is similar to that estimated for Swift J1644 \citep{yoon+15}, and may also resemble the host of AT 2022cmc, which has yet to be detected \citep{andreoni+22} and may be another atypical example. While the BH masses for both prior events were initially estimated at $> 10^6~\rm{M_\odot}$ \citep{Burrows+11, andreoni+22}, recent studies have suggested BH masses for both events in the range of $\sim 10^5~M_\odot$ \citep{Eftekhari2024}, which agree with our estimate for AT 2019ijn. 



We note that alternate interpretations have been proposed for jetted TDE candidates, such as a central engine-powered transient following a core-collapse SN for AT 2022cmc, wherein the engine deposits half its energy into heating of the SN ejecta and half into the relativistic jet \citep{matsumoto+metzger23}. The commonalities between AT 2019ijn and AT 2022cmc do not allow us to exclude this possibility from AT 2019ijn, and instead suggest that identifying AT 2022cmc as an exotic class of stellar explosion may necessarily encompass both events. 

We conclude that AT 2019ijn is most likely a BH-powered transient with a relativistic jet and a strong off-axis analog to AT 2022cmc. This most naturally explains AT 2019ijn as a jetted TDE from a low-mass BH, though it may also suggest a novel extension of current BH-powered transient models. Though unique, the identification of AT 2019ijn through radio surveys is not overly surprising, as surveys such as VLASS and that of the upcoming Deep Synoptic Array (DSA) are particularly well-suited to the detection of relativistic jets both on and off-axis. These surveys will therefore be crucial in expanding the population of AT 2019ijn-like transients and discriminating between possible interpretations of these events.

\section{Methods}\label{sec:methods}

As seen in VLASS, AT 2019ijn is located at a Right Ascension of 13h12m40.54s and a Declination of +21d13m39.62s, with a positional uncertainty of 0.2''. We use the VLASS location of the source throughout the paper, which is 0.16'' offset from the reported ZTF location (positional uncertainty 55 mas). All analysis in this paper assumes a flat $\Lambda\rm{CDM}$ cosmology with $h = 0.696$, $\Omega_M = 0.286$, and $\Omega_{\Lambda} = 0.714$ \citep{Bennett+14}. When describing the temporal and spectral behavior of the emission in this paper, we adopt the convention $F_{\nu} \propto t^{-\alpha}\nu^{-\beta}$.

\subsection{Archival data}\label{sec:archive}

\subsubsection{Radio}\label{subsec:rad}
AT 2019ijn was first detected in radio with the Karl G. Jansky Very Large Array Sky Survey (VLASS) \citep{lacy+20} on July 21, 2020, during VLASS Epoch 2.1 (Extended Data Figure \ref{fig:discovery}). This location was also observed during the three other VLASS Epochs:  on July 25, 2017 in Epoch 1.1;  on January 23, 2023 in Epoch 3.1, and on October 17, 2025 in Epoch 4.1. These observations were made in S-band (2-4 GHz), and the fluxes were extracted using the PyBDSF package \footnote{Mohan \& Rafferty 2015, \url{https://ui.adsabs.harvard.edu/abs/2015ascl.soft02007M}}, which fits a Gaussian to identified sources of emission and computes the source flux and background root-mean-squared error (RMS). VLASS reports a non-detection of this transient at a 3$\sigma$ threshold of $F_{\nu} = 0.4$ mJy in Epoch 1.1, followed by three detections of the transient at a flux density of $F_\nu = 6.2 \pm 0.2$ mJy in Epoch 2.1, $F_\nu = 3.8 \pm 0.1$ mJy in Epoch 3.1, and $F_\nu = 2.3 \pm 0.1$ mJy in Epoch 4.1.

Data from RACS \citep{mcconell+20,hale+21,duchesne+23,duchesne+24,duchesne+25} was obtained from the CSIRO ASKAP Science Data Archive (CASDA; \url{https://research.csiro.au/casda/}) and source fluxes were again extracted using PyBDSF. The location of AT 2019ijn was observed in the RACS-low bands (888 MHz,  943 MHz) on March 26, 2022 (1030 d; observer frame) and December 28, 2023 (1673 d) 
We note that the data quality of the observations at 943 MHz are flagged as ``Uncertain'' due to above-threshold errors in positional offset, spectral index, and RMS. All other observations were flagged as ``Good'' quality. The location was observed and detected again with the RACS-mid band (1368 MHz) on December 28, 2020 (578 d) and November 1, 2024 (1981 d) and the RACS-high band (1656 MHz) on January 2, 7, and 10, 2022 (948-956 d). 
The measured flux densities from both the VLASS and RACS observations are given in Supplementary Table 1 and plotted in Extended Data Figure \ref{fig:radioLCs}. 

The Low-Frequency Array (LOFAR) observed this location as part of the LOFAR Two-metre Sky Survey (LoTSS), which contains observations from 2014-2024 at a central frequency of 144 MHz \cite{vanhaarlem+13, shimwell+19, shimwell+22, shimwell+26_axv}. LOFAR pointings near the location of AT 2019ijn were taken in 2019 and 2021. The source is not detected down to the $1\sigma$ median RMS sensitivity of 92 uJy/beam. 

\subsubsection{Optical}\label{subsec:opt}
AT 2019ijn was first announced by the Zwicky Transient Facility (ZTF) \citep{Bellm+19, Graham+19, Dekany+20, Masci+19} on June 24, 2019 \citep{Nordin+19}. The initial report contained two photometric detections in $g$ and $r$ bands at $g = 19.8 \pm 0.1$ and $r = 19.8 \pm 0.2$, with no official classification. Forced photometry was obtained from the ZTF forced-photometry service \citep{Masci+19} for the full duration of the survey from March 17, 2018 to present and processed via recommended procedures (\url{https://irsa.ipac.caltech.edu/data/ZTF/docs/ztf_forced_photometry.pdf}). The transient emission was detected with ZTF from May 31, 2019 to July 30, 2019. We adopt the detections on May 31, 2019 as the date of initial detection $t_0$ hereafter. The ZTF light curve is given in Supplementary Table 2. Hereafter, optical magnitudes are corrected for Galactic extinction with $R_{\rm{V}} = 3.1$ \citep{fitzpatrick99} and $E(B-V) = 0.03$ \citep{schlegel+98}, with extinction coefficients given by \cite{Schlafly2011}. Host extinction is considered to be negligible.

The forced photometry light curves include data in the g, r, and i bands. For ease of comparison with other TDE-like transients, the g and r data were binned by day and fit with a Gaussian rise and exponential decay, following the methods of \cite{vanvelzen+21, yao+23}: 
\begin{equation}
    L_{\nu}(t) = L_{\nu,g}\frac{B_\nu}{B_{\nu,g}} \times \begin{cases}
    e^{-(t-t_p)^2/2\sigma^2} & t \leq t_p \\
    e^{-(t-t_p)^2/\tau} & t > t_p
    \end{cases}
\end{equation}
where $L_{\nu,g}$ is the peak g-band luminosity, $B_{\nu, g}$ is the g-band blackbody flux, $B_{\nu}$ is the blackbody flux in the observed band, and $t_p$ is the time of peak luminosity in the light curve.

We obtain $B_{\nu}$, $B_{\nu, g}$ by fitting a blackbody to the optical spectral energy distribution (SED). Due to the scarcity of i-band photometric measurements in the ZTF light curve, we are only able to obtain an SED in all three bands at three epochs near the peak of the optical emission - one at 4 d (rest-frame) prior to the peak, one at peak, and one at 8 d post-peak. The fitting was done using a Markov-chain Monte Carlo (MCMC) method with the EnsembleSampler in the \texttt{emcee} Python package \citep{Foreman-Mackey+13} assuming a Gaussian likelihood and using 50 chains and 5000 steps, with the first 1/3 of steps discarded to ensure sufficient convergence. We adopt conservative uniform priors with $\rm{log}~T \in \left(3, 6\right)~\rm{K}$, $\rm{log}~R \in \left(10, 20\right)~\rm{cm}$. We fit these three-point SEDs with a blackbody function to derive the blackbody temperature $T_{bb}$, the photospheric radius $R_{bb}$, and the blackbody luminosity $L_{bb}$. At peak, we obtain values of $\rm{log}~T_{bb} = 4.2 \pm 0.1~\rm{K}$ and $\rm{log}~R_{bb} = 15.3 \pm 0.1~\rm{cm}$, which corresponds to a blackbody luminosity of $\rm{log}~L_{bb} = 44.4 \pm 0.6~\rm{erg/s}$, with emission on the Rayleigh-Jeans tail of the blackbody (Extended Data Figure \ref{fig:models}b). These values are consistent with those found for many optical TDEs in \cite{yao+23}. We assume these values for further analysis and light curve fitting. 

We then use the fitted blackbody temperature to compute the normalization of the the light curve fitting function. We again use an MCMC fitting routine with the same parameters as above and conservative uniform priors of $\sigma \in \left(0.01, 100\right)~\rm{d}$, $\tau \in \left(1, 100\right)~\rm{d}$, and $t_p \in \left(0.01, 50\right)~\rm{d}$. We obtain fit parameters in the observer frame as given in Extended Data Table \ref{tab:optLCfit}. In the rest frame, we find that the g band light curve peaks at $6.9\pm 0.4$ days from half-maximum, and is visible above half-maximum for $\geq 45$ days before the observing constraints. The r band peaks 1 day later than the g band, though is visible above half-maximum for the same duration. There are no photometric measurements in i-band during the rise, so we cannot accurately constrain the peak. The i-band light curve is first measured 4 rest-frame days before the g-band peak and remains above half-maximum for at least 10 days after the g-band peak, with the last i-band measurement of the transient occurring at 15 days after $t_0$.


\subsubsection{X-ray}\label{subsec:xray}
We do not find observations of an X-ray source within 0.5' of the location of AT 2019ijn in any publicly-available archival X-ray data, including Chandra \citep{evans+24}, Swift \citep{Gehrels+04,Burrows+05,roming+05}, or XMM-Newton \citep{jansen+01}. eROSITA \citep{predehl+21,merloni+24, tubinarenas+24} observed the location of the radio source between December 15-22, 2019, approximately 6 months after the ZTF detection of the optical transient and $\sim 8$ months before the initial radio detection with VLASS. Though there were no confirmed detections, the upper limit on the X-ray flux at the location of the source is $F_X < 1.4\times10^{-13} ~ \rm{erg/s/cm^2}$ in the 0.2-5 keV band, corresponding to a luminosity of $L_X < 3.2\times10^{43}$ erg/s. We note that there is a known quasar (SDSS J131243.17+211336.5) 36.6" from AT 2019ijn that does have archival X-ray detections with eROSITA.

We similarly do not find any $\gamma$-ray detections of the source in any publicly-available archival data, including Swift and Fermi \citep{atwood+09}. There were no recorded GRBs within 1' of AT 2019ijn within 1 year of the initial ZTF detection. 

To assess possible scenarios for X-ray emission from AT 2019ijn, we compute the expected X-ray emission from inverse Compton scattering by extrapolating from the optical and radio emission, as
\begin{equation}
    \frac{L_{\rm{X}}}{L_{\rm{rad}}} \sim \frac{u_{\rm{ph}}}{u_{\rm{B}}},
\end{equation}
where $u_{\rm{ph}}=\frac{L_{\rm{opt}}}{4\pi R^3/3}$ with $R$ defined as $R = \frac{L_{\rm{opt}}}{4\pi \sigma_{\rm{B}} T_{\rm{eff}}^4}$, and $u_{\rm{B}} = \frac{B^2}{8\pi}$. We assume $T_{\rm{eff}}\sim 2\times10^4~\rm{K}$ as is common for similar transients in the literature. 
At the time of the earliest radio detection (412 d post-optical transient, nearest to the eROSITA measurement at 204 d post-optical transient) the expected X-ray luminosity would be $L_{\rm{X}}\sim 1.5\times10^{43}~\rm{erg/s}$, which agrees with the eROSITA limit.


\subsubsection{Host galaxy}\label{subsec:host}
Aside from the DESI-Legacy \citep{Dey+19} detection, the host galaxy of AT 2019ijn was undetected in all bands in the PanSTARRS-1 Survey \citep{chambers+16, Waters+20, Magnier+20a, Magnier+20b, Magnier+20c, Flewelling+20} down to the PanSTARRS limiting magnitudes of $g, r, i, z, y = 23.3, 23.2, 23.1, 22.3, 21.4$ \citep{chambers+16}. 

We additionally cross-match with major infrared surveys including the Wide-field Infrared Survey Explorer (WISE) \citep{wright+10,skrutskie+06}, the Two-Micron All-Sky Survey (2MASS) \citep{skrutskie+06}, and the United Kingdom Infrared Telescope (UKIRT) Infrared Deep Sky Survey (UKIDSS) \citep{lawrence+07} and Hemisphere Survey (UHS) \citep{dye+18}. While the field containing the transient location was observed in WISE, 2MASS, and UHS, there was no source detected at the location of AT 2019ijn. The field containing the transient location was not observed by UKIDSS.


\subsection{False-alarm probability}\label{sec:false}
We examine the probability that the VLASS-detected radio transient and the ZTF-detected optical transient are associated by chance. The radio transient was identified using a catalog of VLASS radio sources from VLASS Epochs 1, 2, and 3.1, which was the most recent epoch at the time of catalog creation. These sources were undetected at a $3 \sigma$ threshold in Epoch 1, then detected at a $>7\sigma$ level in one or both of Epochs 2 and 3. This catalog consists of 3783 sources over the VLASS survey area of $\sim 34000 ~ \rm{deg}^2$, which is a density of 0.11 transients$~\rm{deg}^{-2}$. To determine if there were any associated optical transients detected in other surveys, we cross-match this catalog with the Transient Name Server (TNS) catalog of transients, searching for matches within 5'' of the radio source position. We find 41 radio transients with TNS counterparts as a result of this search. AT 2019ijn was reported as part of the TNS catalog of 153374 transients distributed over the whole sky ($41253 ~ \rm{deg^2}$), giving a density of TNS transients on-sky of $3.72~\rm{deg}^{-2}$. The probability that any TNS transient, over the entire survey, falls within the total area defined by a 5'' radius from any VLASS radio transient is therefore (TNS density) $\times$ (\# VLASS sources $\times$ area) $= 3.72~\rm{deg}^{-2} \times 3783 \times (6\times10^{-6}~deg^2) = 0.1$. In the single-object case, the reported optical and radio positions of AT 2019ijn have a separation of 0.16'', with a positional uncertainty of 0.2'' in VLASS and $\sim$ 55 mas in ZTF. The false-alarm probability that any TNS transient falls, by chance, within the region defined by a 0.16'' radius from the VLASS position of AT 2019ijn is $3.72~\rm{deg}^{-2} \times (6\times10^{-9}~deg^2) = 2.3\times10^{-8}$. We therefore determine the radio and optical sources to be associated. 

\subsection{Observations and data reduction}\label{sec:observations}

\subsubsection{Radio}\label{subsec:obsrad}
Follow-up observations of the VLASS source were obtained with the VLA on multiple occasions from March 2021 to November 2024. An observation in S-band was taken on March 2, 2021 as part of program 20B-393 (PI: Dong) during the switch between the A and D array configurations. These observations were reduced and calibrated with Common Astronomy Software Applications (CASA) \citep{casa+22} v6.6.1.17, using the source 3C286 as a flux calibrator, and J1327+221 as a phase calibrator. Fluxes of the source were extracted using PyBDSF. This measurement indicates that the radio transient brightened to an S-band peak of $F_{\nu} = 8.1 \pm 0.1$ mJy 641 d (observer frame) post-detection. The observation was additionally broken into sub-bands using independently-calibrated spectral windows within the band to improve spectral resolution.

Two multi-band observations in L-, S-, C-, and X-bands, spanning from 1-12 GHz, were taken on February 22, 2022 (998 d post-discovery), as part of program 21B-322 (PI: Hallinan) in the BnA configuration, and November 16, 2024 (1997 d post-discovery), as part of program 24B-448 (PI: Dong) in the A configuration. Observations were again reduced with CASA, broken into sub-bands by spectral window, and fluxes extracted using PyBDSF.  Due to high noise from the nearby quasar, self-calibration was employed for the L band observations at 1997 d, and fluxes were extracted with the CASA routine \texttt{imfit}. The complete set of flux density measurements is given in Supplementary Table 1.

After imaging and flux extraction, the radio spectra were fit with a synchrotron spectrum as given in \cite{granot+02}:
\begin{equation}
    F_\nu = F_{\nu_p}\left[\left(\frac{\nu}{\nu_p}\right)^{-s\beta_1} + \left(\frac{\nu}{\nu_p}\right)^{-s\beta_2}\right], 
\end{equation}
where $\nu_p$ is the frequency of the spectral peak, $\beta_{1, 2}$ are the slopes of the spectrum before and after the peak, and $s$ is a smoothness parameter. An MCMC fitting was performed using 50 chains with 5000 steps each, with the first $1/3$ of the steps discarded to ensure sufficient convergence. We use uniform priors of $\nu_p \in \left(0.1, 30\right)$, $s, \beta_1, \beta_2 \in \left(-3, 3\right)$, and $F_{\nu_p} \in \left(0.05, 20\right)$. The best fit for each epoch is shown in Extended Data Figure \ref{fig:radspec} and the parameters are given in Extended Data Table \ref{tab:radfit}.

The VLASS observations were additionally broken down into in-band spectra by Stokes I, Q, U, and V parameters in four sub-bands within the S-band (2.16 GHz, 2.58 GHz, 3.05 GHz, 3.69 GHz) and fit with the SSA model. Being single-band observations, the fit is unconstrained at other frequencies, and the spectral indices cannot be confidently stated. 

\subsubsection{Optical}\label{subsec:obsopt}
Three optical spectra of the host galaxy of AT 2019ijn were obtained with the Low Resolution Imaging Spectrometer (LRIS) \citep{oke+95, mccarthy+98} at the W. M. Keck Observatory, spanning a wavelength range of 3075~\r{A} - 10317~\r{A}. For all observations, the instrument was configured with the 1'' longslit in parallactic position, with the 400/3400 grism, 400/8500 grating, central wavelength 7830~\r{A}, and the 560 dichroic. This gives a resolution of $R\sim700$. The spectra were taken as a 2100s exposure on December 6th, 2023 (1650 d post-optical transient onset); a 900s exposure on December 1, 2024 (2011 d); and a 5400s exposure on April 28, 2025 (2160 d). The data were reduced using the \texttt{LPipe} reduction software \citep{Perley+19} and the PypeIt reduction software \citep{prochaska+20}. Reduced spectra were corrected for Galactic extinction prior to fitting \citep[as][]{fitzpatrick99} with $R_V = 3.1$ and $E(B-V) = 0.03$ \citep{schlegel+98}. A plot of all reduced spectra is shown in Extended Data Figure \ref{fig:alloptspec}. 

We determine the redshift by performing simple Gaussian fits to all detected emission lines in the highest-quality spectrum at 2160 d. The fits were again performed via MCMC, assuming a Gaussian likelihood and using 100 chains and 1500 steps, with the first 1/3 of steps discarded to ensure sufficient convergence. The redshift was taken as the mean of those determined from each of the individual fitted lines, which yielded a value of $z = 0.2729 \pm 0.0004$. All further quantities derived from the host optical spectra were derived from the spectrum at 2160 d. 

Detailed fits were then performed for all detected emission lines in the observed spectra following the MCMC procedure above and as in \cite{dong+21}. The H$\alpha$ line was fit as a complex with both broad- and narrow-line contributions from H$\alpha$ and narrow-line contributions from [NII] 6548, 6583; all others were fit as narrow lines only. The narrow lines were fixed at their rest wavelength, while the broad line center wavelength was allowed to vary. Widths of lines known to be doublets were tied during fitting. We adopt conservative uniform priors of $\lambda_{\rm{broad}} \in \left(\lambda_{\rm{rest}}-1000, \lambda_{\rm{rest}}+1000\right)$, $A \in \left(0, 10^{-15}\right)$, $\sigma_{\rm{narrow}} \in \left(\sigma_{\rm{inst}}, 2\sigma_{\rm{inst}}\right)$, $\sigma_{\rm{broad}} \in \left(\sigma_{\rm{inst}}, 0.015\lambda_{\rm{rest}}\right)$, where $\sigma_{\rm{inst}}$ is the $7~\rm\mathring{A}$ full-width half-maximum (FWHM) instrumental resolution. For each local fit, we also include a zero-point offset and wavelength shift for the spectrum, as well as the slope of the local continuum emission. The wavelength shift is estimated to contribute an uncertainty of $< 0.002$ to the redshift. The quality of the fits were evaluated using the Bayesian Information Criterion (BIC). The parameters of all fitted lines are given in Extended Data Table \ref{tab:optfit}. 

For the spectrum at 1650 d, the default reduction produced by \texttt{LPipe} did not produce a valid flexure correction and wavelength solution on the red side of the spectrum. We therefore applied a manual wavelength correction to the red spectrum using the known positions of the emission lines in the subsequent spectra at 2011 d and 2060 d. We then applied a manual rescaling to the red side before connection with the blue spectrum. We also note that there is a strong sky line present near the [OIII] 4959 line in the spectrum at 2011 d, which may have impacted the line fit during that epoch.

The spectra show primarily narrow lines, with the only broad feature present in the H$\alpha$ line in all epochs. The H$\alpha$ line is dominated by the broad feature at 1650 d, while the narrow feature becomes more apparent at later times. Due to the high noise in the earlier observations, the broad-line velocities of the H$\alpha$ lines in all epochs agree within error bars, and we therefore cannot confidently state any evolution in the broad line. Evolution of any other observed lines also cannot be confidently stated. 

The fitted spectrum at 2160 d shows a Balmer decrement of $\rm{H}\alpha_{\rm{narrow}}/\rm{H}\beta = 2.5 \pm 0.4$, similar to that expected from star-forming galaxies. To evaluate a constraint on the presence of a broad component in the H$\beta$ line, as many transients with broad H$\alpha$ components also show this feature in the other Balmer lines, we perform another MCMC fit wherein we fix the central wavelength of the narrow component to the value of H$\beta$ and the widths of both the broad and narrow components to the fitted values of the broad and narrow components in the H$\alpha$ line. The broad line centroid was fixed at the wavelength shift seen in the H$\alpha$ line, and amplitudes were set at an initial guess of the H$\alpha$ line amplitudes and allowed to vary freely. This process yields an upper limit on the luminosity of a broad H$\beta$ component of $L = (4.9 \pm 0.6)\times10^{39}~\rm{erg/s}$, which corresponds to a Balmer decrement of $\rm{H}\alpha_{\rm{broad}}/\rm{H}\beta_{\rm{broad}} = 9\pm 1$. This is consistent with observations of some radio-selected TDEs \citep{somalwar+25b}, and has also been seen in the spectra of interacting SNe \citep[e.g. SN 2019oys, $\sim 8.5$;][]{sollerman+20} and some AGNs \citep[see][and references therein]{wu+23}. The steep Balmer decrements in these cases are often attributed to high-density material in the emitting region.

The host can be placed on a Baldwin, Phillips \& Terlevich (BPT) diagram \citep{baldwin+81} to evaluate the line ratios in relation to AGN and typical star-forming galaxies. The host displays ratios of $\rm{[NII]}~6583/\rm{H}\alpha_{\rm{narrow}} = 0.02 \pm 0.02$ and $\rm{[OIII]}~5007/\rm{H}\beta = 1.8 \pm 0.3$, which places it within the region typically occupied by star-forming galaxies (Extended Data Figure \ref{fig:alloptspec}b). The values obtained from earlier spectra are also shown in Extended Data Figure \ref{fig:alloptspec}, though we caution that high noise and possible transient contamination may misrepresent the properties of the host galaxy at times earlier than 2160 d. 


\subsubsection{Near-infrared}\label{subsec:nir}
Near-infrared images of the host were taken on three separate occasions from January-May 2025. A 900s Ks-band image was obtained with the Wide Field Infrared Camera (WIRC) \citep{Wilson+03} on the 200-in telescope at Palomar Observatory on January 31, 2025. The data were reduced using wircPipe \citep{de+20}, which makes use of standard procedures for handling dark current, flat-fielding, sky-subtraction, and astrometric calibration. A 600s Ks-band image was obtained with the FourStar Infrared Camera \citep{Persson+13} on the 6.5-m Magellan telescope at Las Campanas Observatory on February 7, 2025, with data reduction performed using FourCLift \citep{kelson+14}. The source was not detected in either image. To estimate the magnitude level of the non-detection, we use forced photometry at the location of the source. An aperture with a 10-pixel radius (corresponding to 2.5'' for the WIRC images, 1.6'' for the FourStar images) was placed at the source location, and the sky background was estimated using a 20- to 25-pixel annulus around the source. The aperture counts were converted to magnitudes and calibrated using sources in the field with photometric measurements, flagged as highest quality, in 2MASS. 
The magnitude conversion was computed as $m = -2.5 \rm{log_{10}}(\frac{f_{\rm{samp}}}{f_{\rm{ref}}}) + m_{\rm{ref}}$, where $f_{\rm{samp}}$ is the sum of the counts in the aperture at the sampled location, $f_{\rm{ref}}$ is the sum of the counts in the aperture for the 2MASS reference source, and $m_{\rm{ref}}$ is the 2MASS-measured Ks-band magnitude of the reference source. To estimate the $5\sigma$ non-detection, we take the counts in the aperture to be $5 \times \rm{noise}$, where the noise is the standard deviation in the aperture times the aperture area. This computation yields a limiting magnitude of $m_{Ks} > 15.5$ for the WIRC image and $m_{Ks} > 16.8$ for the FourStar image. We note that both nights were affected by poor seeing and mild cloud coverage. 

A 3600s Ks-band image was taken with the Multi-Object Spectrometer for Infrared Exploration (MOSFIRE) \citep{mclean+10,mclean+12} on the 10-m Keck-I telescope at W. M. Keck Observatory on May 14, 2025. The data were reduced using mosPipe \citep{de+20}, adapted from wircPipe. We estimate the $5\sigma$ limiting magnitude of the source to be $m_{Ks} > 16.9$. However, due to the poor observing conditions on the MOSFIRE night contributing to high sky variability, the reduced data has an artificially high level of noise in the sky background.  We attempt to mitigate the high noise levels by randomly sampling the local background near the location of the host (within 36'' of the source location) with the same aperture/annulus combination as used in the prior images. The average of the computed magnitudes at the locations of the apertures is then taken as the magnitude level of the sky at the location of the source. Using this method, we estimate a magnitude limit at the location of the source of $m_{Ks} \geq 21.6 \pm 0.8$. 

A 3150s Ks-band image was taken with MOSFIRE on December 14, 2025, under improved observing conditions. The data were again reduced using mosPipe, revealing a clear detection of the source in the image. The source magnitude was obtained using traditional aperture photometry with a 9-pixel (1.6'') aperture, adjusted to match the zeropoint estimates derived in the data reduction pipeline, and a 20- to 25-pixel (3.6'' - 4.5'') annulus around the source. The derived instrumental magnitude was calibrated using three 2MASS sources in the field that are not saturated in the MOSFIRE image and have highest-quality photometry in the Ks band in the 2MASS catalog. We thus obtain a magnitude of $m_{Ks} = 21.2 \pm 0.2$ for the source. 

 \subsection{K-corrections to optical and radio luminosities and total radiated energies}\label{sec:kcorr}
All analysis, unless otherwise noted, was performed in the observer frame. However, we present rough cosmological k-corrections for the radio and optical luminosities for more accurate comparison with transients at other redshifts. 
In the radio, the k-correction was performed as 
\begin{equation}
L' = L(1+z)^{\alpha-\beta-1}
\end{equation}
\citep{bloom+01,chandra+frail12,ho+22b} where $\alpha_R = 0.98\pm 0.02$ is the temporal index, taken as the power-law slope of the declining 2-4 GHz light curve, while $\beta_R = -0.2 \pm 0.1$ is the spectral index, taken as the power-law slope of the optically-thin spectrum at the time of the peak luminosity, 998 d. This yields a k-corrected peak luminosity of $L_{\rm{R,peak}} = 1.8\times10^{41}~\rm{erg/s}$.

The k-correction on the peak absolute magnitude of the optical transient was estimated as $M_{\rm{g,rest}} = m_{\rm{g,obs}} - 5\rm{log_{10}} \left(\frac{D_L}{10~\rm{pc}}\right) + 2.5\rm{log_{10}}(1+z)$ following \cite{whitesides+17}, which may introduce errors of up to 0.1 mag. This gives peak absolute magnitudes of $M_{\rm{g, peak}} = -21.2$, $M_{\rm{r, peak}} = -21.0$, $M_{\rm{i, peak}} = -20.9$. This can be converted to a k-corrected peak luminosity as $L/L_{\odot} = 10^{0.4(M_{\odot} - M_{\rm{peak}})}$, which gives $L_{\rm{g,peak}} = 7.8\times10^{43}~\rm{erg/s}$, $L_{\rm{r,peak}} = 5.3\times10^{43}~\rm{erg/s}$, $L_{\rm{i,peak}} = 4.0\times10^{43}~\rm{erg/s}$. 

The host magnitudes were k-corrected by the same method, which yields optical absolute magnitudes of $M_{\rm{g,host}} = -16.0$, $M_{\rm{r,host}} = -16.7$, $M_{\rm{i,host}} = -17.2$, $M_{\rm{z,host}} = -16.7$, and a NIR absolute magnitude of $M_{\rm{Ks,host}} = -19.3$. 



We also determine the total radiated energy in both optical and radio bands. In radio, integration of the 3 GHz light curve (as $\nu L_{\nu}$), allows us to estimate the total radiated energy in the observed radio band to be to be $E_{\rm{rad, rad}} > 4.0 \times 10^{48}$ erg. Similarly, we compute a rough estimate of the bolometric optical light curve with the limited spectral data available by constructing three-point SEDs from the binned light curve and integrating to estimate the bolometric luminosity. Integrating this bolometric light curve allows us to estimate the total radiated energy in the optical bands as $E_{\rm{rad, opt}} > 1.6\times10^{50}~\rm{erg}$. 

\subsection{Dust echo as a source of IR emission}\label{sec:dust}
Some TDE-like transients have shown long-lived dust echoes, which typically appear as mid-IR (MIR) emission produced by the reprocessing of UV emission by dust surrounding the black hole progenitor \citep{vanvelzen+21b}. These can last for years post-optical/UV transient and appear as MIR luminosities in the range $L_{\rm{MIR}} \sim 10^{41}-10^{42}~\rm{erg/s}$. We therefore briefly evaluate the possibility that the NIR detection of the host of AT 2019ijn is partially due to a dust echo, rather than host galaxy emission alone. 

The emission from dust echoes can be approximated as a blackbody, expected to peak in the MIR at a frequency $\nu_{\rm{peak}} \simeq 9.3\times10^{13}~\rm{Hz} \left(T/1600~\rm{K}\right)$ \citep{lu+16}. As we only have one NIR measurement for AT 2019ijn, we cannot perform a proper blackbody fit to the IR emission. However, assuming a blackbody temperature of $T_{\rm{bb}} \sim 1600~\rm{K}$, the Ks-band measurement is consistent with this spectrum for a blackbody radius of $R_{\rm{bb}} \sim 1.5\times10^{16}~\rm{cm}$. The peak luminosity of the dust at the time of the Ks-band measurement can thus be taken as the peak of this blackbody, or $L_{\rm{dust}} \sim 6.5\times10^{41}~\rm{erg/s}$. This implies a dust covering fraction of $L_{\rm dust}/L_{\rm{bb, opt}} \sim 0.3\%$, taking $L_{\rm{bb, opt}} \sim 2\times10^{44}~\rm{erg/s}$ as the peak luminosity of the optical blackbody emission, derived in Section \ref{subsec:opt}. We can also estimate the radius of the dust producing this echo, under the assumption that the UV emission that heated the dust was concurrent with the optical flare. The Ks-band measurement was taken at $\Delta t \sim 2390$ days post-$t_0$, implying a dust radius $R_{\rm{dust}} \simeq ct_{\rm{IR,rise}}/2 \simeq 1~\rm{pc}$. 

Many optical TDEs with detected dust echoes have dust covering fractions of $f_c \sim 1-10\%$ and typical dust radii of $R_{\rm{dust}} \sim 0.5~\rm{pc}$ \citep{lu+16, vanvelzen+21b, somalwar+22, somalwar+25a}, though many optical TDEs do not have detected dust echoes, with estimated dust covering fractions of $f_c \leq 1\%$ \citep{jiang+21}. The values derived from AT 2019ijn are broadly consistent with optical TDEs without detected dust echoes - however, given the sparse spectral coverage and large uncertainties in the estimates of the blackbody properties of both the optical and the IR emission, we cannot confidently state the absence of a dust echo in AT 2019ijn.

\subsection{Color evolution of the optical transient}\label{sec:color}
To evaluate the optical color and evolution of the transient, the g-r color over time was estimated from the ZTF forced-photometry light curves. As the g and r data were not taken simultaneously during transient observations, the color evolution was computed by interpolation of the available g and r data over the duration of the light curve. A comparison of the color evolution for AT 2019ijn and that of other luminous transients in the literature is shown in Extended Data Figure \ref{fig:colev}. 



\subsection{Host galaxy and associated black hole masses}\label{sec:mass}
The host galaxy mass was initially estimated using the mass-to-light ($M/L$) ratio of the host and the formulation of \cite{bell+03}, which describes the correlation between galaxy stellar $M/L$ ratios and the galaxy SDSS/2MASS color photometry. We use the extinction corrected multi-band photometry from the DESI Legacy $g$-band ($g = 24.5 \pm 0.2$) and $r$-band ($r = 23.8 \pm 0.2$) measurements and the observed MOSFIRE Ks-band magnitude ($K_s = 21.2 \pm 0.2$). The use of a combination of optical and NIR photometry is more robust to extinction and diverse stellar populations within the host than either optical or NIR measurements alone. The photometric measurements in each band were k-corrected as outlined in section \ref{sec:kcorr} before conversion to solar luminosities as $L_{\rm{\lambda}} = 10^{0.4(M_{\rm{\lambda, \odot}} - M_{\rm{\lambda, 0}})} \nu_{\rm{\lambda}}$. 

The mass was then computed using the Legacy $g-r$ color of $(g-r) = 0.7 \pm 0.3$ and the Ks-band luminosity of $(1.0 \pm 0.2) \times 10^9 ~\rm{L_\odot}$ as 
\begin{equation}
    \rm{log}_{10}(M_*/L_{Ks}) = a_K + (b_K * (g-r)) = -0.209 + (0.197 * (g-r)),
\end{equation}
which has a scatter of 0.2 dex on the blue end. This gives a host stellar mass of $M_* \sim (9 \pm 5)\times10^8 ~ M_{\odot}$.


We adopt the \cite{bell+03} relations for best comparison with host mass estimates in the literature. However, we note two caveats with this method - one, that the galaxy sample used to derive these relations did not include galaxies with masses less than $M_* < 10^9~M_\odot$, and two, that we cannot fully rule out possible transient contamination in the NIR in the form of a dust echo, and therefore the above estimate may only be an upper limit. For completeness, we therefore also evaluate the method of \cite{delosreyes+25}, a more recent study that 
provides an updated correlation using optical colors based on mock observations of simulated galaxies from $10^{4}~M_\odot < M_* < 10^{10.8}~M_\odot$. This relation gives an estimate of 
\begin{equation}
    \rm{log}_{10}(M_*/M_{\odot}) = 1.598 + 1.347(g-r) - 0.4M_{g,0} = (1 \pm 2) \times 10^9~M_\odot, 
\end{equation}
where $M_{g,0}$ is the k-corrected host magnitude (see \citep{delosreyes+25} and references therein). This agrees with the \cite{bell+03} estimate within $1\sigma$. 

We additionally consider the possible masses of a black hole hosted by a galaxy of this size using the scaling relations between galactic stellar mass and black hole mass $M_* - M_{\rm{BH}}$. We specifically consider relations in the literature for dwarf galaxy stellar masses such as that in \cite{Greene+20}, which fits both early- and late-type galaxies and for a mass regime of $10^8 < M_* (\rm{M_\odot}) < 10^{12}$. The galaxies used to fit the low end of this mass range are primarily upper limits. 

Using our host mass estimate of $M_* \sim (9 \pm 5)\times10^8~\rm{M_\odot}$ and assuming the host lies on the relation given in \cite{Greene+20},
{\small
\begin{multline}
        \rm{log}(M_{\rm{BH}}) = \alpha + \beta~\rm{log}\left(\frac{M_*}{M_0}\right) \pm \epsilon \\
        = (7.43\pm0.09) + (1.61\pm0.12)~\rm{log}\left(\frac{M_*}{3\times10^{10}~M_{\odot}}\right) \pm (0.81\pm0.06),
\end{multline}}
we obtain a $1\sigma$ range of possible black hole masses of between $1.1\times10^4~\rm{M_\odot} < M_{\rm{BH}}< 7.3\times10^5~\rm{M_\odot}$. This lies primarily in the intermediate-mass black hole (IMBH) mass range, typically $10^2 < M_{\rm{BH}} < 10^5 M_{\odot}$. 
We confirm this estimate with the relations of \cite{yao+23}, which are derived from a sample of host galaxies of optically-selected TDEs. This may be a more relevant population for AT 2019ijn, though the mass range of hosts used to derive this relation ($10^{9.17} < M_* (\rm{M_\odot}) < 10^{11.03}$) does not include the host mass of AT 2019ijn. 
With this relation,
\begin{equation}
    \rm{log}\left(\frac{M_{\rm{BH}}}{10^9~M_\odot}\right) = -(1.75\pm0.13) + (1.73\pm0.23)~\rm{log}\left(\frac{M_*}{3\times10^{10}~M_{\odot}}\right),
\end{equation}
with an intrinsic scatter of $\epsilon = 0.17$, we obtain a $1\sigma$ range of possible black hole masses of between $2.4\times10^4~\rm{M_\odot} < M_{\rm{BH}}< 3.4\times10^5~\rm{M_\odot}$, which agrees with our earlier estimate and the IMBH interpretation.

\subsection{Star-formation rate and metallicity of the host}\label{sec:sfr}
The star-formation rate (SFR) of the host was determined using the H$\alpha$ narrow-line luminosity as $\rm{SFR} = 7.9\times10^{-42}L_{\rm{H\alpha,narrow}} \leq 0.12 \pm 0.01 ~ M_\odot/yr$ \citep{kennicutt+94}. We note that the transient may have contributed to the H$\alpha$ luminosity, and this SFR may therefore be an upper limit. The specific SFR (sSFR) was then computed as $\rm{sSFR = SFR/M_*} \sim (1.4 \pm 0.8)\times10^{-10} ~\rm{yr^{-1}}$. 

Standard methods of computing galactic metallicity require the detection of [OII] 4363, which was not detected in the optical spectrum of the host. Instead, we compute the host metallicity using strong line ratios, specifically the $\rm{R}_{23}$ parameter \citep{Pagel+79}, defined as $\rm{R}_{23} = \frac{\rm{F([OII]}3727) + \rm{F([OIII]}4959) + \rm{F([OIII]}5007)}{\rm{F(H}\beta)}$. The $\rm{R}_{23}$ parameter is calibrated to solar metallicity using the relation and coefficients from Table 5 of \cite{nagao+06}: 
\begin{equation}
\rm{log}~ R_{23} = a_0 + a_1x + a_2 x^2 + a_3 x^3 = 0.718 - 0.695x - 0.622 x^2 - 0.632x^3, 
\end{equation}
where $x \equiv \rm{log}(Z/Z_{\odot}) \equiv 12 + log(O/H) - 8.69$. These equations yield multiple plausible solutions; to break the degeneracy we use another strong line ratio between the [OII] and [OIII] lines, $\rm{R} = \frac{\rm{F([OIII]}5007)}{\rm{F([OII]}3727)}$. Using these, we determine the metallicity of the host galaxy to be $\sim 75\%$ of solar values, or $12 + \rm{log}(O/H) \sim 8.572 \pm 0.001$.

\subsection{Modeling of the radio afterglow} \label{sec:radenergy}
The spectrum of the radio emission fits a synchrotron source. We assume that the synchrotron emission is due to a blast wave that propagates into a static medium, using the standard afterglow and radio supernova theory \citep[e.g.][]{chevalier98,granot+02}. In this model, the blast wave accelerates electrons to a power-law, $dN_e/d\gamma \propto \gamma^{-p}$ for $\gamma>\gamma_m$, and the electrons and the magnetic field take fractions $\varepsilon_e$ and $\varepsilon_B$ of the post-shock internal energy, respectively. The model fits radio frequencies where we can identify $\nu_a$ and/or $\nu_m$, where $\nu_m$ is the synchrotron frequency of electrons with $\gamma_m$ and $\nu_a$ is the synchrotron self-absorption frequency. The cooling frequency is assumed to be much larger, so cooling does not affect the radio data.

First, we derive equations that are valid in a more general model, where a general source (not necessarily a blast wave), with a power-law distribution of relativistic electrons and a magnetic field, is moving at some angle with respect to the observer. This is a simplified one-zone model, which is valid for both relativistic and Newtonian emitting regions. To find the constraints set by the model, we start from three equations that relate the following five free parameters to the observations: $N_e$, the number of radiating electrons; $B$, the magnetic field in the emitting region; $A$, the area of the source as projected on the observed plane; $\Gamma$ ($\beta$) the Lorentz factor (velocity) of the source and the corresponding Doppler factor $\delta_D=[\Gamma(1-\beta\theta_{obs})]^{-1}$; and $\gamma_e$, the Lorentz factor of the electrons that dominate the radio emission at the peak $F_\nu$. These equations are similar to those derived by \cite{barniolduran+13,Matsumoto2023}, notated here as:
\begin{equation}\label{eq:ge}
    \gamma_e \approx 1700 [F_{p,mJy} d^2_{L,28} \nu_{p,10}^{-2} (1+z)^{-3}] \eta^{\frac{5}{3}} \frac{1}{A \delta_D} 
\end{equation}
\begin{equation}\label{eq:Ne}
    N_e \approx 2 \times 10^{54 } [F_{p,mJy}^3 d_{L,28}^6 \nu_{p,10}^{-5} \eta^{\frac{10}{3}} (1+z)^{-8}] \frac{1}{A^2 \delta_D^4} 
\end{equation}
\begin{equation}\label{eq:B}
    B \approx 1.3 \times 10^{-3} [F_{p,mJy}^{-2} d_{L,28}^{-4} \nu_{p,10}^{5} \eta^{-\frac{10}{3}} (1+z)^{7}] A^2 \delta_D 
\end{equation}
where $\eta=\max\{\frac{\nu_m}{\nu_a},1\}$. The observables are written within the brackets,`[ ]', with the following notations: $d_{L,28}$ is the luminosity distance in units of $10^{28}$ cm, $\nu_{p,10}$ is the peak frequency in units of $10$ GHz, $F_{p,mJy}$ is the specific flux at $\nu_p$ in units of mJy, and $z$ is the redshift. Note that when $\nu_m>\nu_a$ then $\nu_p=\nu_m$, $\gamma_e=\gamma_m$ and $\eta=\nu_m/\nu_a$. When $\nu_m<\nu_a$ then $\nu_p=\nu_a$, $\gamma_e=\gamma_a$, and $\eta=1$. 

The assumption that the relativistic electrons and magnetic field are accelerated by a blast wave introduces the following relations:
\begin{equation}\label{eq:gm}
    \gamma_m \approx \frac{p-2}{p-1} \frac{m_p}{m_e}\varepsilon_e (\Gamma-1) ~,
\end{equation}
introducing a sixth free parameter $\varepsilon_e$, where $m_p$ is the proton mass and $m_e$ is the electron mass. If $\gamma_m<1$ according to this equation, then we set $\gamma_m=1$. We consider $p$ in the usual observed range $2.5-3$, finding that the results depend weakly on $p$ in this range. 
The magnetic field energy fraction can be extracted from the other parameters,
\begin{equation}
    \varepsilon_B \approx \frac{B^2}{32\pi \Gamma(\Gamma-1) n m_p c^2} ~,
\end{equation}
where $c$ is the speed of light and $n$ is the density at the location of the blast wave. $n \approx N_e/V$, where $V$ is the volume over which the blast wave collects the emitting electrons. Note that here we approximated the shock velocity to be the velocity of the shocked medium and the density to be constant. The correction for the exact shock velocity and various density profiles is less than an order of magnitude. Since at least half of the energy is in protons, the energy fractions are limited:
\begin{equation}
    \varepsilon_B+\varepsilon_e < 1/2~.
\end{equation}
Finally, the time, $t$, since the shock is launched into the medium is 
\begin{equation}
    t \sim \frac{R}{c} \frac{1-\beta \rm{cos(\theta_{obs})}}{\beta} 
\end{equation}
where we approximated the shock velocity to be constant and comparable to the velocity of the shocked medium. Accounting for deceleration, such as in the case of a constant energy shock, results in a time that is smaller by a factor of a few; on the other hand, accounting for delays due to light travel time can increase this time by a factor of a few, but only in the relativistic regime. Finally, the energy of the system can be approximated as 
\begin{equation}\label{eq:Etot}
    E \sim N_e m_p c^2 \Gamma (\Gamma-1)
\end{equation}

We first consider a spherical blast wave at a radius $R$, for which $\delta_D\approx \Gamma$, $A \approx \pi R^2 \Gamma^{-2}$ and $V\approx R^3 \Gamma^{-2}$ (the emitting electrons are confined to an angle $1/\Gamma$ with respect to the line of sight). Under these relations Eqs. \ref{eq:ge}-\ref{eq:B} above are reduced to  Eqs. 14-16  of \cite{barniolduran+13}.
We start with the most constraining observation at 998 d. The spectrum at this epoch shows a flat $F_\nu$ between 2 and 12 GHz which may extend to higher frequencies. In the case of emission from a blast wave that propagates into a medium, this is expected if $\nu_m \approx 10$ GHz and $\nu_a \approx 3$ GHz. This observation therefore provides three observables: $\nu_a$, $\nu_m$, and $F_p \approx 6$ mJy. Applying a spherical model to these observables, we find the following constraints. The minimum energy is $\sim 5 \times 10^{50}$ erg, obtained for $\varepsilon_e \approx \varepsilon_B \approx 1/4$. The radius is in the range $3 \times 10^{17} - 1.5 \times 10^{18}~\rm{cm}$, and the shock velocity must be larger than about $0.7c$. 
The electrons must have $\varepsilon_e \gtrsim 0.01$, otherwise $\varepsilon_B$ is nonphysical. For the canonical values of $\varepsilon_e =0.1$ and $\varepsilon_B=0.01$ often considered,  the parameters are determined as $E \approx 3 \times 10^{51}$ erg, $R \approx 5 \times 10^{17}~\rm{cm}$, $\Gamma \approx 3$, and $n \approx 1 ~{\rm cm^{-3}}$. Interestingly, the maximal value of $t$ in this model is about 100 d for $\varepsilon_e \approx \varepsilon_B \approx 1/4$ and 50 d for the canonical values of $\varepsilon_e = 0.1$ and $\varepsilon_B = 0.01$. We will discuss the implications of this result later.

Next, we apply the spherical model to observations at 1997 d. These observations show a relatively wide peak with $F_p \approx 3$ mJy in the range 1-3 GHz, followed by a steep optically thin spectrum that suggests $p \approx 3$. The wide peak suggests $\nu_m \approx \nu_a \approx 1-2$ GHz, although we cannot rule out that the peak corresponds to only one of these characteristic frequencies and the other is significantly smaller. Assuming  $\nu_m \approx \nu_a$ we obtain the following constraints: $E_{min} \sim 5 \times 10^{50}$ erg, $\varepsilon_e \gtrsim 0.005$, $R\approx 10^{17} - 4 \times 10^{18}~\rm{cm}$ and $\beta>0.7$. The maximum time after ejection for this epoch is approximately 300 d, obtained for the minimum shock velocity.  For $\varepsilon_e=0.1$ and $\varepsilon_B=0.01$, the model parameters are $E \approx 3 \times 10^{51}$ erg, $R \approx 10^{18}~\rm{cm}$, $\Gamma \approx 2.3$, $n \approx 0.1~ {\rm cm^{-3}}$, and $t \approx 100$ d.

These results show that the model fits the data with reasonable parameters, with the one caveat being that the time since blast wave launch does not agree with the timescale of the radio data. The time of the two epochs occur about 1000 and 2000 days since the optical transient, and about 500 and 1500 days since the first radio detection. Therefore, although our estimate of $t$ is correct within an order of magnitude, it is still in tension with a spherical model of a blast wave that was launched before the first radio detection. However, our reasonable solutions to the data for mildly relativistic velocities imply that there are solutions for off-axis emission from a narrow jet with reasonable jet and medium parameters, and times that fit the observations, as we show below. This scenario is actually more plausible than a spherical blast wave, since relativistic outflows are usually collimated, and as shown recently, narrowly-collimated relativistic jets remain narrow when they decelerate to mildly relativistic velocities \citep{GovreenSegal2024}.

A narrow jet introduces two free parameters: the jet opening angle $\theta_j$, and the observing angle $\theta_{obs}$, where we assume $\theta_{obs} \gg \theta_j$. Under this assumption, $A \approx \pi R^2 \theta_j^2 \rm{cos(\theta_{obs})}$ and $V \approx R^3 \theta_j^2$. Given that this model has more freedom, we expect to find solutions for a range of observing and jet angles. Since the primary goal is to demonstrate that the model can explain the observations and to identify constraints on the parameters, we select reasonable viewing and jet angles, $\theta_j=0.1$ rad and $\theta_{obs}=1$ rad, and determine the constraints for these angles. For the observations at 998 d we find the following constraints: $E_{min} \sim  10^{51}$ erg, $\varepsilon_e \gtrsim 0.03$, $R \approx 3-8 \times 10^{18}$ cm and $\Gamma \approx 2-10$, where a higher Lorentz factor requires much higher energies. The time range for these solutions is $t \approx 600-1800$ d, which, given the approximated model, is in agreement with a jet that is launched anytime between the optical transient and the first radio detection. The model parameters for $\varepsilon_e=0.1$ and $\varepsilon_B=0.01$ are  $E \approx 3 \times 10^{52}$ erg, $R \approx 5 \times 10^{18}$, $\Gamma \approx 4.5$, $n \approx 0.4~ {\rm cm^{-3}}$ and $t \approx 1100$ d. For the observations at 1997 d, $E_{min} \sim  10^{51}$ erg, $\varepsilon_e \gtrsim 0.02$, $R \approx 5-15 \times 10^{18}$ cm, $\Gamma \approx 1.7-10$, and $t \approx 1200-4000$ d. The model parameters for $\varepsilon_e=0.1$ and $\varepsilon_B=0.01$ are  $E \approx 1.5 \times 10^{52}$ erg, $R \approx 10^{19}~\rm{cm}$, $\Gamma \approx 3.3$, $n \approx 0.04~ {\rm cm^{-3}}$ and $t \approx 2300$ d.

The main conclusion of the modeling is that a mildly relativistic jet seen off-axis provides the most natural explanation for the observed radio emission. The jet may have been launched simultaneously with the optical emission, or it may have been launched later, close to the first radio detection - the data cannot discriminate between these options. We also cannot constrain the initial velocity of the jet, which might have been ultra-relativistic. However, by the time of the observed radio emission, the jet must have $\Gamma<10$, and most likely (to keep a more realistic energy budget) $\Gamma<5$. We show an example of one possible model fit to the radio data in Figure \ref{fig:radiojet}, though we note that the hydrodynamics of the jet evolution are not completely captured in our model, and the exact density profile of the surrounding medium cannot be fully constrained. Finally, a comparison of the parameters of the jetted model to the spherical model shows that the blast wave energy does not depend strongly on the geometry. We therefore conclude that the minimal energy is $\sim 10^{51}$ erg, and for canonical microphysical parameters, it is more likely to be of order $\sim 10^{52}$ erg.

\subsection{Origin of the optical emission}\label{sec:optmodel}

We consider potential sources of the optical emission of AT 2019ijn. These scenarios are evaluated on the properties of the optical emission alone, and are evaluated in context with the multiwavelength data in section \ref{sec:options}. 
 
\subsubsection{Observed properties of TDE-like transients}
We discuss the typical observed properties of TDE-like transients, as this represents a strong possible explanation for AT 2019ijn. The observed properties of TDE emission \citep[e.g.][]{rees+88, Phinney89} are commonly parameterized in terms of timescales, mass fallback rate, and available energies \citep[see e.g.][and references therein]{strubbe+quataert09, stone+metzger16}. 
 
The optical emission of TDEs is often assumed to trace the fallback rate of stellar material onto the BH post-disruption, which occurs on a timescale of 
\begin{equation}\label{eq:tfall}
    t_{\rm{fall}} \sim 3.5\times10^6\left(\frac{M_{\rm{BH}}}{10^6~M_\odot}\right)^{1/2} \left(\frac{M_*}{M_\odot}\right)^{-1} \left(\frac{R_*}{R_\odot}\right)^{3/2}~\rm{s}, 
\end{equation}
where $M_{\rm{BH}}$ is the mass of the black hole, and $M_*$ and $R_*$ are the mass and radius of the disrupted star, respectively. The peak mass fallback rate at this time can be described as 
\begin{equation}\label{eq:Mdot}
    \frac{\dot{M}_{\rm{peak}}}{\dot{M}_{\rm{Edd}}} \sim 133 \left(\frac{\eta}{0.1}\right) \left(\frac{M_{\rm{BH}}}{10^6~M_\odot}\right)^{-3/2} \left(\frac{M_*}{M_\odot}\right)^{2} \left(\frac{R_*}{R_\odot}\right)^{-3/2}, 
\end{equation}
where $\eta$ is the accretion efficiency and $\dot{M}_{\rm{Edd}}$ is the Eddington accretion rate, equivalent to $\dot{M}_{\rm{Edd}} \sim \frac{L_{\rm{Edd}}}{\eta c^2} \sim 1.5\times10^{46} \left(\frac{M_{\rm{BH}}}{10^8~M_\odot}\right) \left(\frac{1}{\eta c^2}\right)$. An initially super-Eddington accretion rate remains so for a time 
\begin{equation}\label{eq:tedd}
     t_{\rm{Edd}} \sim 6.6\times10^7 \left(\frac{\eta}{0.1}\right)^{3/5} \left(\frac{M_{\rm{BH}}}{10^6~M_\odot}\right)^{-2/5} \left(\frac{M_*}{M_\odot}\right)^{1/5} \left(\frac{R_*}{R_\odot}\right)^{3/5}~\rm{s}. 
\end{equation}
If we therefore assume that the luminosity of the TDE can be described by $L_{\rm{Edd}}$ if $t < t_{\rm{Edd}}$ and $L \sim \dot{M}\eta c^2$ for $t > t_{\rm{Edd}}$ the total radiated energy of the TDE can be described as 
\begin{equation}\label{eq:Erad}
    E_{\rm{rad}} \sim L_{\rm{Edd}} \left(\tilde{t}_{\rm{Edd}} - t_{\rm{fall}}\right) + \frac{M_* \eta c^2}{2} \left(\frac{t_{\rm{fall}}}{\tilde{t}_{\rm{Edd}}}\right)^{3/2}~\rm{erg}, 
\end{equation}
where $\tilde{t}_{\rm{Edd}} = \rm{max(t_{\rm{fall}}, t_{\rm{Edd}})}$. 

In the following sections, we employ these to parameterize the emission of AT 2019ijn using the derived black hole mass, $M_{\rm{BH}} \sim 10^5~M_\odot$, and the common assumptions of $\eta = 0.1$ and $R_* \sim M_*^{0.8}$. As the mass of the disrupted star is unknown, we compute these for two cases within reasonable values of $M_* = 0.1-1~M_\odot$ and assume our values lie in this range. 



\subsubsection{Accretion-powered flare}
At very early times, the emission due to fallback of material onto the accreting black hole is expected to dominate the observed luminosities \citep{kasen+ramirezruiz2010}. The rise time of AT 2019ijn (6.9 d in the rest frame) generally agrees with the expected fallback timescale for a $\sim 10^5~\rm{M_\odot}$ black hole, $t_{\rm{fall}}\sim 8.1~\rm{d} - 12.8~\rm{d}$ for $M_* = 0.1, 1~\rm{M_\odot}$, respectively, indicating the rise of the light curve may be consistent with the onset of accretion. AT 2019ijn would initially accrete at super-Eddington rates ($\dot{M}_{\rm{peak}}\sim 670 - 4200~\dot{M}_{\rm{Edd}}$ for $M_* = 0.1, 1~\rm{M_\odot}$ and the Eddington luminosity $L_{\rm{Edd}} \sim 1.5\times10^{43}~\rm{erg/s}$), and would remain in the super-Eddington phase for year(s)-long timescales ($t_{\rm{Edd}}\sim 400~\rm{d} - 1900~\rm{d}$), which is atypical of previously-observed optical TDEs \citep[e.g.][]{yao+23} but is expected for lower-mass black holes, scaling roughly as $\frac{\dot{M}_{\rm{peak}}}{\dot{M}_{\rm{Edd}}} \sim \left(\frac{M_{\rm{BH}}}{10^6~M_\odot}\right)^{-3/2}$ \citep{stone+metzger16, wu+18}. This long-lasting super-Eddington phase may be responsible for the high peak luminosities seen in AT 2019ijn, which peaks at $L_{\rm{g, peak, corr}} \sim 4L_{\rm{Edd}}$. 

We attempt to model this scenario with the oft-used light curve fitting code \texttt{MOSFiT} \cite{guillochon+18, mockler+19}. The model assumes that the TDE engine comes from the fallback of material onto the BH post-disruption, which is simulated with a grid of hydrodynamic simulations assuming $M_{\rm{BH}} = 10^6~\rm{M_{\odot}}$ and $M_* = 1~\rm{M_{\odot}}$, and is then scaled to other masses. The luminosity is assumed to follow the fallback rate. The model includes time delays for circularizaton and accretion of the forming disk, and a simple blackbody reprocessing layer (photosphere) is assumed. 

The \texttt{MOSFiT tde} module was initially designed to model emission only within the sub-Eddington regime. As the super-Eddington phase of AT 2019ijn far exceeds the duration of the optical emission, we modify the \texttt{tde} module to allow luminosities exceeding the Eddington limit by artificially increasing the allowed limit \citep[as in][]{perley+19a, somalwar+25}. We note that this modification only allows for a rough estimate of the super-Eddington accretion scenario in TDEs and cannot adequately capture the intricacies of the physics involved. We allow for black hole masses within the range $5\times10^3~\rm{M_\odot} \leq M_{\rm{BH}} \leq 5\times10^8~\rm{M_\odot}$. We find that \texttt{MOSFiT} is able to reproduce the observed emission with a black hole of mass $M_{\rm{BH}} = 2^{+11}_{-2}\times10^5~\rm{M_{\odot}}$ disrupting a star of mass $M_* = 0.1^{+0.2}_{-0.1}~\rm{M_{\odot}}$ (Extended Data Figure \ref{fig:models}a). The \texttt{MOSFiT}-derived black hole mass agrees with the black hole mass derived from galactic scaling relations within error bars. With the caveat of incomplete handling of super-Eddington accretion, we conclude that as \texttt{MOSFiT} can reproduce the observed emission within reasonable parameters,  a super-Eddington accretion-powered flare may describe the optical emission of AT 2019ijn.

\subsubsection{Stream-stream collisions}


Stream-stream collisions, occurring when streams of debris from the disrupted star self-intersect during the initial circularization of the forming disk, have been predicted to produce optical flares \citep{piran+15a, jiang+16, chen+shen21}. Recent studies have shown that this can reproduce the observed blackbody spectra, including temperature, photospheric radius, and peak luminosities, observed from typical TDEs with $M_{\rm{BH}}\sim10^6~\rm{M_\odot}$, with optical emission that peaks at $\nu L_\nu \sim 10^{42}-10^{43}~\rm{erg/s}$ \citep{guo+25, steinberg+24, huang+25_axv}. \cite{chen+shen21} have also considered the optical evolution resulting from stream-stream collisions of both full and partial TDEs. 

In this scenario, the circularization of the debris occurs over a timescale $t_{\rm{circ}}$, defined as 
\begin{equation}
    t_{\rm{circ}} = 8 \beta^{-1} \left(\frac{M_{\rm{BH}}}{10^6~\rm{M_\odot}}\right)^{-\frac{5}{3}} \left(\frac{M_{*}}{\rm{M_\odot}}\right)^{-\frac{1}{3}} \left(\frac{R_{*}}{\rm{R_\odot}}\right)^{2} t_{\rm{fb}}~\rm{s}, 
\end{equation}
where $M_{\rm{BH}}$ is the mass of the BH, $M_*$ and $R_*$ are the mass and radius of the disrupted star, $t_{\rm{fb}}$ is the fallback time, and $\beta$ is a parameter describing the ``completeness'' of the tidal disruption - $\beta = 1$ denotes a full disruption, while $\beta < 1$ denotes a partial disruption. 

The peak optical luminosity generated by stream-stream collisions is given by \cite{chen+shen21} as 
\begin{equation}
    L_{p} = 6\times 10^{42} \left(\frac{\Delta M}{0.01~\rm{M_\odot}}\right) \beta^{\frac{9}{2}} \left(\frac{M_{\rm{BH}}}{10^6~\rm{M_\odot}}\right)^2 \left(\frac{M_{*}}{\rm{M_\odot}}\right)^{\frac{3}{2}} \left(\frac{R_{*}}{\rm{R_\odot}}\right)^{-\frac{9}{2}}~\rm{erg/s}
\end{equation}
where $\Delta M$ is the total fallback mass. 

Following our earlier assumptions for these parameters for AT 2019ijn, along with a nominal value of $\Delta M = 0.01~\rm{M_\odot}$ and a full tidal disruption, we find that the circularization timescale and peak luminosity that can be generated by stream-stream collisions for a $\sim 10^5~\rm{M_\odot}$ BH is $t_{\rm{circ}} \sim 160~\rm{d}$, $L_{p}\sim 7.5 \times10^{42}~\rm{erg/s}$ for a disrupted star mass of $M_* = 0.1~\rm{M_{\odot}}$ and $t_{\rm{circ}} \sim 4700~\rm{d}$, $L_{p}\sim 6\times10^{40}~\rm{erg/s}$ for a disrupted star mass of $M_* = 1~\rm{M_{\odot}}$. This indicates that under this formulation, reproducing the peak luminosity of AT 2019ijn via stream-stream collision would require a disrupted star mass of $M_* < 0.1~\rm{M_{\odot}}$. 
We note that these models predict a steeper decline post-peak than is observed in AT 2019ijn, indicating that if stream-stream collision contributes to the optical emission, there must be an additional source of emission to explain the full optical light curve.

\subsubsection{Super-Eddington outflows and shock-driven emission}
The disks of TDEs accreting at highly super-Eddington rates have been predicted to be able to drive optically thick, mildly relativistic outflows, which have also been seen in recent simulations of TDEs from $10^6~M_\odot$ BHs \citep{price+24}. In initial predictions from \cite{strubbe+quataert09, stone+metzger16}, the emitting photosphere, at early times, is limited by the size of the outflow, during which time the luminosity on the Rayleigh-Jeans tail increases as $\nu L_\nu \sim t^{65/36}$.
Eventually, the photosphere recedes within the edge of the outflow and the luminosity on the Rayleigh-Jeans tail declines as $\nu L_\nu \sim t^{-95/36}$, with a peak luminosity of $\nu L_{\nu, \rm{SE}} \sim 2.2\times10^{41}~\rm{erg/s}~\left(\frac{M_{\rm{BH}}}{10^6~M_\odot}\right)^{-79/72}$ \citep{stone+metzger16}. These predict far steeper optical light curves in both rise and decline scales  than that observed in AT 2019ijn and fail to reproduce the peak optical luminosity for the relevant black hole masses, disfavoring super-Eddington outflows as a strong contributor to the observed optical emission.


Instead, more recent simulations of TDEs from IMBHs \citep[e. g.][]{martire+25_axv} have determined that unlike the typical $10^6~M_\odot$ BH scenario, the outflows from IMBH TDEs are expected to be unbound and form a wind rather than a bound envelope as in \cite{strubbe+quataert09, stone+metzger16, price+24}. The emission during the rise of the transient is expected to be shock-driven, with the primary mechanism of energy dissipation being the so-called ``nozzle shock'' at super-Eddington rates, with accretion becoming a more dominant factor at later times. The emission is also expected to be Eddington-limited, with the bolometric luminosity peaking at $L_{\rm{bol, peak}} \sim 2L_{\rm{Edd}}$ on timescales of $t \sim 1.5 t_{\rm{fb}}$ for a $10^4~M_\odot$ BH before settling to approximately $L_{\rm{Edd}}$. 
The expected bolometric luminosity at timescales of $t \sim t_{\rm{fb}}$ as predicted by \cite{martire+25_axv} for a $10^5~M_\odot$ BH is approximately $L_{\rm{bol, peak}} \sim 2.1L_{\rm{Edd}}$, which is broadly comparable to the observed g-band optical peak of $L_{\rm{g, peak, corr}} \sim 4L_{\rm{Edd}}$ on timescales of $t \sim 0.9 t_{\rm{fb}} -0.5t_{\rm{fb}}$ for a $M_* = 0.1- 1~\rm{M_\odot}$ disrupted star. This may indicate that the shock-driven emission predicted from IMBH TDEs could be a strong contributor to the optical light curve of AT 2019ijn.


\subsubsection{Off-axis synchrotron jet}
The off-axis jet predicted in AT 2019ijn may also be a source of optical emission. The shock produced by the jet interaction with the surrounding medium is expected to emit with a synchrotron spectrum, for which the optical emission can be assumed to lie between the peak synchrotron frequency, $\nu_m$, typically in radio bands, and the characteristic cooling frequency, $\nu_c$, which can range from infrared to ultraviolet wavelengths \citep{stone+metzger16}. After deceleration to mildly relativistic speeds, the optical luminosity from the jet can be expected to decline as $\nu L_\nu \sim t^{-1.65}$. The peak optical luminosity produced by the jet can be taken as 
{\small
\begin{multline}
    \nu L_{\nu, \rm{jet}} \sim 2.5\times10^{43} \left(\frac{n}{1~\rm{cm^{-3}}}\right)^{0.33} \left(\frac{\epsilon_B}{0.01}\right)^{0.88} \left(\frac{\epsilon_e}{0.1}\right)^{1.5} \left(\frac{M_{\rm{BH}}}{10^6~\rm{M_\odot}}\right)^{-0.8} \left(\frac{E}{10^{52}~\rm{erg}}\right)^{1.55} \\
    \times \left(\frac{M_*}{M_{\odot}}\right) ^{-0.32} t^{-1.65}~\rm{erg/s},
\end{multline}}
where $t = \frac{t_{nr}}{2t_{\rm{fall}}}$ if the reverse shock is strong and quickly decelerates the jet, or $t = t_{\rm{nr}}$ if the jet remains relativistic after the reverse shock. Following the assumed values in \cite{stone+metzger16} ($n = 1~\rm{cm^{-3}},~\epsilon_B = 0.01,~ \epsilon_e = 0.1,~ E_{52} = 10^{52}~\rm{erg},~ t_{\rm{nr}} = 10 t_{\rm{fall}}$), we obtain a peak luminosity of $\nu L_{\nu, \rm{jet}} \sim 2\times10^{43}~\rm{erg/s}, 8\times10^{42}~\rm{erg/s}$ for $M_* = 0.1, 1~\rm{M_\odot}$, respectively, for jets with a strong reverse shock. If the jet is assumed to remain relativistic after the reverse shock crossing, these estimates drop to $\nu L_{\nu, \rm{jet}} \sim 10^{33}~\rm{erg/s}$. 

The observed radio emission does not allow us to place a strong constraint on the medium density at the radius of the optical emission, but reproducing the observed peak luminosities in AT 2019ijn under these assumptions would require $n > 10~\rm{cm^{-3}}$. The expected decay for the jet emission is, again, much steeper than the decay we observe. Additionally, recent simulations of IMBH TDEs have found circularization of stellar debris for lower-mass black holes to be relatively inefficient \citep{martire+25_axv}, which disfavors a strong contribution from the jet to the optical emission, as this would require the jet to have been launched almost immediately after disruption. Though our radio modeling does not rule out a jet launched immediately after disruption, the off-axis jet alone cannot explain the slower decline following the optical peak, and we therefore disfavor the off-axis jet as a strong contributor to the optical light curve.

\subsubsection{Shock breakout}

Many transient classes that produce high luminosity optical emission are thought to do so through shock breakout following core-collapse of a massive star. Here we consider the possibilities of shock breakout of supernova ejecta powering the optical light curve of AT 2019ijn following the analysis of \cite{khatami+kasen24}, and shock breakout of a jet + cocoon structure following the models of \cite{Gottlieb2022, gottlieb+metzger24}. 

\cite{khatami+kasen24} argue that a light curve powered by circumstellar medium (CSM) interaction undergoes four distinct phases: 1) the shock breakout, in which the shock has propagated to a region with optical depth low enough for photons to escape ahead of the shock front; 2) the continued interaction phase, wherein the shock interacts with the CSM ahead of the breakout radius; 3) the shock emergence, when the shock passes the edge of the CSM and the luminosity drops, and 4) the shock cooling, where earlier photons and alternate sources of heating continue to power faint emission. The presence and duration of each of these phases depends on the location of the breakout and the properties of the CSM - shocks which only break out at the edge of the CSM will not have a continued interaction phase, for example, where shocks which breakout interior to the CSM do exhibit this phase. In these models, the peak of the light curve corresponds to the time of shock breakout, and the morphology and duration of the following phases depend on the properties of the CSM. 

CSM-interaction light curves can therefore be broken into four regimes: ``edge'' vs. `` interior'' breakout, and ``heavy'' vs. ``light'' CSM. Light curves which experience ``edge breakout'' are expected to show a sharp decline in luminosity following the peak, which is not seen in AT 201ijn; we therefore favor an interior breakout scenario. Interior breakout light curves in a heavy CSM tend to show a slower rise and a steeper drop-off following the peak before flattening in the continued interaction phase, where lighter CSM light curves show a fast evolution and an immediate gentle decline. As this is more in line with the morphology of AT 2019ijn's optical light curve, we favor a light CSM scenario. In this case, we assume the light curve peak at 6.9 d corresponds to the time of the shock breakout (therefore $L_{\rm{bo}} = L_{\rm{g, peak}} = 7\times10^{43}~\rm{erg/s}$, $t_{\rm{bo}} = t_{\rm{g, peak}} = 6.9~\rm{d}$), and the following decline corresponds to the continued interaction phase. As we do not observe a second break in the light curve, we cannot fully constrain the time of the shock emergence, and must assume a lower limit corresponding to the termination of the light curve due to observing constraints $ t_{\rm{se}} > 45~\rm{d}$.

We can get a rough understanding of the expected evolution of the light curve in this scenario by making some assumptions about the CSM density and ejecta density profiles. \cite{khatami+kasen24} assume the CSM density profile to follow a power law $\rho(r) \propto r^{-s}$, and the ejecta density profile to follow a power law as $\rho_{\rm{ej}}(r) \propto r^{-n}$ where typically $7 < n < 10$. We follow their assumption of $s = 2$ and $n = 10$. The luminosity in the continued interaction phase is expected to evolve as 
\begin{equation}
    L_{\rm{CI}} = L_{\rm{bo}}\xi^{-\sigma_{\rm{s}}}\left(\frac{t}{t_{\rm{se}}}\right)^{\frac{\sigma_s}{k_0}}
\end{equation}
where $\xi$ is the breakout parameter ($\xi < 1$ for an interior breakout; here we adopt a value of $\xi = 0.5$); and power-law exponents $k_0 = 0.6$ per their assumption, and $\sigma_s = \frac{(n-3)(n-5)}{n-s}-3 = -0.23$ for $s = 2, n = 10, k_0 = 0.6$. We find that under these assumptions, the luminosity is expected to decline as $L_{\rm{CI}} \propto t^{-0.38}$, and a rough power-law fit to the decline of the g-band light curve yields $L_{\rm{g}} \propto t^{-0.37}$ - however, we caution that this is strongly dependent on the assumptions of the CSM and ejecta density profiles (see especially Figure 12 in \cite{khatami+kasen24}), which cannot be constrained by our observations. Assuming the duration of the breakout is $\Delta t_{\rm{bo}} = t_{\rm{bo}}/2$, a light curve of this morphology is possible with a CSM of radius $R_{\rm{CSM}} \sim 10^{14}~\rm{cm}$ and mass $M_{\rm{CSM}} \sim 10^{-2} - 10^{-1}~\rm{M_\odot}$, which, given the large uncertainties in the blackbody fitting, may be in agreement with the photospheric radius expected from the blackbody emission. 

We additionally caution that the analysis in \cite{khatami+kasen24} tracks the evolution of the bolometric luminosity, which we cannot effectively approximate with our limited spectral coverage and may behave differently than our monochromatic light curves. However, as this model can reproduce the observed optical emission within reasonable parameters, we cannot rule out a shock breakout for the origin of the optical emission.

Another possibility for a shock breakout is the jet emerging from the expanding ejecta, which has been suggested to power the UV/optical emission from both SLSNe and luminous fast blue optical transients (LFBOTs) \citep{gottlieb+metzger24,Gottlieb2022}. In these models, successful jet breakout creates a shocked cocoon that can power luminous UV and optical emission, reaching bolometric luminosities of $L_{\rm{bol}} \geq 10^{44}~\rm{erg/s}$ within several days of the explosion. After jet shutoff, the luminosity will decrease, after which point the light curve will become dominated by emission from the main SN ejecta, which is expected to peak on timescales of weeks. 

There are a few difficulties in reconciling this emission with that of AT 2019ijn. For typical parameters of kinetic energy $E_{\rm{K}} \sim 3\times 10^{51}~\rm{erg}$ and jet luminosity $L_j \sim 10^{45}~\rm{erg/s}$ \citep{gottlieb+metzger24}, along with a jet opening angle of $\theta_j \sim 0.1~\rm{rad}$ as assumed in the radio modeling (Section \ref{sec:radenergy}), the jet would be expected to break out on a timescale of 
\begin{equation}
    t_{\rm{bo}} = 5\times\left(\frac{E_{\rm{K}}}{3\times10^{51}~\rm{erg}}\right) \left(\frac{10^{45}~\rm{erg/s}}{L_j}\right) \left(\frac{\theta_j}{10^{\circ}}\right)^4~\rm{s} \sim 0.5~\rm{d},
\end{equation}
which is an order of magnitude shorter than the rise time of AT 2019ijn. Though the bolometric luminosity is expected to peak at $ \geq 10^{44}~\rm{erg/s}$, the monochromatic light curves are expected to peak at luminosities 1-2 orders of magnitude less than that seen in AT 2019ijn for an on-axis viewing angle. We do not see a second peak in the light curve on timescales of ~weeks, which would indicate that the main SN emission is either missed due to observing constraints, or the entire light curve is powered by the main SN emission, with little to no contribution from the jet breakout. Differences in viewing angle will also contribute to the observed emission - though there have not been detailed models of the off-axis emission from jet breakout in these scenarios, one can assume that the optical light curve would peak at later times than the nominal shock breakout time, which may align with our observed rise time, but at a lower luminosity than the on-axis predictions, which would represent an almost negligible contribution to the optical light curve of AT 2019ijn. These models would therefore require significant modification to accurately describe our observations, and we thus disfavor the jet breakout interpretation for the optical emission.

\subsubsection{Spin-down luminosity from newborn millisecond magnetar}\label{sec:magnetar}

Millisecond magnetars have been proposed to power the luminous optical emission from SLSNe \citep{Kasen2010, Woosley2010, metzger+15,nicholl+17,margalit+18} through spin-down shortly after their birth. These models produce peak luminosities of near $\sim 10^{44}~\rm{erg/s}$, though they typically predict more symmetrical light curves than seen in AT 2019ijn, as SLSNe rarely produce optical light curves with rise times of $t< 10~\rm{d}$ and typically exhibit much slower declines \citep{nicholl+15, lunnan+18}. However, given the broad similarity of AT 2019ijn's optical flare to typical SLSNe, we attempt to fit the optical light curve with the SLSN-magnetar model of \cite{nicholl+17} via \texttt{MOSFiT}. We find the emission can be reproduced given a magnetar with magnetic field $B = 3^{+2}_{-1}\times10^{14}~\rm{G}$, spin period $P = 2.7^{+0.5}_{-0.7}~\rm{ms}$, and ejecta mass $M_{\rm{ej}}= 0.6^{+0.3}_{-0.3}~\rm{M_\odot}$. 
We do not rule out a magnetar origin for AT 2019ijn by the optical emission alone. 





\subsection{Alternate possible transient scenarios}\label{sec:options}

\subsubsection{BH-stellar mergers/Luminous Fast Blue Optical Transients (LFBOTs)}
Luminous Fast Blue Optical Transients (LFBOTs) are an emerging category of transient, with the ``canonical'' LFBOT being AT 2018cow \citep{prentice+18, perley+19a, ho+19, margutti+19}. These transients have short durations ($t_{1/2} < 12~\rm{d}$) and blue colors at peak light \citep[$g-r < -0.2$;][]{ho+23}, with the ``luminous'' subset of FBOTs having peak luminosities of $ > 10^{44}~\rm{erg/s}$ \citep{metzger+22a}. These transients have been suggested to originate from a variety of scenarios, including core-collapse of a massive star giving birth to a compact object \citep[e.g.][]{prentice+18, perley+19a, margutti+19}, and TDEs originating from an IMBH \citep[e.g.][]{perley+19a, Kuin2019}. Recently, scenarios involving common-envelope mergers between a compact object and a stellar companion have been suggested as alternative explanations for observed properties of LFBOTs \citep{metzger+22a, tsuna+lu25, klencki+metzger25_axv}. As LFBOTs are an observational class with many proposed explanations, here we focus on the theoretical common-envelope merger scenario and comparisons to other observed LFBOTs, with the other scenarios later discussed as their own class.

Nominally, AT 2019ijn evolves too slowly to be considered a typical LFBOT, with an optical light curve duration of $t_{\rm{rest}} > 45~\rm{d}$. Paired with the fast rise, this creates an asymmetric light curve atypical of most LFBOTs (see Figure \ref{fig:phase}). However, the optical peak magnitude and peak color are similar to other observed LFBOTs \citep{ho+23}. The radio emission is more luminous, energetic, and long-lasting than most known LFBOTs (Figure \ref{fig:LC_panel}c), and also displays a year-delayed onset not seen, to our knowledge, in any known LFBOT to date. The host galaxy also aligns well with the properties of other LFBOT host galaxies, though on the lower end of typical star-formation rates \citep{nugent+26_axv}.

In the common-envelope merger scenarios, a SN-like transient is generated when a compact object merges with and tidally disrupts a stellar companion. 
These models generate high peak optical luminosities, quickly-evolving light curves, and blue colors consistent with observed LFBOTs, as well as energetic radio emission and a dense, extended CSM via mass loss from the companion star prior to the merger. The observed properties of AT 2019ijn are most consistent with predictions from a BH disrupting a main-sequence companion, as suggested by \cite{tsuna+lu25}, or a stripped He-core star, as suggested by \cite{klencki+metzger25_axv}, which produce similar optical peak luminosities and may produce relativistic jets resulting from accretion onto the BH. The long-lasting radio emission is also consistent with predictions of an extended CSM, and the broad H$\alpha$ feature in the late-time optical spectrum would be consistent with CSM interaction from a slow outflow/wind nebula following the merger \citep[e.g.][]{tsuna+lu25}. However, both models predict more symmetrical optical light curves than that of AT 2019ijn, where a short rise also corresponds to a rapid decline. Longer-duration light curves also correspond with higher ejecta mass ($M_{\rm{ej}} > 10~M_\odot$), while light curves with a rapid rise require very low ejecta mass ($M_{\rm{ej}} < 5~M_\odot$). This indicates that reproducing the slowly-declining optical emission in AT 2019ijn following the rapid rise would likely require some source of continued energy injection post-merger. 

The CSM nominally predicted from these models is very dense \citep[$10^3-10^5~\rm{cm^{-3}}$;][]{klencki+metzger25_axv} at radii of $\sim 10^{17}~\rm{cm}$ before a steep dropoff at radii of $\sim 10^{18}-10^{19}~\rm{cm}$, 
though the exact properties of the CSM are strongly dependent on the properties of the stellar companion and the mechanism of mass loss before and during the merger. Our radio modeling similarly predicts a very low density medium ($n \sim 0.4-0.04~\rm{cm}$; section \ref{sec:radenergy}) at radii of $\sim 10^{18}-10^{19}~\rm{cm}$, though our data does not give strong constraints on the density at radii of $< 10^{18}~\rm{cm}$. The radio emission of AT 2019ijn, being more energetic and longer-lasting than most LFBOTs, would likely require a combination of a very powerful jet and favorable CSM properties that are not commonly seen in LFBOT scenarios. However, given the dependence of the resulting emission on specific properties of the system (i.e. the evolutionary stage of the stellar companion, the timescales of mass loss prior to the merger, the timescale of the merger and resulting tidal disruption of the companion itself), an extension of this model that is not tuned to LFBOT observables may reproduce the observed properties of AT 2019ijn. 


\subsubsection{SLSN-magnetar}

The peak optical luminosity of AT 2019ijn agrees with that commonly seen in SLSNe, a class of supernovae that show peak optical luminosities of $> -21 $ mag \citep{gal-yam2012}. One of the most popular explanations for the origins of SLSNe is the spin-down luminosity from a newborn magnetar \citep{Kasen2010, Woosley2010, metzger+15,nicholl+17,margalit+18} with strong magnetic fields \citep[$B > 10^{13}~\rm{G}$;][]{nicholl+15} and extremely short spin periods ($P\sim 1~\rm{ms}$), which have also been suggested to power LFBOTs and GRBs. In typical magnetar scenarios, the optical emission is produced via thermal heating of the SN ejecta, with very late-time radio emission possible from a pulsar-wind nebula post-SN, peaking on timescales of $\sim 10~\rm{yrs}$ post-SN \citep{metzger+15, margalit+18, nicholl+17}. It has also been suggested that some magnetars may be capable of producing relativistic jets \citep{gottlieb+metzger24, desai+26_axv}, in which case shocks due to jet breakout may contribute to the optical emission, and more luminous early-time radio emission may arise from the jet interaction with the CSM. 

In the case of AT 2019ijn, the luminosity and color at peak light are broadly consistent with expectations from magnetar-powered SLSNe. The asymmetry of the light curve again presents a challenge, as SLSNe rarely produce optical light curves with rise times of $t< 10~\rm{d}$ and typically exhibit much slower declines \citep{nicholl+15, lunnan+18}. Models of optical light curves produced by shock breakout of a jet + cocoon predict short rise times with longer decay times \citep[e.g.][]{gottlieb+metzger24}, though these predict peak emission that is still more quickly decaying and far less luminous than AT 2019ijn. The optical light curve can, however, be reproduced by a more classic shock breakout of SN ejecta (see Section \ref{sec:optmodel}), or following the model of \cite{nicholl+17} (section \ref{sec:magnetar}). The late-time broad H$\alpha$ line in this scenario would again be expected from interaction of a slow-moving outflow from the SN with the CSM. The host galaxy of AT 2019ijn is broadly similar to SLSNe hosts, though SLSN hosts tend to be less massive, more star-forming, and less metal-rich than that of AT 2019ijn, which is instead more similar to LFBOT hosts \citep[e.g.][]{schulze+21, nugent+26_axv}. 

The observed radio emission is challenging to explain via this model, even in the case of jet production. The observed radio luminosity and energies are much greater than the two observed radio SLSNe to date \citep{eftekhari+19, margutti+23}, though both have delayed-onset radio emission observable years post-explosion (see Figure \ref{fig:LC_panel}c). Jets anticipated in a SLSN-magnetar scenario are expected to be less energetic than GRB jets \citep{Gottlieb2022, gottlieb+metzger24, desai+26_axv}, which are already generally less luminous and have shorter durations than the radio emission of AT 2019ijn. Powering AT 2019ijn's long-lived and luminous radio emission would require an immense reserve of rotational energy from the magnetar and likely a period of continuous energy injection following the initial jet launch, which is not naturally predicted in the millisecond magnetar scenario. Despite this, \cite{matsumoto+metzger23} have proposed a millisecond magnetar origin or accreting stellar-mass BH for AT 2022cmc, which has similar radio energetics to AT 2019ijn, where the central engine shares its energy equally between SN heating and the relativistic jet. Considering the magnetar interpretation, we note that AT 2022cmc's jet was viewed on-axis, which is in line with the expectation of radio emission from magnetar jets to only be visible for observers within a very narrow range of the narrow opening angle of the jet and inconsistent with our off-axis interpretation for AT 2019ijn. A magnetar would also be expected to launch the relativistic jet near-immediately after its birth, which we cannot constrain for AT 2019ijn based on available observations (Section \ref{sec:radenergy}). Though we cannot completely rule out this scenario, we disfavor it as compared to scenarios involving BH accretion. Considering the accreting stellar-mass BH interpretation, we cannot rule out this possibility for AT 2019ijn.

\subsection{Comparison to prior jetted TDEs in the literature} \label{sec:TDEs}

Jetted TDEs are an incredibly rare class of transient, with only a few candidates observed to date. The most oft-cited examples include Swift J1644 \citep{Burrows+11, Zauderer+11}, Swift J2058 \citep{cenko+12}, Swift J1112, \citep{brown+15}, which are X-ray-selected events, and the optically-selected AT 2022cmc \citep{andreoni+22}, all believed to have been viewed on-axis. Recently, possible off-axis jetted TDE candidates have also been proposed via radio observations, including AT 2018hyz \citep{short+20, gomez+20, cendes+22, sfaradi+24a, cendes+26}, a TDE with a delayed radio flare now thought to be an off-axis jet, and radio-selected candidates such as VT J0243 \citep{Somalwar2023} and the events in the radio TDE sample from \cite{Dykaar2024}. We discuss the similarities of AT 2019ijn to these examples.

\textit{Optical:} Both AT 2022cmc and AT 2018hyz produced bright optical transients. AT 2018hyz produced an optical flare first observed at its peak with a g-band luminosity of $L_g = 3.6\times10^{43}~\rm{erg/s}$ \citep{gomez+20, vanvelzen+21}, which then declined with an exponential decay timescale of $\tau_{\rm{dec}} = 52.5$ days. This emission is consistent with that of other optical, non-jetted TDEs \citep{vanvelzen+21, yao+23, Hammerstein+23a}, which is thought to originate from the circularization of stellar debris and accretion onto the black hole. The bolometric light curve features a ``bump'' around 50 d post-$t_0$, which has been suggested to be due to accretion from a thin disk, and flattens at late ($> 250$ d) times, thought to be due to a sudden outflow of material \citep{gomez+20}. AT 2022cmc produced a fast, red initial optical flare before transitioning to a more slowly-evolving ``plateau'' phase that was initially blue but reddened over time \citep{andreoni+22, hammerstein+26}. The initial $\sim 1$ d red flare, with a peak luminosity of $L_g = 3.6\times10^{45}~\rm{erg/s}$, is thought to be from the forward shock from the on-axis jet as it collides with the surrounding medium. The ``plateau'' phase, with an exponential decay timescale of $\tau_{\rm{dec}} = 1.5$ d \citep{hammerstein+26}, is thought to be due to thermal emission from optically thick outflows and the accretion of stellar debris onto the black hole. The reddening of this phase at late times is attributed to increasing contribution from the host galaxy as the transient emission fades. Both AT 2018hyz and AT 2022cmc display blackbody properties similar to the population of optical TDEs, with blackbody temperatures of $T_{\rm{bb}} \sim 10^4 - 10^5~\rm{K}$ and luminosities of $L_{\rm{bb}} \sim 10^{44} - 10^{45}~\rm{erg/s}$. Swift J2058, Swift J1112, and VT J0243 also displayed optical flares, which peaked between $10^{42}-10^{43}~\rm{erg/s}$. 

The optical emission from AT 2019ijn also displays peak luminosities similar to optical, non-jetted TDEs (Figure \ref{fig:phase}) and blackbody properties expected of the optical TDE population. If the initial red flare from AT 2022cmc originates from the relativistic jet and requires an on-axis viewing angle, then the lack of this phase in AT 2019ijn may be due to the predicted off-axis viewing angle. The following plateau phase in AT 2022cmc is more similar to the light curve in AT 2019ijn, including the initial blue phase that reddens with time. The morphology of AT 2019ijn's light curve is also similar to that of AT 2018hyz, albeit more luminous. The similarity of these properties to AT 2018hyz, the plateau phase of AT 2022cmc, and the general population of optically-selected TDEs may suggest a similar origin for the emission, i.e. circularization of stellar debris and accretion onto the black hole. 

The broad H$\alpha$ feature in AT 2019ijn has not been seen at late times in prior jetted TDEs, but is sometimes seen in non-jetted TDEs \citep{roth+kasen18, Charalampopoulos+22, somalwar+25a, somalwar+25b}. AT 2018hyz displayed a broad, double-peaked H$\alpha$ line at early times, while AT 2022cmc showed featureless optical spectra at early times, though a robust comparison cannot be made as spectra at the late epochs of our LRIS spectra are not available for these sources. As we do not have early-time spectra for AT 2019ijn, we cannot rule out possible similarities to prior transients at early times. Continued monitoring of these events will reveal if they undergo similar spectral evolution to AT 2019ijn.

\textit{Radio:} The radio emission of AT 2019ijn is incredibly similar to that seen in prior jetted TDEs. Both Swift J1644 and AT 2022cmc produced radio emission with a peak luminosity of $\nu L_{\nu} > 10^{41}~\rm{erg/s}$, which remained visible for years post-disruption. Late-time modeling of J1644 and 2022cmc have suggested circum-transient densities of $n \sim 0.1-1~\rm{cm^{-3}}$ at radii of $R \sim 10^{18}~\rm{cm}$ \citep{eftekhari+18, rhodes+25}, which agrees with our estimates of the medium surrounding AT 2019ijn. AT 2018hyz produced a delayed radio flare that was last measured at a luminosity of $\nu L_{\nu} \sim 10^{40}~\rm{erg/s}$ and is still rising, with similar inferred densities of $n \sim 0.01-1~\rm{cm^{-3}}$ at a radius of $R \sim 10^{18}~\rm{cm}$ \citep{sfaradi+24a, cendes+26}. Prior events with sufficient radio coverage for detailed modeling assume or infer energies of near $E\sim 10^{52}~\rm{erg/s}$, as in AT 2019ijn. Swift J2058 was also noted to have a flat radio spectral index at $\sim 40-60$ days, possibly attributed to the jet structure and surrounding environment \citep{cenko+12}. 

\textit{X-ray:} All prior on-axis jetted TDEs showed luminous X-ray emission beginning at early times. Some prior off-axis candidates also have early detections of X-ray emission, including AT 2018hyz, which was initially detected up to 86 d and faded by a factor of 2 by 1200 d \citep{cendes+22}. VT J0243 was nominally undetected in X-ray until $\sim 1000~\rm{d}$ after the first radio detection, at an isotropic luminosity of $\sim 10^{42}~\rm{erg/s}$, though there were not enough observations to constrain the true onset of the X-ray emission. As we only have one X-ray measurement of AT 2019ijn, we cannot constrain the X-ray behavior, which may be consistent with that of either AT 2018hyz or VT J0243. 

\textit{Host galaxy and black hole mass:} The properties of the host galaxies of past jetted TDEs vary between cases. The host of Swift J1644 has been estimated at a mass of $\rm{log}\frac{M_*}{M_\odot} = 9.14$ with an SFR of $0.03~\rm{M_\odot/yr}$ \citep{yoon+15}, which is similar in mass to AT 2019ijn ($\rm{log}\frac{M_*}{M_\odot} \sim 9.05$), though less star-forming ($\sim 0.1~\rm{M_\odot/yr}$ for AT 2019ijn). Other candidates have host masses ranging from $10^9 - 10^{11}~\rm{M_\odot}$, with star formation rates ranging from very low (AT 2018hyz, in an E+A galaxy) to $< 5~\rm{M_\odot/yr}$ (VT J0243). The host of AT 2022cmc has yet to be detected, which may indicate similarity with the faint host galaxy of AT 2019ijn. The black hole masses for each prior candidate are nominally in the SMBH regime of $10^6 - 10^8~\rm{M_\odot}$, though recent work has suggested black hole masses of $\leq \rm{few} \times 10^5~\rm{M_\odot}$ for Swift J1644 and Swift J2058, and $\leq 10^5~\rm{M_\odot}$ for AT 2022cmc \citep{Eftekhari2024}, which are well in agreement with our estimate for AT 2019ijn.

\subsection{Data used in figures} \label{sec:figs}

Data used in Figures \ref{fig:LC_panel}, \ref{fig:phase}, and Extended Data Figure \ref{fig:colev} were taken from the literature as follows: 

\textbf{Figure \ref{fig:LC_panel}b}: Comparison of g-band optical light curves of AT 2019ijn and other energetic transients in the literature. All light curves were corrected for Galactic extinction, and k-corrections were applied as in Section \ref{sec:kcorr}. Upper limits not shown.

\noindent \textit{SLSNe}: Light curves shown for PTF 10hgi \citep{decia+18}, SN 2017ens \citep{margutti+23}, and SN 2018bgv \citep{lunnan+20}.

\noindent \textit{GRB-SNe}: Light curve shown for SN2011kl/GRB 111209A \citep{greiner+15}. 

\noindent \textit{TDEs}: Light curves shown for ASASSN-14li \citep{Holoien2014}, AT 2018hyz \citep{gomez+20}, and AT 2019dsg \citep{Cannizzaro2021}. 

\noindent \textit{LFBOTs}: Light curves shown for AT 2018cow \citep{Perley+19}, ZTF18abvkwla \citep{ho+20} and AT 2020xnd \citep{perley+21}. 

\noindent \textit{Jetted TDEs}: Light curve shown for AT 2022cmc \citep{andreoni+22, hammerstein+26}.

\noindent \textit{AT 2024puz}: Light curve taken from \citep{somalwar+25}.

\textbf{Figure \ref{fig:LC_panel}c}: Comparison of radio light curves of AT 2019ijn and other energetic transients in the literature. Light curves were selected for the greatest temporal coverage at a single frequency below 10 GHz for best comparison with the VLA observations of AT 2019ijn. Light curves are not k-corrected due to lack of available spectral information for all sources. Specifically: 

\noindent \textit{Swift J1644}: 5 GHz light curve shown, with data taken from \cite{Zauderer+11, berger+12, zauderer+13,eftekhari+18}. 

\noindent \textit{AT 2022cmc}: 5 GHz light curve shown, with data taken from \cite{Rhodes2023, rhodes+25}. Early-time data is available at higher frequencies. $3\sigma$ upper limits are denoted downward triangles and dashed lines. 

\noindent \textit{ASASSN-14li}: 5 GHz light curve shown, with data taken from \cite{alexander+16}. 

\noindent \textit{AT 2018hyz}: 5 GHz light curve shown, with data taken from \cite{sfaradi+24a, cendes+26}. 

\noindent \textit{AT 2019dsg}: 9 GHz light curve shown, with data taken from \cite{stein+21, cendes+21}. 

\noindent \textit{AT 2018cow}: 9 GHz light curve shown, with data taken from \cite{ho+19, margutti+19}. 

\noindent \textit{ZTF18abvkwla}: 10 GHz light curve shown, with data taken from \cite{ho+20}. 

\noindent \textit{AT 2020xnd}: 5 GHz light curve shown, with data taken from \cite{bright+22}. $3\sigma$ upper limits are denoted downward triangles and dashed lines. 

\noindent \textit{GRB 030329A}: 5 GHz light curve shown, with data taken from \cite{berger+03, vanderhorst+07}. 

\noindent \textit{GRB 130427A}: 5 GHz light curve shown, with data taken from \cite{perley+14, vanderHorst2014}. 

\noindent \textit{GRB 221009A}: 5 GHz light curve shown, with data taken from \cite{laskar+23}. 

\noindent \textit{SN 1998bw}: 6 GHz light curve shown, with data taken from \cite{kulkarni+98}. 

\noindent \textit{VT1210+4956}: 3 GHz light curve shown, with data taken from \cite{dong+21}. Upper limit by FIRST at 20.6 years before VLASS detection not shown. 

\noindent \textit{PTF 10hgi}: Single measurement at 6 GHz shown, with data taken from \cite{eftekhari+19}. 

\noindent \textit{SN 2017ens}: 6 GHz light curve shown, with data taken from \cite{margutti+23}. $3\sigma$ upper limits are denoted downward triangles and dashed lines.

\textbf{Figure \ref{fig:phase}}: Figure adapted from \cite{ho+22b, andreoni+22}. Sample of FBOTs taken from \cite{ho+23}, sample of SLSNe taken from \cite{perley+20}, and sample of TDEs taken from \cite{yao+23}. LFBOTs taken from \cite{ho+22a, ho+23, ho+23b, perley+21, Pursiainen+25, chrimes+24}. Jetted TDE AT 2022cmc taken from \cite{andreoni+22}.  ``Fast'' flare is taken as optical emission for $t < 4$ days from detection in observer frame, while the ``slow'' flare is taken as optical emission for $t \geq 4$ days from detection. AT 2024puz taken from \cite{somalwar+25}. Peak luminosity taken from k-corrected g-band absolute magnitude for all classes; when the k-corrected luminosity is not available from the literature, an estimated k-correction is applied as in Section \ref{sec:kcorr}. Rise and decline timescales are both taken in the rest frame.

\textbf{Extended Data Figure \ref{fig:colev}}: Comparison of g-r color evolution with time of AT 2019ijn and other energetic transients in the literature. For all objects, g and r light curves were corrected for Galactic extinction and interpolated (rather than fit with any model, to avoid making undue assumptions about the underlying physics) from available data over the duration of the light curve. Specifically: 

\noindent \textit{SN 2018bgv}: g, r band data taken from \cite{lunnan+20}. Upper limits not included in the analysis. Chosen for comparison as a SLSN with short optical rise. 

\noindent \textit{SN 2017ens}: g, r band data taken from \cite{chen+18}. 

\noindent \textit{PTF 10hgi}: g, r band data taken from \cite{decia+18}. 

\noindent \textit{AT 2022cmc}: g, r band data taken from \cite{andreoni+22, hammerstein+26}. Light curves made from combining g, r and ZTF g, ZTF r measurements. Upper limits not included in the analysis. 

\noindent \textit{AT 2018hyz}: g, r band data taken from \cite{gomez+20}. AT 2018hyz is color-coded as a non-relativistic TDE in this figure, though it is thought to have an off-axis jet, as described above. 

\noindent \textit{AT 2019dsg}: g, r band data taken from \cite{stein+21}. Upper limits not included in the analysis. 

\noindent \textit{AT 2018cow}: g, r band data taken from \cite{prentice+18}. Light curves made from combining g, g' and r, r' measurements. 

\noindent \textit{ZTF18abvkwla}: g, r band data taken from \cite{ho+20}. Upper limits not included in the analysis. 

\noindent \textit{AT 2020xnd}: g, r, band data taken from \cite{perley+21}. Upper limits not included in the analysis. 

\noindent \textit{SN 2011kl/GRB 111209A}: g, r band data taken from \cite{greiner+15}. 

\noindent \textit{GRB 130427A}: g', r' band data taken from \cite{vestrand+14}.

\noindent \textit{AT 2024puz}: g, r, band data taken from \cite{somalwar+25} and ZTF forced photometry service. Upper limits not included in the analysis. 

\textbf{Additional information for Figure \ref{fig:radiojet} is as follows:} 
Models derived in this work (Methods Section \ref{sec:radenergy}) and shown with the following parameters for all epochs: $E = 2\times10^{52}~\rm{erg/s}$, $\theta_{\rm{obs}} = 0.7~\rm{rad}$, $\theta_j = 0.1~\rm{rad}$, $\epsilon_B = 0.1$, $p=3$. $\epsilon_e = 0.1$ for all epochs except $\Delta t = 1997~\rm{d}$, where it is instead assumed to be $\epsilon_e = 0.08$. We assumed a jet with power-law wings, $E_{\rm{iso}}(\theta) \propto \theta^{-10}$ for $\theta>\theta_j$ and our modeling did not include jet spreading. Since the code is set for either a constant or a wind density profile, while the data does not fit any of them, we simulate each epoch using a different normalization of a wind profile $\rho=Ar^{-2}$. This is justified since the emission at any epoch is dominated by the hydrodynamic evolution in the last dynamical time of the blast wave. The values of $A$ that we find for the different epochs are  $A = [8, 7, 4.8, 3.6, 3, 2.8]A_*$ (where $A_* = 5\times10^{11}~\rm{g/cm^3}$ as in \citep{Chevalier2000}) for $\Delta t = [412, 641, 998, 1333, 1997, 2331]~\rm{d}$, respectively. This slow continuous decay in the value of $A$ suggests a density profile of the form $Ar^{-k}$, where $k \approx 2.5$. 

\newpage
\subsection{Supplementary Tables}

Supplementary tables are available as separate files via the submission system. 

Supplementary Table 1: Log of radio observations of AT 2019ijn

Supplementary Table 2: ZTF forced photometry of AT 2019ijn

\bmhead{Acknowledgements}

The authors would like to thank Brian Metzger, Sterl Phinney, Shri Kulkarni, and Dale Frail for assistance in discussion and interpretation of the source. We also thank Adolfo Carvalho for assistance with the infrared data reduction. 

The National Radio Astronomy Observatory and Green Bank Observatory are facilities of the U.S. National Science Foundation operated under cooperative agreement by Associated Universities, Inc.

Some of the data presented herein were obtained at Keck Observatory, which is a private 501(c)3 non-profit organization operated as a scientific partnership among the California Institute of Technology, the University of California, and the National Aeronautics and Space Administration. The Observatory was made possible by the generous financial support of the W. M. Keck Foundation. The authors wish to recognize and acknowledge the very significant cultural role and reverence that the summit of Maunakea has always had within the Native Hawaiian community. We are most fortunate to have the opportunity to conduct observations from this mountain. 

This research has made use of the Keck Observatory Archive (KOA), which is operated by the W. M. Keck Observatory and the NASA Exoplanet Science Institute (NExScI), under contract with the National Aeronautics and Space Administration.

Based on observations obtained with the Samuel Oschin Telescope 48-inch and the 60-inch Telescope at the Palomar Observatory as part of the Zwicky Transient Facility project. ZTF is supported by the National Science Foundation under Grant No. AST-2034437 and a collaboration including Caltech, IPAC, the Weizmann Institute for Science, the Oskar Klein Center at Stockholm University, the University of Maryland, Deutsches Elektronen-Synchrotron and Humboldt University, the TANGO Consortium of Taiwan, the University of Wisconsin at Milwaukee, Trinity College Dublin, Lawrence Livermore National Laboratories, and IN2P3, France. Operations are conducted by COO, IPAC, and UW.

This paper includes data gathered with the 6.5 meter Magellan Telescopes located at Las Campanas Observatory, Chile.

The Pan-STARRS1 Surveys (PS1) and the PS1 public science archive have been made possible through contributions by the Institute for Astronomy, the University of Hawaii, the Pan-STARRS Project Office, the Max-Planck Society and its participating institutes, the Max Planck Institute for Astronomy, Heidelberg and the Max Planck Institute for Extraterrestrial Physics, Garching, The Johns Hopkins University, Durham University, the University of Edinburgh, the Queen's University Belfast, the Harvard-Smithsonian Center for Astrophysics, the Las Cumbres Observatory Global Telescope Network Incorporated, the National Central University of Taiwan, the Space Telescope Science Institute, the National Aeronautics and Space Administration under Grant No. NNX08AR22G issued through the Planetary Science Division of the NASA Science Mission Directorate, the National Science Foundation Grant No. AST–1238877, the University of Maryland, Eotvos Lorand University (ELTE), the Los Alamos National Laboratory, and the Gordon and Betty Moore Foundation.

The Legacy Surveys consist of three individual and complementary projects: the Dark Energy Camera Legacy Survey (DECaLS; Proposal ID \#2014B-0404; PIs: David Schlegel and Arjun Dey), the Beijing-Arizona Sky Survey (BASS; NOAO Prop. ID \#2015A-0801; PIs: Zhou Xu and Xiaohui Fan), and the Mayall z-band Legacy Survey (MzLS; Prop. ID \#2016A-0453; PI: Arjun Dey). DECaLS, BASS and MzLS together include data obtained, respectively, at the Blanco telescope, Cerro Tololo Inter-American Observatory, NSF’s NOIRLab; the Bok telescope, Steward Observatory, University of Arizona; and the Mayall telescope, Kitt Peak National Observatory, NOIRLab. Pipeline processing and analyses of the data were supported by NOIRLab and the Lawrence Berkeley National Laboratory (LBNL). The Legacy Surveys project is honored to be permitted to conduct astronomical research on Iolkam Du’ag (Kitt Peak), a mountain with particular significance to the Tohono O’odham Nation.

NOIRLab is operated by the Association of Universities for Research in Astronomy (AURA) under a cooperative agreement with the National Science Foundation. LBNL is managed by the Regents of the University of California under contract to the U.S. Department of Energy.

This project used data obtained with the Dark Energy Camera (DECam), which was constructed by the Dark Energy Survey (DES) collaboration. Funding for the DES Projects has been provided by the U.S. Department of Energy, the U.S. National Science Foundation, the Ministry of Science and Education of Spain, the Science and Technology Facilities Council of the United Kingdom, the Higher Education Funding Council for England, the National Center for Supercomputing Applications at the University of Illinois at Urbana-Champaign, the Kavli Institute of Cosmological Physics at the University of Chicago, Center for Cosmology and Astro-Particle Physics at the Ohio State University, the Mitchell Institute for Fundamental Physics and Astronomy at Texas A\&M University, Financiadora de Estudos e Projetos, Fundacao Carlos Chagas Filho de Amparo, Financiadora de Estudos e Projetos, Fundacao Carlos Chagas Filho de Amparo a Pesquisa do Estado do Rio de Janeiro, Conselho Nacional de Desenvolvimento Cientifico e Tecnologico and the Ministerio da Ciencia, Tecnologia e Inovacao, the Deutsche Forschungsgemeinschaft and the Collaborating Institutions in the Dark Energy Survey. The Collaborating Institutions are Argonne National Laboratory, the University of California at Santa Cruz, the University of Cambridge, Centro de Investigaciones Energeticas, Medioambientales y Tecnologicas-Madrid, the University of Chicago, University College London, the DES-Brazil Consortium, the University of Edinburgh, the Eidgenossische Technische Hochschule (ETH) Zurich, Fermi National Accelerator Laboratory, the University of Illinois at Urbana-Champaign, the Institut de Ciencies de l’Espai (IEEC/CSIC), the Institut de Fisica d’Altes Energies, Lawrence Berkeley National Laboratory, the Ludwig Maximilians Universitat Munchen and the associated Excellence Cluster Universe, the University of Michigan, NSF’s NOIRLab, the University of Nottingham, the Ohio State University, the University of Pennsylvania, the University of Portsmouth, SLAC National Accelerator Laboratory, Stanford University, the University of Sussex, and Texas A\&M University.

BASS is a key project of the Telescope Access Program (TAP), which has been funded by the National Astronomical Observatories of China, the Chinese Academy of Sciences (the Strategic Priority Research Program “The Emergence of Cosmological Structures” Grant \# XDB09000000), and the Special Fund for Astronomy from the Ministry of Finance. The BASS is also supported by the External Cooperation Program of Chinese Academy of Sciences (Grant \# 114A11KYSB20160057), and Chinese National Natural Science Foundation (Grant \# 12120101003, \# 11433005).

The Legacy Survey team makes use of data products from the Near-Earth Object Wide-field Infrared Survey Explorer (NEOWISE), which is a project of the Jet Propulsion Laboratory/California Institute of Technology. NEOWISE is funded by the National Aeronautics and Space Administration.

The Legacy Surveys imaging of the DESI footprint is supported by the Director, Office of Science, Office of High Energy Physics of the U.S. Department of Energy under Contract No. DE-AC02-05CH1123, by the National Energy Research Scientific Computing Center, a DOE Office of Science User Facility under the same contract; and by the U.S. National Science Foundation, Division of Astronomical Sciences under Contract No. AST-0950945 to NOAO.

\bmhead{Funding}
D.L. and G. H. disclose support from Schmidt Sciences and the United States Israel Binational Science Foundation (Grant number 2020203). A.H. is grateful for the support by the Israel Science Foundation (ISF grant 1679/23). This research was also supported in part by grant NSF PHY-2309135 to the Kavli Institute for Theoretical Physics (KITP).

\bmhead{Author Contributions}
D.L. and G.H. determined the search strategy that led to contextual identification of the source. D.D. implemented the source extraction pipeline for identification of the source in VLASS with assistance from J.S. J.S. and D.D. led early radio and optical follow-up, with assistance from G.H., V.R., and J.M. Later radio, optical, and MOSFIRE near-infrared follow-up was obtained by D.L., with assistance from G.H., D.D., J.M., and V.R. Near-infrared follow-up with WIRC and FourStar was obtained by S.O. and D.L. VLA and Keck data were reduced and analyzed by D.L. with assistance from G.H., J.S., D.D, and J.M. FourStar data was reduced by D.K. and S.O. Radio modeling of the source was done by E.N. and K.H. VLASS in-band spectra were prepared by S.M. All further analysis was performed by D.L. J.S., A.H., V.R., and C.L. contributed significant discussion to the interpretation of the source. Manuscript was prepared by D.L. and G.H., with contributions from E.N. and K.H., and reviewed by all authors. 

\bmhead{Competing Interest Declaration}
The authors declare no competing interests. 

\bmhead{Correspondence}
Correspondence and requests for materials should be addressed to Delina Levine at dlevine2@caltech.edu. 

\bmhead{Code Availability}
Python code used to produce figures can be provided upon request. 

\bmhead{Data Availability}
Reduced radio SEDs and ZTF photometry are provided in Supplementary Tables 1 and 2. VLASS, PanSTARRS, and DESI-Legacy data are available from their respective public databases. VLA and Keck data are publicly available from their respective data archives. 

\bibliography{at2019ijn}

\begin{thebibliography}{100}
\expandafter\ifx\csname url\endcsname\relax
  \def\url#1{\burl{#1}}\fi
\expandafter\ifx\csname urlprefix\endcsname\relax\def\urlprefix{URL }\fi
\providecommand{\bibinfo}[2]{#2}
\providecommand{\eprint}[2][]{\url{#2}}
\providecommand{\doi}[1]{\url{https://doi.org/#1}}
\bibcommenthead

\bibitem{gal-yam2012}
\bibinfo{author}{{Gal-Yam}, A.}
\newblock \bibinfo{title}{{Luminous Supernovae}}.
\newblock \emph{\bibinfo{journal}{Science}} \textbf{\bibinfo{volume}{337}}, \bibinfo{pages}{927} (\bibinfo{year}{2012}).

\bibitem{tsuna+lu25}
\bibinfo{author}{{Tsuna}, D.} \& \bibinfo{author}{{Lu}, W.}
\newblock \bibinfo{title}{{Stellar Tidal Disruptions by Newborn Neutron Stars or Black Holes: A Mechanism for Hydrogen-poor (Super)luminous Supernovae and Fast Blue Optical Transients}}.
\newblock \emph{\bibinfo{journal}{\apj}} \textbf{\bibinfo{volume}{986}}, \bibinfo{pages}{84} (\bibinfo{year}{2025}).

\bibitem{rees+88}
\bibinfo{author}{{Rees}, M.~J.}
\newblock \bibinfo{title}{{Tidal disruption of stars by black holes of {}10$^{6}$-{}10$^{8}$ solar masses in nearby galaxies}}.
\newblock \emph{\bibinfo{journal}{\nat}} \textbf{\bibinfo{volume}{333}}, \bibinfo{pages}{523--528} (\bibinfo{year}{1988}).

\bibitem{perley+19a}
\bibinfo{author}{{Perley}, D.~A.} \emph{et~al.}
\newblock \bibinfo{title}{{The fast, luminous ultraviolet transient AT2018cow: extreme supernova, or disruption of a star by an intermediate-mass black hole?}}
\newblock \emph{\bibinfo{journal}{\mnras}} \textbf{\bibinfo{volume}{484}}, \bibinfo{pages}{1031--1049} (\bibinfo{year}{2019}).

\bibitem{yao+23}
\bibinfo{author}{{Yao}, Y.} \emph{et~al.}
\newblock \bibinfo{title}{{Tidal Disruption Event Demographics with the Zwicky Transient Facility: Volumetric Rates, Luminosity Function, and Implications for the Local Black Hole Mass Function}}.
\newblock \emph{\bibinfo{journal}{\apjl}} \textbf{\bibinfo{volume}{955}}, \bibinfo{pages}{L6} (\bibinfo{year}{2023}).

\bibitem{ho+23}
\bibinfo{author}{{Ho}, A. Y.~Q.} \emph{et~al.}
\newblock \bibinfo{title}{{A Search for Extragalactic Fast Blue Optical Transients in ZTF and the Rate of AT2018cow-like Transients}}.
\newblock \emph{\bibinfo{journal}{\apj}} \textbf{\bibinfo{volume}{949}}, \bibinfo{pages}{120} (\bibinfo{year}{2023}).

\bibitem{ho+19}
\bibinfo{author}{{Ho}, A. Y.~Q.} \emph{et~al.}
\newblock \bibinfo{title}{{AT2018cow: A Luminous Millimeter Transient}}.
\newblock \emph{\bibinfo{journal}{\apj}} \textbf{\bibinfo{volume}{871}}, \bibinfo{pages}{73} (\bibinfo{year}{2019}).

\bibitem{Zauderer+11}
\bibinfo{author}{{Zauderer}, B.~A.} \emph{et~al.}
\newblock \bibinfo{title}{{Birth of a relativistic outflow in the unusual {\ensuremath{\gamma}}-ray transient Swift J164449.3+573451}}.
\newblock \emph{\bibinfo{journal}{\nat}} \textbf{\bibinfo{volume}{476}}, \bibinfo{pages}{425--428} (\bibinfo{year}{2011}).

\bibitem{andreoni+22}
\bibinfo{author}{{Andreoni}, I.} \emph{et~al.}
\newblock \bibinfo{title}{{A very luminous jet from the disruption of a star by a massive black hole}}.
\newblock \emph{\bibinfo{journal}{\nat}} \textbf{\bibinfo{volume}{612}}, \bibinfo{pages}{430--434} (\bibinfo{year}{2022}).

\bibitem{Burrows+11}
\bibinfo{author}{{Burrows}, D.~N.} \emph{et~al.}
\newblock \bibinfo{title}{{Relativistic jet activity from the tidal disruption of a star by a massive black hole}}.
\newblock \emph{\bibinfo{journal}{\nat}} \textbf{\bibinfo{volume}{476}}, \bibinfo{pages}{421--424} (\bibinfo{year}{2011}).

\bibitem{klencki+metzger25_axv}
\bibinfo{author}{{Klencki}, J.} \& \bibinfo{author}{{Metzger}, B.~D.}
\newblock \bibinfo{title}{{Luminous Fast Blue Optical Transients as ``Failed'' Gravitational Wave Sources: Helium Core$-$Black Hole Mergers Following Delayed Dynamical Instability}}.
\newblock \emph{\bibinfo{journal}{arXiv e-prints}} \bibinfo{pages}{arXiv:2510.09745} (\bibinfo{year}{2025}).

\bibitem{teboul+23}
\bibinfo{author}{{Teboul}, O.} \& \bibinfo{author}{{Metzger}, B.~D.}
\newblock \bibinfo{title}{{A Unified Theory of Jetted Tidal Disruption Events: From Promptly Escaping Relativistic to Delayed Transrelativistic Jets}}.
\newblock \emph{\bibinfo{journal}{\apjl}} \textbf{\bibinfo{volume}{957}}, \bibinfo{pages}{L9} (\bibinfo{year}{2023}).

\bibitem{lacy+20}
\bibinfo{author}{{Lacy}, M.} \emph{et~al.}
\newblock \bibinfo{title}{{The Karl G. Jansky Very Large Array Sky Survey (VLASS). Science Case and Survey Design}}.
\newblock \emph{\bibinfo{journal}{\pasp}} \textbf{\bibinfo{volume}{132}}, \bibinfo{pages}{035001} (\bibinfo{year}{2020}).

\bibitem{Nordin+19}
\bibinfo{author}{{Nordin}, J.} \emph{et~al.}
\newblock \bibinfo{title}{{ZTF Transient Discovery Report for 2019-06-28}}.
\newblock \emph{\bibinfo{journal}{Transient Name Server Discovery Report}} \textbf{\bibinfo{volume}{2019-1095}}, \bibinfo{pages}{1} (\bibinfo{year}{2019}).

\bibitem{Perley+11}
\bibinfo{author}{{Perley}, R.~A.}, \bibinfo{author}{{Chandler}, C.~J.}, \bibinfo{author}{{Butler}, B.~J.} \& \bibinfo{author}{{Wrobel}, J.~M.}
\newblock \bibinfo{title}{{The Expanded Very Large Array: A New Telescope for New Science}}.
\newblock \emph{\bibinfo{journal}{\apjl}} \textbf{\bibinfo{volume}{739}}, \bibinfo{pages}{L1} (\bibinfo{year}{2011}).

\bibitem{ho+22b}
\bibinfo{author}{{Ho}, A. Y.~Q.} \emph{et~al.}
\newblock \bibinfo{title}{{Cosmological Fast Optical Transients with the Zwicky Transient Facility: A Search for Dirty Fireballs}}.
\newblock \emph{\bibinfo{journal}{\apj}} \textbf{\bibinfo{volume}{938}}, \bibinfo{pages}{85} (\bibinfo{year}{2022}).

\bibitem{hammerstein+26}
\bibinfo{author}{{Hammerstein}, E.} \emph{et~al.}
\newblock \bibinfo{title}{{The Jetted Tidal Disruption Event AT 2022cmc: Investigating Connections to the Optical Tidal Disruption Event Population and Spectral Subclasses through Late-time Follow-up}}.
\newblock \emph{\bibinfo{journal}{\apj}} \textbf{\bibinfo{volume}{996}}, \bibinfo{pages}{143} (\bibinfo{year}{2026}).

\bibitem{sfaradi+24}
\bibinfo{author}{{Sfaradi}, I.} \emph{et~al.}
\newblock \bibinfo{title}{{The dense and non-homogeneous circumstellar medium revealed in radio wavelengths around the Type Ib SN 2019oys}}.
\newblock \emph{\bibinfo{journal}{\aap}} \textbf{\bibinfo{volume}{686}}, \bibinfo{pages}{A129} (\bibinfo{year}{2024}).

\bibitem{falcke+00}
\bibinfo{author}{{Falcke}, H.}, \bibinfo{author}{{Nagar}, N.~M.}, \bibinfo{author}{{Wilson}, A.~S.} \& \bibinfo{author}{{Ulvestad}, J.~S.}
\newblock \bibinfo{title}{{Radio Sources in Low-Luminosity Active Galactic Nuclei. II. Very Long Baseline Interferometry Detections of Compact Radio Cores and Jets in a Sample of LINERs}}.
\newblock \emph{\bibinfo{journal}{\apj}} \textbf{\bibinfo{volume}{542}}, \bibinfo{pages}{197--200} (\bibinfo{year}{2000}).

\bibitem{nagar+02}
\bibinfo{author}{{Nagar}, N.~M.}, \bibinfo{author}{{Falcke}, H.}, \bibinfo{author}{{Wilson}, A.~S.} \& \bibinfo{author}{{Ulvestad}, J.~S.}
\newblock \bibinfo{title}{{Radio sources in low-luminosity active galactic nuclei. III. ``AGNs`` in a distance-limited sample of ``LLAGNs''}}.
\newblock \emph{\bibinfo{journal}{\aap}} \textbf{\bibinfo{volume}{392}}, \bibinfo{pages}{53--82} (\bibinfo{year}{2002}).

\bibitem{fender+01}
\bibinfo{author}{{Fender}, R.~P.}
\newblock \bibinfo{title}{{Powerful jets from black hole X-ray binaries in low/hard X-ray states}}.
\newblock \emph{\bibinfo{journal}{\mnras}} \textbf{\bibinfo{volume}{322}}, \bibinfo{pages}{31--42} (\bibinfo{year}{2001}).

\bibitem{fender06}
\bibinfo{author}{{Fender}, R.}
\newblock \bibinfo{title}{ in \textit{{Jets from X-ray binaries}}} (eds \bibinfo{editor}{{Lewin}, W. H.~G.} \& \bibinfo{editor}{{van der Klis}, M.}) \emph{\bibinfo{booktitle}{Compact stellar X-ray sources}}, Vol.~\bibinfo{volume}{39} \bibinfo{pages}{381--419} (\bibinfo{publisher}{Cambridge University Press}, \bibinfo{year}{2006}).

\bibitem{kaiser06}
\bibinfo{author}{{Kaiser}, C.~R.}
\newblock \bibinfo{title}{{The flat synchrotron spectra of partially self-absorbed jets revisited}}.
\newblock \emph{\bibinfo{journal}{\mnras}} \textbf{\bibinfo{volume}{367}}, \bibinfo{pages}{1083--1094} (\bibinfo{year}{2006}).

\bibitem{chevalier98}
\bibinfo{author}{{Chevalier}, R.~A.}
\newblock \bibinfo{title}{{Synchrotron Self-Absorption in Radio Supernovae}}.
\newblock \emph{\bibinfo{journal}{\apj}} \textbf{\bibinfo{volume}{499}}, \bibinfo{pages}{810--819} (\bibinfo{year}{1998}).

\bibitem{gottlieb+metzger24}
\bibinfo{author}{{Gottlieb}, O.} \& \bibinfo{author}{{Metzger}, B.~D.}
\newblock \bibinfo{title}{{Late Jets, Early Sparks: Illuminating the Premaximum Bumps in Superluminous Supernovae}}.
\newblock \emph{\bibinfo{journal}{\apjl}} \textbf{\bibinfo{volume}{974}}, \bibinfo{pages}{L9} (\bibinfo{year}{2024}).

\bibitem{desai+26_axv}
\bibinfo{author}{{Desai}, D.~K.}, \bibinfo{author}{{Combi}, L.}, \bibinfo{author}{{Siegel}, D.~M.} \& \bibinfo{author}{{Metzger}, B.~D.}
\newblock \bibinfo{title}{{Relativistic jets from millisecond proto-magnetars}}.
\newblock \emph{\bibinfo{journal}{arXiv e-prints}} \bibinfo{pages}{arXiv:2601.07918} (\bibinfo{year}{2026}).

\bibitem{eftekhari+19}
\bibinfo{author}{{Eftekhari}, T.} \emph{et~al.}
\newblock \bibinfo{title}{{A Radio Source Coincident with the Superluminous Supernova PTF10hgi: Evidence for a Central Engine and an Analog of the Repeating FRB 121102?}}
\newblock \emph{\bibinfo{journal}{\apjl}} \textbf{\bibinfo{volume}{876}}, \bibinfo{pages}{L10} (\bibinfo{year}{2019}).

\bibitem{margutti+23}
\bibinfo{author}{{Margutti}, R.} \emph{et~al.}
\newblock \bibinfo{title}{{Luminous Radio Emission from the Superluminous Supernova 2017ens at 3.3 yr after Explosion}}.
\newblock \emph{\bibinfo{journal}{\apjl}} \textbf{\bibinfo{volume}{954}}, \bibinfo{pages}{L45} (\bibinfo{year}{2023}).

\bibitem{perley+20}
\bibinfo{author}{{Perley}, D.~A.} \emph{et~al.}
\newblock \bibinfo{title}{{The Zwicky Transient Facility Bright Transient Survey. II. A Public Statistical Sample for Exploring Supernova Demographics}}.
\newblock \emph{\bibinfo{journal}{\apj}} \textbf{\bibinfo{volume}{904}}, \bibinfo{pages}{35} (\bibinfo{year}{2020}).

\bibitem{somalwar+25}
\bibinfo{author}{{Somalwar}, J.~J.} \emph{et~al.}
\newblock \bibinfo{title}{{A luminous and hot infrared through X-ray transient at a 5 kpc offset from a dwarf galaxy}}.
\newblock \emph{\bibinfo{journal}{arXiv e-prints}} \bibinfo{pages}{arXiv:2505.11597} (\bibinfo{year}{2025}).

\bibitem{Vinko2015}
\bibinfo{author}{{Vink{\'o}}, J.} \emph{et~al.}
\newblock \bibinfo{title}{{A Luminous, Fast Rising UV-transient Discovered by ROTSE: A Tidal Disruption Event?}}
\newblock \emph{\bibinfo{journal}{\apj}} \textbf{\bibinfo{volume}{798}}, \bibinfo{pages}{12} (\bibinfo{year}{2015}).

\bibitem{yan+15}
\bibinfo{author}{{Yan}, L.} \emph{et~al.}
\newblock \bibinfo{title}{{Detection of Broad H{\ensuremath{\alpha}} Emission Lines in the Late-time Spectra of a Hydrogen-poor Superluminous Supernova}}.
\newblock \emph{\bibinfo{journal}{\apj}} \textbf{\bibinfo{volume}{814}}, \bibinfo{pages}{108} (\bibinfo{year}{2015}).

\bibitem{inserra+18}
\bibinfo{author}{{Inserra}, C.} \emph{et~al.}
\newblock \bibinfo{title}{{On the nature of hydrogen-rich superluminous supernovae}}.
\newblock \emph{\bibinfo{journal}{\mnras}} \textbf{\bibinfo{volume}{475}}, \bibinfo{pages}{1046--1072} (\bibinfo{year}{2018}).

\bibitem{somalwar+25b}
\bibinfo{author}{{Somalwar}, J.~J.}, \bibinfo{author}{{Ravi}, V.} \& \bibinfo{author}{{Lu}, W.}
\newblock \bibinfo{title}{{VLASS Tidal Disruption Events with Optical Flares. II. Discovery of Two TDEs with Intermediate Width Balmer Emission Lines and Connections to the Ambiguous Extreme Coronal Line Emitters}}.
\newblock \emph{\bibinfo{journal}{\apj}} \textbf{\bibinfo{volume}{983}}, \bibinfo{pages}{159} (\bibinfo{year}{2025}).

\bibitem{metzger+15}
\bibinfo{author}{{Metzger}, B.~D.}, \bibinfo{author}{{Margalit}, B.}, \bibinfo{author}{{Kasen}, D.} \& \bibinfo{author}{{Quataert}, E.}
\newblock \bibinfo{title}{{The diversity of transients from magnetar birth in core collapse supernovae}}.
\newblock \emph{\bibinfo{journal}{\mnras}} \textbf{\bibinfo{volume}{454}}, \bibinfo{pages}{3311--3316} (\bibinfo{year}{2015}).

\bibitem{margalit+18}
\bibinfo{author}{{Margalit}, B.}, \bibinfo{author}{{Metzger}, B.~D.}, \bibinfo{author}{{Thompson}, T.~A.}, \bibinfo{author}{{Nicholl}, M.} \& \bibinfo{author}{{Sukhbold}, T.}
\newblock \bibinfo{title}{{The GRB-SLSN connection: misaligned magnetars, weak jet emergence, and observational signatures}}.
\newblock \emph{\bibinfo{journal}{\mnras}} \textbf{\bibinfo{volume}{475}}, \bibinfo{pages}{2659--2674} (\bibinfo{year}{2018}).

\bibitem{Gottlieb2022}
\bibinfo{author}{{Gottlieb}, O.}, \bibinfo{author}{{Tchekhovskoy}, A.} \& \bibinfo{author}{{Margutti}, R.}
\newblock \bibinfo{title}{{Shocked jets in CCSNe can power the zoo of fast blue optical transients}}.
\newblock \emph{\bibinfo{journal}{\mnras}} \textbf{\bibinfo{volume}{513}}, \bibinfo{pages}{3810--3817} (\bibinfo{year}{2022}).

\bibitem{Greene+20}
\bibinfo{author}{{Greene}, J.~E.}, \bibinfo{author}{{Strader}, J.} \& \bibinfo{author}{{Ho}, L.~C.}
\newblock \bibinfo{title}{{Intermediate-Mass Black Holes}}.
\newblock \emph{\bibinfo{journal}{\araa}} \textbf{\bibinfo{volume}{58}}, \bibinfo{pages}{257--312} (\bibinfo{year}{2020}).

\bibitem{metzger+22a}
\bibinfo{author}{{Metzger}, B.~D.}
\newblock \bibinfo{title}{{Luminous Fast Blue Optical Transients and Type Ibn/Icn SNe from Wolf-Rayet/Black Hole Mergers}}.
\newblock \emph{\bibinfo{journal}{\apj}} \textbf{\bibinfo{volume}{932}}, \bibinfo{pages}{84} (\bibinfo{year}{2022}).

\bibitem{nugent+26_axv}
\bibinfo{author}{{Nugent}, A.~E.} \emph{et~al.}
\newblock \bibinfo{title}{{The Environments of Luminous Fast Blue Optical Transients: Evidence for a Compact Object and Wolf-Rayet Star Merger Origin}}.
\newblock \emph{\bibinfo{journal}{arXiv e-prints}} \bibinfo{pages}{arXiv:2603.23597} (\bibinfo{year}{2026}).

\bibitem{cenko+12}
\bibinfo{author}{{Cenko}, S.~B.} \emph{et~al.}
\newblock \bibinfo{title}{{Swift J2058.4+0516: Discovery of a Possible Second Relativistic Tidal Disruption Flare?}}
\newblock \emph{\bibinfo{journal}{\apj}} \textbf{\bibinfo{volume}{753}}, \bibinfo{pages}{77} (\bibinfo{year}{2012}).

\bibitem{brown+15}
\bibinfo{author}{{Brown}, G.~C.} \emph{et~al.}
\newblock \bibinfo{title}{{Swift J1112.2-8238: a candidate relativistic tidal disruption flare}}.
\newblock \emph{\bibinfo{journal}{\mnras}} \textbf{\bibinfo{volume}{452}}, \bibinfo{pages}{4297--4306} (\bibinfo{year}{2015}).

\bibitem{strubbe+quataert09}
\bibinfo{author}{{Strubbe}, L.~E.} \& \bibinfo{author}{{Quataert}, E.}
\newblock \bibinfo{title}{{Optical flares from the tidal disruption of stars by massive black holes}}.
\newblock \emph{\bibinfo{journal}{\mnras}} \textbf{\bibinfo{volume}{400}}, \bibinfo{pages}{2070--2084} (\bibinfo{year}{2009}).

\bibitem{stone+metzger16}
\bibinfo{author}{{Stone}, N.~C.} \& \bibinfo{author}{{Metzger}, B.~D.}
\newblock \bibinfo{title}{{Rates of stellar tidal disruption as probes of the supermassive black hole mass function}}.
\newblock \emph{\bibinfo{journal}{\mnras}} \textbf{\bibinfo{volume}{455}}, \bibinfo{pages}{859--883} (\bibinfo{year}{2016}).

\bibitem{Dai+18}
\bibinfo{author}{{Dai}, L.}, \bibinfo{author}{{McKinney}, J.~C.}, \bibinfo{author}{{Roth}, N.}, \bibinfo{author}{{Ramirez-Ruiz}, E.} \& \bibinfo{author}{{Miller}, M.~C.}
\newblock \bibinfo{title}{{A Unified Model for Tidal Disruption Events}}.
\newblock \emph{\bibinfo{journal}{\apjl}} \textbf{\bibinfo{volume}{859}}, \bibinfo{pages}{L20} (\bibinfo{year}{2018}).

\bibitem{eftekhari+18}
\bibinfo{author}{{Eftekhari}, T.}, \bibinfo{author}{{Berger}, E.}, \bibinfo{author}{{Zauderer}, B.~A.}, \bibinfo{author}{{Margutti}, R.} \& \bibinfo{author}{{Alexander}, K.~D.}
\newblock \bibinfo{title}{{Radio Monitoring of the Tidal Disruption Event Swift J164449.3+573451. III. Late-time Jet Energetics and a Deviation from Equipartition}}.
\newblock \emph{\bibinfo{journal}{\apj}} \textbf{\bibinfo{volume}{854}}, \bibinfo{pages}{86} (\bibinfo{year}{2018}).

\bibitem{rhodes+25}
\bibinfo{author}{{Rhodes}, L.} \emph{et~al.}
\newblock \bibinfo{title}{{Thermal Electrons in the Radio Afterglow of Relativistic Tidal Disruption Event ZTF22aaajecp/AT 2022cmc}}.
\newblock \emph{\bibinfo{journal}{\apj}} \textbf{\bibinfo{volume}{992}}, \bibinfo{pages}{146} (\bibinfo{year}{2025}).

\bibitem{horesh+21a}
\bibinfo{author}{{Horesh}, A.}, \bibinfo{author}{{Cenko}, S.~B.} \& \bibinfo{author}{{Arcavi}, I.}
\newblock \bibinfo{title}{{Delayed radio flares from a tidal disruption event}}.
\newblock \emph{\bibinfo{journal}{Nature Astronomy}} \textbf{\bibinfo{volume}{5}}, \bibinfo{pages}{491--497} (\bibinfo{year}{2021}).

\bibitem{horesh+21b}
\bibinfo{author}{{Horesh}, A.} \emph{et~al.}
\newblock \bibinfo{title}{{Are Delayed Radio Flares Common in Tidal Disruption Events? The Case of the TDE iPTF 16fnl}}.
\newblock \emph{\bibinfo{journal}{\apjl}} \textbf{\bibinfo{volume}{920}}, \bibinfo{pages}{L5} (\bibinfo{year}{2021}).

\bibitem{yoon+15}
\bibinfo{author}{{Yoon}, Y.} \emph{et~al.}
\newblock \bibinfo{title}{{Host Galaxy Properties and Black Hole Mass of Swift J164449.3+573451 from Multi-wavelength Long-term Monitoring and HST Data}}.
\newblock \emph{\bibinfo{journal}{\apj}} \textbf{\bibinfo{volume}{808}}, \bibinfo{pages}{96} (\bibinfo{year}{2015}).

\bibitem{Eftekhari2024}
\bibinfo{author}{{Eftekhari}, T.} \emph{et~al.}
\newblock \bibinfo{title}{{Late-time X-Ray Observations of the Jetted Tidal Disruption Event AT2022cmc: The Relativistic Jet Shuts Off}}.
\newblock \emph{\bibinfo{journal}{\apj}} \textbf{\bibinfo{volume}{974}}, \bibinfo{pages}{149} (\bibinfo{year}{2024}).

\bibitem{matsumoto+metzger23}
\bibinfo{author}{{Matsumoto}, T.} \& \bibinfo{author}{{Metzger}, B.~D.}
\newblock \bibinfo{title}{{Synchrotron afterglow model for AT 2022cmc: jetted tidal disruption event or engine-powered supernova?}}
\newblock \emph{\bibinfo{journal}{\mnras}} \textbf{\bibinfo{volume}{522}}, \bibinfo{pages}{4028--4037} (\bibinfo{year}{2023}).

\bibitem{Bennett+14}
\bibinfo{author}{{Bennett}, C.~L.}, \bibinfo{author}{{Larson}, D.}, \bibinfo{author}{{Weiland}, J.~L.} \& \bibinfo{author}{{Hinshaw}, G.}
\newblock \bibinfo{title}{{The 1\% Concordance Hubble Constant}}.
\newblock \emph{\bibinfo{journal}{\apj}} \textbf{\bibinfo{volume}{794}}, \bibinfo{pages}{135} (\bibinfo{year}{2014}).

\bibitem{mcconell+20}
\bibinfo{author}{{McConnell}, D.} \emph{et~al.}
\newblock \bibinfo{title}{{The Rapid ASKAP Continuum Survey I: Design and first results}}.
\newblock \emph{\bibinfo{journal}{\pasa}} \textbf{\bibinfo{volume}{37}}, \bibinfo{pages}{e048} (\bibinfo{year}{2020}).

\bibitem{hale+21}
\bibinfo{author}{{Hale}, C.~L.} \emph{et~al.}
\newblock \bibinfo{title}{{The Rapid ASKAP Continuum Survey Paper II: First Stokes I Source Catalogue Data Release}}.
\newblock \emph{\bibinfo{journal}{\pasa}} \textbf{\bibinfo{volume}{38}}, \bibinfo{pages}{e058} (\bibinfo{year}{2021}).

\bibitem{duchesne+23}
\bibinfo{author}{{Duchesne}, S.~W.} \emph{et~al.}
\newblock \bibinfo{title}{{The Rapid ASKAP Continuum Survey IV: continuum imaging at 1367.5 MHz and the first data release of RACS-mid}}.
\newblock \emph{\bibinfo{journal}{\pasa}} \textbf{\bibinfo{volume}{40}}, \bibinfo{pages}{e034} (\bibinfo{year}{2023}).

\bibitem{duchesne+24}
\bibinfo{author}{{Duchesne}, S.~W.} \emph{et~al.}
\newblock \bibinfo{title}{{The Rapid ASKAP Continuum Survey V: Cataloguing the sky at 1 367.5 MHz and the second data release of RACS-mid}}.
\newblock \emph{\bibinfo{journal}{\pasa}} \textbf{\bibinfo{volume}{41}}, \bibinfo{pages}{e003} (\bibinfo{year}{2024}).

\bibitem{duchesne+25}
\bibinfo{author}{{Duchesne}, S.} \emph{et~al.}
\newblock \bibinfo{title}{{The Rapid ASKAP Continuum Survey (RACS) VI: The RACS-high 1655.5 MHz images and catalogue.}}
\newblock \emph{\bibinfo{journal}{\pasa}} \textbf{\bibinfo{volume}{42}}, \bibinfo{pages}{38} (\bibinfo{year}{2025}).

\bibitem{vanhaarlem+13}
\bibinfo{author}{{van Haarlem}, M.~P.} \emph{et~al.}
\newblock \bibinfo{title}{{LOFAR: The LOw-Frequency ARray}}.
\newblock \emph{\bibinfo{journal}{\aap}} \textbf{\bibinfo{volume}{556}}, \bibinfo{pages}{A2} (\bibinfo{year}{2013}).

\bibitem{shimwell+19}
\bibinfo{author}{{Shimwell}, T.~W.} \emph{et~al.}
\newblock \bibinfo{title}{{The LOFAR Two-metre Sky Survey. II. First data release}}.
\newblock \emph{\bibinfo{journal}{\aap}} \textbf{\bibinfo{volume}{622}}, \bibinfo{pages}{A1} (\bibinfo{year}{2019}).

\bibitem{shimwell+22}
\bibinfo{author}{{Shimwell}, T.~W.} \emph{et~al.}
\newblock \bibinfo{title}{{The LOFAR Two-metre Sky Survey. V. Second data release}}.
\newblock \emph{\bibinfo{journal}{\aap}} \textbf{\bibinfo{volume}{659}}, \bibinfo{pages}{A1} (\bibinfo{year}{2022}).

\bibitem{shimwell+26_axv}
\bibinfo{author}{{Shimwell}, T.~W.} \emph{et~al.}
\newblock \bibinfo{title}{{The LOFAR Two-metre Sky Survey: VII. Third Data Release}}.
\newblock \emph{\bibinfo{journal}{arXiv e-prints}} \bibinfo{pages}{arXiv:2602.15949} (\bibinfo{year}{2026}).

\bibitem{Bellm+19}
\bibinfo{author}{{Bellm}, E.~C.} \emph{et~al.}
\newblock \bibinfo{title}{{The Zwicky Transient Facility: System Overview, Performance, and First Results}}.
\newblock \emph{\bibinfo{journal}{\pasp}} \textbf{\bibinfo{volume}{131}}, \bibinfo{pages}{018002} (\bibinfo{year}{2019}).

\bibitem{Graham+19}
\bibinfo{author}{{Graham}, M.~J.} \emph{et~al.}
\newblock \bibinfo{title}{{The Zwicky Transient Facility: Science Objectives}}.
\newblock \emph{\bibinfo{journal}{\pasp}} \textbf{\bibinfo{volume}{131}}, \bibinfo{pages}{078001} (\bibinfo{year}{2019}).

\bibitem{Dekany+20}
\bibinfo{author}{{Dekany}, R.} \emph{et~al.}
\newblock \bibinfo{title}{{The Zwicky Transient Facility: Observing System}}.
\newblock \emph{\bibinfo{journal}{\pasp}} \textbf{\bibinfo{volume}{132}}, \bibinfo{pages}{038001} (\bibinfo{year}{2020}).

\bibitem{Masci+19}
\bibinfo{author}{{Masci}, F.~J.} \emph{et~al.}
\newblock \bibinfo{title}{{The Zwicky Transient Facility: Data Processing, Products, and Archive}}.
\newblock \emph{\bibinfo{journal}{\pasp}} \textbf{\bibinfo{volume}{131}}, \bibinfo{pages}{018003} (\bibinfo{year}{2019}).

\bibitem{fitzpatrick99}
\bibinfo{author}{{Fitzpatrick}, E.~L.}
\newblock \bibinfo{title}{{Correcting for the Effects of Interstellar Extinction}}.
\newblock \emph{\bibinfo{journal}{\pasp}} \textbf{\bibinfo{volume}{111}}, \bibinfo{pages}{63--75} (\bibinfo{year}{1999}).

\bibitem{schlegel+98}
\bibinfo{author}{{Schlegel}, D.~J.}, \bibinfo{author}{{Finkbeiner}, D.~P.} \& \bibinfo{author}{{Davis}, M.}
\newblock \bibinfo{title}{{Maps of Dust Infrared Emission for Use in Estimation of Reddening and Cosmic Microwave Background Radiation Foregrounds}}.
\newblock \emph{\bibinfo{journal}{\apj}} \textbf{\bibinfo{volume}{500}}, \bibinfo{pages}{525--553} (\bibinfo{year}{1998}).

\bibitem{Schlafly2011}
\bibinfo{author}{{Schlafly}, E.~F.} \& \bibinfo{author}{{Finkbeiner}, D.~P.}
\newblock \bibinfo{title}{{Measuring Reddening with Sloan Digital Sky Survey Stellar Spectra and Recalibrating SFD}}.
\newblock \emph{\bibinfo{journal}{\apj}} \textbf{\bibinfo{volume}{737}}, \bibinfo{pages}{103} (\bibinfo{year}{2011}).

\bibitem{vanvelzen+21}
\bibinfo{author}{{van Velzen}, S.} \emph{et~al.}
\newblock \bibinfo{title}{{Seventeen Tidal Disruption Events from the First Half of ZTF Survey Observations: Entering a New Era of Population Studies}}.
\newblock \emph{\bibinfo{journal}{\apj}} \textbf{\bibinfo{volume}{908}}, \bibinfo{pages}{4} (\bibinfo{year}{2021}).

\bibitem{Foreman-Mackey+13}
\bibinfo{author}{{Foreman-Mackey}, D.}, \bibinfo{author}{{Hogg}, D.~W.}, \bibinfo{author}{{Lang}, D.} \& \bibinfo{author}{{Goodman}, J.}
\newblock \bibinfo{title}{{emcee: The MCMC Hammer}}.
\newblock \emph{\bibinfo{journal}{\pasp}} \textbf{\bibinfo{volume}{125}}, \bibinfo{pages}{306} (\bibinfo{year}{2013}).

\bibitem{evans+24}
\bibinfo{author}{{Evans}, I.~N.} \emph{et~al.}
\newblock \bibinfo{title}{{The Chandra Source Catalog Release 2 Series}}.
\newblock \emph{\bibinfo{journal}{\apjs}} \textbf{\bibinfo{volume}{274}}, \bibinfo{pages}{22} (\bibinfo{year}{2024}).

\bibitem{Gehrels+04}
\bibinfo{author}{{Gehrels}, N.} \emph{et~al.}
\newblock \bibinfo{title}{{The Swift Gamma-Ray Burst Mission}}.
\newblock \emph{\bibinfo{journal}{\apj}} \textbf{\bibinfo{volume}{611}}, \bibinfo{pages}{1005--1020} (\bibinfo{year}{2004}).

\bibitem{Burrows+05}
\bibinfo{author}{{Burrows}, D.~N.} \emph{et~al.}
\newblock \bibinfo{title}{{The Swift X-Ray Telescope}}.
\newblock \emph{\bibinfo{journal}{\ssr}} \textbf{\bibinfo{volume}{120}}, \bibinfo{pages}{165--195} (\bibinfo{year}{2005}).

\bibitem{roming+05}
\bibinfo{author}{{Roming}, P. W.~A.} \emph{et~al.}
\newblock \bibinfo{title}{{The Swift Ultra-Violet/Optical Telescope}}.
\newblock \emph{\bibinfo{journal}{\ssr}} \textbf{\bibinfo{volume}{120}}, \bibinfo{pages}{95--142} (\bibinfo{year}{2005}).

\bibitem{jansen+01}
\bibinfo{author}{{Jansen}, F.} \emph{et~al.}
\newblock \bibinfo{title}{{XMM-Newton observatory. I. The spacecraft and operations}}.
\newblock \emph{\bibinfo{journal}{\aap}} \textbf{\bibinfo{volume}{365}}, \bibinfo{pages}{L1--L6} (\bibinfo{year}{2001}).

\bibitem{predehl+21}
\bibinfo{author}{{Predehl}, P.} \emph{et~al.}
\newblock \bibinfo{title}{{The eROSITA X-ray telescope on SRG}}.
\newblock \emph{\bibinfo{journal}{\aap}} \textbf{\bibinfo{volume}{647}}, \bibinfo{pages}{A1} (\bibinfo{year}{2021}).

\bibitem{merloni+24}
\bibinfo{author}{{Merloni}, A.} \emph{et~al.}
\newblock \bibinfo{title}{{The SRG/eROSITA all-sky survey. First X-ray catalogues and data release of the western Galactic hemisphere}}.
\newblock \emph{\bibinfo{journal}{\aap}} \textbf{\bibinfo{volume}{682}}, \bibinfo{pages}{A34} (\bibinfo{year}{2024}).

\bibitem{tubinarenas+24}
\bibinfo{author}{{Tub{\'\i}n-Arenas}, D.} \emph{et~al.}
\newblock \bibinfo{title}{{The eROSITA upper limits. Description and access to the data}}.
\newblock \emph{\bibinfo{journal}{\aap}} \textbf{\bibinfo{volume}{682}}, \bibinfo{pages}{A35} (\bibinfo{year}{2024}).

\bibitem{atwood+09}
\bibinfo{author}{{Atwood}, W.~B.} \emph{et~al.}
\newblock \bibinfo{title}{{The Large Area Telescope on the Fermi Gamma-Ray Space Telescope Mission}}.
\newblock \emph{\bibinfo{journal}{\apj}} \textbf{\bibinfo{volume}{697}}, \bibinfo{pages}{1071--1102} (\bibinfo{year}{2009}).

\bibitem{Dey+19}
\bibinfo{author}{{Dey}, A.} \emph{et~al.}
\newblock \bibinfo{title}{{Overview of the DESI Legacy Imaging Surveys}}.
\newblock \emph{\bibinfo{journal}{\aj}} \textbf{\bibinfo{volume}{157}}, \bibinfo{pages}{168} (\bibinfo{year}{2019}).

\bibitem{chambers+16}
\bibinfo{author}{{Chambers}, K.~C.} \emph{et~al.}
\newblock \bibinfo{title}{{The Pan-STARRS1 Surveys}}.
\newblock \emph{\bibinfo{journal}{arXiv e-prints}} \bibinfo{pages}{arXiv:1612.05560} (\bibinfo{year}{2016}).

\bibitem{Waters+20}
\bibinfo{author}{{Waters}, C.~Z.} \emph{et~al.}
\newblock \bibinfo{title}{{Pan-STARRS Pixel Processing: Detrending, Warping, Stacking}}.
\newblock \emph{\bibinfo{journal}{\apjs}} \textbf{\bibinfo{volume}{251}}, \bibinfo{pages}{4} (\bibinfo{year}{2020}).

\bibitem{Magnier+20a}
\bibinfo{author}{{Magnier}, E.~A.} \emph{et~al.}
\newblock \bibinfo{title}{{The Pan-STARRS Data-processing System}}.
\newblock \emph{\bibinfo{journal}{\apjs}} \textbf{\bibinfo{volume}{251}}, \bibinfo{pages}{3} (\bibinfo{year}{2020}).

\bibitem{Magnier+20b}
\bibinfo{author}{{Magnier}, E.~A.} \emph{et~al.}
\newblock \bibinfo{title}{{Pan-STARRS Pixel Analysis: Source Detection and Characterization}}.
\newblock \emph{\bibinfo{journal}{\apjs}} \textbf{\bibinfo{volume}{251}}, \bibinfo{pages}{5} (\bibinfo{year}{2020}).

\bibitem{Magnier+20c}
\bibinfo{author}{{Magnier}, E.~A.} \emph{et~al.}
\newblock \bibinfo{title}{{Pan-STARRS Photometric and Astrometric Calibration}}.
\newblock \emph{\bibinfo{journal}{\apjs}} \textbf{\bibinfo{volume}{251}}, \bibinfo{pages}{6} (\bibinfo{year}{2020}).

\bibitem{Flewelling+20}
\bibinfo{author}{{Flewelling}, H.~A.} \emph{et~al.}
\newblock \bibinfo{title}{{The Pan-STARRS1 Database and Data Products}}.
\newblock \emph{\bibinfo{journal}{\apjs}} \textbf{\bibinfo{volume}{251}}, \bibinfo{pages}{7} (\bibinfo{year}{2020}).

\bibitem{wright+10}
\bibinfo{author}{{Wright}, E.~L.} \emph{et~al.}
\newblock \bibinfo{title}{{The Wide-field Infrared Survey Explorer (WISE): Mission Description and Initial On-orbit Performance}}.
\newblock \emph{\bibinfo{journal}{\aj}} \textbf{\bibinfo{volume}{140}}, \bibinfo{pages}{1868--1881} (\bibinfo{year}{2010}).

\bibitem{skrutskie+06}
\bibinfo{author}{{Skrutskie}, M.~F.} \emph{et~al.}
\newblock \bibinfo{title}{{The Two Micron All Sky Survey (2MASS)}}.
\newblock \emph{\bibinfo{journal}{\aj}} \textbf{\bibinfo{volume}{131}}, \bibinfo{pages}{1163--1183} (\bibinfo{year}{2006}).

\bibitem{lawrence+07}
\bibinfo{author}{{Lawrence}, A.} \emph{et~al.}
\newblock \bibinfo{title}{{The UKIRT Infrared Deep Sky Survey (UKIDSS)}}.
\newblock \emph{\bibinfo{journal}{\mnras}} \textbf{\bibinfo{volume}{379}}, \bibinfo{pages}{1599--1617} (\bibinfo{year}{2007}).

\bibitem{dye+18}
\bibinfo{author}{{Dye}, S.} \emph{et~al.}
\newblock \bibinfo{title}{{The UKIRT Hemisphere Survey: definition and J-band data release}}.
\newblock \emph{\bibinfo{journal}{\mnras}} \textbf{\bibinfo{volume}{473}}, \bibinfo{pages}{5113--5125} (\bibinfo{year}{2018}).

\bibitem{casa+22}
\bibinfo{author}{{CASA Team}} \emph{et~al.}
\newblock \bibinfo{title}{{CASA, the Common Astronomy Software Applications for Radio Astronomy}}.
\newblock \emph{\bibinfo{journal}{\pasp}} \textbf{\bibinfo{volume}{134}}, \bibinfo{pages}{114501} (\bibinfo{year}{2022}).

\bibitem{granot+02}
\bibinfo{author}{{Granot}, J.} \& \bibinfo{author}{{Sari}, R.}
\newblock \bibinfo{title}{{The Shape of Spectral Breaks in Gamma-Ray Burst Afterglows}}.
\newblock \emph{\bibinfo{journal}{\apj}} \textbf{\bibinfo{volume}{568}}, \bibinfo{pages}{820--829} (\bibinfo{year}{2002}).

\bibitem{oke+95}
\bibinfo{author}{{Oke}, J.~B.} \emph{et~al.}
\newblock \bibinfo{title}{{The Keck Low-Resolution Imaging Spectrometer}}.
\newblock \emph{\bibinfo{journal}{\pasp}} \textbf{\bibinfo{volume}{107}}, \bibinfo{pages}{375} (\bibinfo{year}{1995}).

\bibitem{mccarthy+98}
\bibinfo{author}{{McCarthy}, J.~K.} \emph{et~al.}
\newblock \bibinfo{editor}{{D'Odorico}, S.} (ed.) \emph{\bibinfo{title}{{Blue channel of the Keck low-resolution imaging spectrometer}}}.
\newblock (ed.\bibinfo{editor}{{D'Odorico}, S.}) \emph{\bibinfo{booktitle}{Optical Astronomical Instrumentation}}, Vol. \bibinfo{volume}{3355} of \emph{\bibinfo{series}{Society of Photo-Optical Instrumentation Engineers (SPIE) Conference Series}}, \bibinfo{pages}{81--92} (\bibinfo{year}{1998}).

\bibitem{Perley+19}
\bibinfo{author}{{Perley}, D.~A.}
\newblock \bibinfo{title}{{Fully Automated Reduction of Longslit Spectroscopy with the Low Resolution Imaging Spectrometer at the Keck Observatory}}.
\newblock \emph{\bibinfo{journal}{\pasp}} \textbf{\bibinfo{volume}{131}}, \bibinfo{pages}{084503} (\bibinfo{year}{2019}).

\bibitem{prochaska+20}
\bibinfo{author}{{Prochaska}, J.} \emph{et~al.}
\newblock \bibinfo{title}{{PypeIt: The Python Spectroscopic Data Reduction Pipeline}}.
\newblock \emph{\bibinfo{journal}{The Journal of Open Source Software}} \textbf{\bibinfo{volume}{5}}, \bibinfo{pages}{2308} (\bibinfo{year}{2020}).

\bibitem{dong+21}
\bibinfo{author}{{Dong}, D.~Z.} \emph{et~al.}
\newblock \bibinfo{title}{{A transient radio source consistent with a merger-triggered core collapse supernova}}.
\newblock \emph{\bibinfo{journal}{Science}} \textbf{\bibinfo{volume}{373}}, \bibinfo{pages}{1125--1129} (\bibinfo{year}{2021}).

\bibitem{sollerman+20}
\bibinfo{author}{{Sollerman}, J.} \emph{et~al.}
\newblock \bibinfo{title}{{Two stripped envelope supernovae with circumstellar interaction. But only one really shows it}}.
\newblock \emph{\bibinfo{journal}{\aap}} \textbf{\bibinfo{volume}{643}}, \bibinfo{pages}{A79} (\bibinfo{year}{2020}).

\bibitem{wu+23}
\bibinfo{author}{{Wu}, J.}, \bibinfo{author}{{Wu}, Q.}, \bibinfo{author}{{Xue}, H.}, \bibinfo{author}{{Lei}, W.} \& \bibinfo{author}{{Lyu}, B.}
\newblock \bibinfo{title}{{Steep Balmer Decrement in Weak AGNs May Not Be Caused by Dust Extinction: Clues from Low-luminosity AGNs and Changing-look AGNs}}.
\newblock \emph{\bibinfo{journal}{\apj}} \textbf{\bibinfo{volume}{950}}, \bibinfo{pages}{106} (\bibinfo{year}{2023}).

\bibitem{baldwin+81}
\bibinfo{author}{{Baldwin}, J.~A.}, \bibinfo{author}{{Phillips}, M.~M.} \& \bibinfo{author}{{Terlevich}, R.}
\newblock \bibinfo{title}{{Classification parameters for the emission-line spectra of extragalactic objects.}}
\newblock \emph{\bibinfo{journal}{\pasp}} \textbf{\bibinfo{volume}{93}}, \bibinfo{pages}{5--19} (\bibinfo{year}{1981}).

\bibitem{Wilson+03}
\bibinfo{author}{{Wilson}, J.~C.} \emph{et~al.}
\newblock \bibinfo{editor}{{Iye}, M.} \& \bibinfo{editor}{{Moorwood}, A. F.~M.} (eds) \emph{\bibinfo{title}{{A Wide-Field Infrared Camera for the Palomar 200-inch Telescope}}}.
\newblock (eds \bibinfo{editor}{{Iye}, M.} \& \bibinfo{editor}{{Moorwood}, A. F.~M.}) \emph{\bibinfo{booktitle}{Instrument Design and Performance for Optical/Infrared Ground-based Telescopes}}, Vol. \bibinfo{volume}{4841} of \emph{\bibinfo{series}{Society of Photo-Optical Instrumentation Engineers (SPIE) Conference Series}}, \bibinfo{pages}{451--458} (\bibinfo{year}{2003}).

\bibitem{de+20}
\bibinfo{author}{{De}, K.} \emph{et~al.}
\newblock \bibinfo{title}{{Palomar Gattini-IR: Survey Overview, Data Processing System, On-sky Performance and First Results}}.
\newblock \emph{\bibinfo{journal}{\pasp}} \textbf{\bibinfo{volume}{132}}, \bibinfo{pages}{025001} (\bibinfo{year}{2020}).

\bibitem{Persson+13}
\bibinfo{author}{{Persson}, S.~E.} \emph{et~al.}
\newblock \bibinfo{title}{{FourStar: The Near-Infrared Imager for the 6.5 m Baade Telescope at Las Campanas Observatory}}.
\newblock \emph{\bibinfo{journal}{\pasp}} \textbf{\bibinfo{volume}{125}}, \bibinfo{pages}{654} (\bibinfo{year}{2013}).

\bibitem{kelson+14}
\bibinfo{author}{{Kelson}, D.~D.} \emph{et~al.}
\newblock \bibinfo{title}{{The Carnegie-Spitzer-IMACS Redshift Survey of Galaxy Evolution since z = 1.5. I. Description and Methodology}}.
\newblock \emph{\bibinfo{journal}{\apj}} \textbf{\bibinfo{volume}{783}}, \bibinfo{pages}{110} (\bibinfo{year}{2014}).

\bibitem{mclean+10}
\bibinfo{author}{{McLean}, I.~S.} \emph{et~al.}
\newblock \bibinfo{editor}{{McLean}, I.~S.}, \bibinfo{editor}{{Ramsay}, S.~K.} \& \bibinfo{editor}{{Takami}, H.} (eds) \emph{\bibinfo{title}{{Design and development of MOSFIRE: the multi-object spectrometer for infrared exploration at the Keck Observatory}}}.
\newblock (eds \bibinfo{editor}{{McLean}, I.~S.}, \bibinfo{editor}{{Ramsay}, S.~K.} \& \bibinfo{editor}{{Takami}, H.}) \emph{\bibinfo{booktitle}{Ground-based and Airborne Instrumentation for Astronomy III}}, Vol. \bibinfo{volume}{7735} of \emph{\bibinfo{series}{Society of Photo-Optical Instrumentation Engineers (SPIE) Conference Series}}, \bibinfo{pages}{77351E} (\bibinfo{year}{2010}).

\bibitem{mclean+12}
\bibinfo{author}{{McLean}, I.~S.} \emph{et~al.}
\newblock \bibinfo{editor}{{McLean}, I.~S.}, \bibinfo{editor}{{Ramsay}, S.~K.} \& \bibinfo{editor}{{Takami}, H.} (eds) \emph{\bibinfo{title}{{MOSFIRE, the multi-object spectrometer for infra-red exploration at the Keck Observatory}}}.
\newblock (eds \bibinfo{editor}{{McLean}, I.~S.}, \bibinfo{editor}{{Ramsay}, S.~K.} \& \bibinfo{editor}{{Takami}, H.}) \emph{\bibinfo{booktitle}{Ground-based and Airborne Instrumentation for Astronomy IV}}, Vol. \bibinfo{volume}{8446} of \emph{\bibinfo{series}{Society of Photo-Optical Instrumentation Engineers (SPIE) Conference Series}}, \bibinfo{pages}{84460J} (\bibinfo{year}{2012}).

\bibitem{bloom+01}
\bibinfo{author}{{Bloom}, J.~S.}, \bibinfo{author}{{Frail}, D.~A.} \& \bibinfo{author}{{Sari}, R.}
\newblock \bibinfo{title}{{The Prompt Energy Release of Gamma-Ray Bursts using a Cosmological k-Correction}}.
\newblock \emph{\bibinfo{journal}{\aj}} \textbf{\bibinfo{volume}{121}}, \bibinfo{pages}{2879--2888} (\bibinfo{year}{2001}).

\bibitem{chandra+frail12}
\bibinfo{author}{{Chandra}, P.} \& \bibinfo{author}{{Frail}, D.~A.}
\newblock \bibinfo{title}{{A Radio-selected Sample of Gamma-Ray Burst Afterglows}}.
\newblock \emph{\bibinfo{journal}{\apj}} \textbf{\bibinfo{volume}{746}}, \bibinfo{pages}{156} (\bibinfo{year}{2012}).

\bibitem{whitesides+17}
\bibinfo{author}{{Whitesides}, L.} \emph{et~al.}
\newblock \bibinfo{title}{{iPTF 16asu: A Luminous, Rapidly Evolving, and High-velocity Supernova}}.
\newblock \emph{\bibinfo{journal}{\apj}} \textbf{\bibinfo{volume}{851}}, \bibinfo{pages}{107} (\bibinfo{year}{2017}).

\bibitem{vanvelzen+21b}
\bibinfo{author}{{van Velzen}, S.}, \bibinfo{author}{{Pasham}, D.~R.}, \bibinfo{author}{{Komossa}, S.}, \bibinfo{author}{{Yan}, L.} \& \bibinfo{author}{{Kara}, E.~A.}
\newblock \bibinfo{title}{{Reverberation in Tidal Disruption Events: Dust Echoes, Coronal Emission Lines, Multi-wavelength Cross-correlations, and QPOs}}.
\newblock \emph{\bibinfo{journal}{\ssr}} \textbf{\bibinfo{volume}{217}}, \bibinfo{pages}{63} (\bibinfo{year}{2021}).

\bibitem{lu+16}
\bibinfo{author}{{Lu}, W.}, \bibinfo{author}{{Kumar}, P.} \& \bibinfo{author}{{Evans}, N.~J.}
\newblock \bibinfo{title}{{Infrared emission from tidal disruption events - probing the pc-scale dust content around galactic nuclei}}.
\newblock \emph{\bibinfo{journal}{\mnras}} \textbf{\bibinfo{volume}{458}}, \bibinfo{pages}{575--581} (\bibinfo{year}{2016}).

\bibitem{somalwar+22}
\bibinfo{author}{{Somalwar}, J.~J.} \emph{et~al.}
\newblock \bibinfo{title}{{The Nascent Milliquasar VT J154843.06+220812.6: Tidal Disruption Event or Extreme Accretion State Change?}}
\newblock \emph{\bibinfo{journal}{\apj}} \textbf{\bibinfo{volume}{929}}, \bibinfo{pages}{184} (\bibinfo{year}{2022}).

\bibitem{somalwar+25a}
\bibinfo{author}{{Somalwar}, J.~J.} \emph{et~al.}
\newblock \bibinfo{title}{{VLASS Tidal Disruption Events with Optical Flares. I. The Sample and a Comparison to Optically Selected TDEs}}.
\newblock \emph{\bibinfo{journal}{\apj}} \textbf{\bibinfo{volume}{982}}, \bibinfo{pages}{163} (\bibinfo{year}{2025}).

\bibitem{jiang+21}
\bibinfo{author}{{Jiang}, N.} \emph{et~al.}
\newblock \bibinfo{title}{{Infrared Echoes of Optical Tidal Disruption Events: {\ensuremath{\sim}}1\% Dust-covering Factor or Less at Subparsec Scale}}.
\newblock \emph{\bibinfo{journal}{\apj}} \textbf{\bibinfo{volume}{911}}, \bibinfo{pages}{31} (\bibinfo{year}{2021}).

\bibitem{bell+03}
\bibinfo{author}{{Bell}, E.~F.}, \bibinfo{author}{{McIntosh}, D.~H.}, \bibinfo{author}{{Katz}, N.} \& \bibinfo{author}{{Weinberg}, M.~D.}
\newblock \bibinfo{title}{{The Optical and Near-Infrared Properties of Galaxies. I. Luminosity and Stellar Mass Functions}}.
\newblock \emph{\bibinfo{journal}{\apjs}} \textbf{\bibinfo{volume}{149}}, \bibinfo{pages}{289--312} (\bibinfo{year}{2003}).

\bibitem{delosreyes+25}
\bibinfo{author}{{de los Reyes}, M. A.~C.} \emph{et~al.}
\newblock \bibinfo{title}{{Stellar Mass Calibrations for Local Low-mass Galaxies}}.
\newblock \emph{\bibinfo{journal}{\apj}} \textbf{\bibinfo{volume}{989}}, \bibinfo{pages}{91} (\bibinfo{year}{2025}).

\bibitem{kennicutt+94}
\bibinfo{author}{{Kennicutt}, R.~C., Jr.}, \bibinfo{author}{{Tamblyn}, P.} \& \bibinfo{author}{{Congdon}, C.~E.}
\newblock \bibinfo{title}{{Past and Future Star Formation in Disk Galaxies}}.
\newblock \emph{\bibinfo{journal}{\apj}} \textbf{\bibinfo{volume}{435}}, \bibinfo{pages}{22} (\bibinfo{year}{1994}).

\bibitem{Pagel+79}
\bibinfo{author}{{Pagel}, B.~E.~J.}, \bibinfo{author}{{Edmunds}, M.~G.}, \bibinfo{author}{{Blackwell}, D.~E.}, \bibinfo{author}{{Chun}, M.~S.} \& \bibinfo{author}{{Smith}, G.}
\newblock \bibinfo{title}{{On the composition of H II regions in southern galaxies - I. NGC 300 and 1365.}}
\newblock \emph{\bibinfo{journal}{\mnras}} \textbf{\bibinfo{volume}{189}}, \bibinfo{pages}{95--113} (\bibinfo{year}{1979}).

\bibitem{nagao+06}
\bibinfo{author}{{Nagao}, T.}, \bibinfo{author}{{Maiolino}, R.} \& \bibinfo{author}{{Marconi}, A.}
\newblock \bibinfo{title}{{Gas metallicity diagnostics in star-forming galaxies}}.
\newblock \emph{\bibinfo{journal}{\aap}} \textbf{\bibinfo{volume}{459}}, \bibinfo{pages}{85--101} (\bibinfo{year}{2006}).

\bibitem{barniolduran+13}
\bibinfo{author}{{Barniol Duran}, R.}, \bibinfo{author}{{Nakar}, E.} \& \bibinfo{author}{{Piran}, T.}
\newblock \bibinfo{title}{{Radius Constraints and Minimal Equipartition Energy of Relativistically Moving Synchrotron Sources}}.
\newblock \emph{\bibinfo{journal}{\apj}} \textbf{\bibinfo{volume}{772}}, \bibinfo{pages}{78} (\bibinfo{year}{2013}).

\bibitem{Matsumoto2023}
\bibinfo{author}{{Matsumoto}, T.} \& \bibinfo{author}{{Piran}, T.}
\newblock \bibinfo{title}{{Generalized equipartition method from an arbitrary viewing angle}}.
\newblock \emph{\bibinfo{journal}{\mnras}} \textbf{\bibinfo{volume}{522}}, \bibinfo{pages}{4565--4576} (\bibinfo{year}{2023}).

\bibitem{GovreenSegal2024}
\bibinfo{author}{{Govreen-Segal}, T.} \& \bibinfo{author}{{Nakar}, E.}
\newblock \bibinfo{title}{{The structure and evolution of relativistic jetted blast waves}}.
\newblock \emph{\bibinfo{journal}{\mnras}} \textbf{\bibinfo{volume}{531}}, \bibinfo{pages}{1704--1720} (\bibinfo{year}{2024}).

\bibitem{Phinney89}
\bibinfo{author}{{Phinney}, E.~S.}
\newblock \bibinfo{editor}{{Morris}, M.} (ed.) \emph{\bibinfo{title}{{Manifestations of a Massive Black Hole in the Galactic Center}}}.
\newblock (ed.\bibinfo{editor}{{Morris}, M.}) \emph{\bibinfo{booktitle}{The Center of the Galaxy}}, Vol. \bibinfo{volume}{136} of \emph{\bibinfo{series}{IAU Symposium}}, \bibinfo{pages}{543} (\bibinfo{year}{1989}).

\bibitem{kasen+ramirezruiz2010}
\bibinfo{author}{{Kasen}, D.} \& \bibinfo{author}{{Ramirez-Ruiz}, E.}
\newblock \bibinfo{title}{{Optical Transients from the Unbound Debris of Tidal Disruption}}.
\newblock \emph{\bibinfo{journal}{\apj}} \textbf{\bibinfo{volume}{714}}, \bibinfo{pages}{155--162} (\bibinfo{year}{2010}).

\bibitem{wu+18}
\bibinfo{author}{{Wu}, S.}, \bibinfo{author}{{Coughlin}, E.~R.} \& \bibinfo{author}{{Nixon}, C.}
\newblock \bibinfo{title}{{Super-Eddington accretion in tidal disruption events: the impactof realistic fallback rates on accretion rates}}.
\newblock \emph{\bibinfo{journal}{\mnras}} \textbf{\bibinfo{volume}{478}}, \bibinfo{pages}{3016--3024} (\bibinfo{year}{2018}).

\bibitem{guillochon+18}
\bibinfo{author}{{Guillochon}, J.} \emph{et~al.}
\newblock \bibinfo{title}{{MOSFiT: Modular Open Source Fitter for Transients}}.
\newblock \emph{\bibinfo{journal}{\apjs}} \textbf{\bibinfo{volume}{236}}, \bibinfo{pages}{6} (\bibinfo{year}{2018}).

\bibitem{mockler+19}
\bibinfo{author}{{Mockler}, B.}, \bibinfo{author}{{Guillochon}, J.} \& \bibinfo{author}{{Ramirez-Ruiz}, E.}
\newblock \bibinfo{title}{{Weighing Black Holes Using Tidal Disruption Events}}.
\newblock \emph{\bibinfo{journal}{\apj}} \textbf{\bibinfo{volume}{872}}, \bibinfo{pages}{151} (\bibinfo{year}{2019}).

\bibitem{piran+15a}
\bibinfo{author}{{Piran}, T.}, \bibinfo{author}{{Svirski}, G.}, \bibinfo{author}{{Krolik}, J.}, \bibinfo{author}{{Cheng}, R.~M.} \& \bibinfo{author}{{Shiokawa}, H.}
\newblock \bibinfo{title}{{Disk Formation Versus Disk Accretion{\textemdash}What Powers Tidal Disruption Events?}}
\newblock \emph{\bibinfo{journal}{\apj}} \textbf{\bibinfo{volume}{806}}, \bibinfo{pages}{164} (\bibinfo{year}{2015}).

\bibitem{jiang+16}
\bibinfo{author}{{Jiang}, Y.-F.}, \bibinfo{author}{{Guillochon}, J.} \& \bibinfo{author}{{Loeb}, A.}
\newblock \bibinfo{title}{{Prompt Radiation and Mass Outflows from the Stream-Stream Collisions of Tidal Disruption Events}}.
\newblock \emph{\bibinfo{journal}{\apj}} \textbf{\bibinfo{volume}{830}}, \bibinfo{pages}{125} (\bibinfo{year}{2016}).

\bibitem{chen+shen21}
\bibinfo{author}{{Chen}, J.-H.} \& \bibinfo{author}{{Shen}, R.-F.}
\newblock \bibinfo{title}{{Light Curves of Partial Tidal Disruption Events}}.
\newblock \emph{\bibinfo{journal}{\apj}} \textbf{\bibinfo{volume}{914}}, \bibinfo{pages}{69} (\bibinfo{year}{2021}).

\bibitem{guo+25}
\bibinfo{author}{{Guo}, H.} \emph{et~al.}
\newblock \bibinfo{title}{{Reverberation Evidence for Stream Collision and Delayed Disk Formation in Tidal Disruption Events}}.
\newblock \emph{\bibinfo{journal}{\apj}} \textbf{\bibinfo{volume}{979}}, \bibinfo{pages}{235} (\bibinfo{year}{2025}).

\bibitem{steinberg+24}
\bibinfo{author}{{Steinberg}, E.} \& \bibinfo{author}{{Stone}, N.~C.}
\newblock \bibinfo{title}{{Stream-disk shocks as the origins of peak light in tidal disruption events}}.
\newblock \emph{\bibinfo{journal}{\nat}} \textbf{\bibinfo{volume}{625}}, \bibinfo{pages}{463--467} (\bibinfo{year}{2024}).

\bibitem{huang+25_axv}
\bibinfo{author}{{Huang}, X.} \emph{et~al.}
\newblock \bibinfo{title}{{X-ray Variability and Photosphere Evolution during Accretion Disk Formation in Tidal Disruption Events}}.
\newblock \emph{\bibinfo{journal}{arXiv e-prints}} \bibinfo{pages}{arXiv:2512.12985} (\bibinfo{year}{2025}).

\bibitem{price+24}
\bibinfo{author}{{Price}, D.~J.} \emph{et~al.}
\newblock \bibinfo{title}{{Eddington Envelopes: The Fate of Stars on Parabolic Orbits Tidally Disrupted by Supermassive Black Holes}}.
\newblock \emph{\bibinfo{journal}{\apjl}} \textbf{\bibinfo{volume}{971}}, \bibinfo{pages}{L46} (\bibinfo{year}{2024}).

\bibitem{martire+25_axv}
\bibinfo{author}{{Martire}, P.} \emph{et~al.}
\newblock \bibinfo{title}{{Wind-mediated Eddington-limited emission in a $10^{4}M_\odot$ Black Hole Tidal Disruption Event}}.
\newblock \emph{\bibinfo{journal}{arXiv e-prints}} \bibinfo{pages}{arXiv:2512.10564} (\bibinfo{year}{2025}).

\bibitem{khatami+kasen24}
\bibinfo{author}{{Khatami}, D.~K.} \& \bibinfo{author}{{Kasen}, D.~N.}
\newblock \bibinfo{title}{{The Landscape of Thermal Transients from Supernovae Interacting with a Circumstellar Medium}}.
\newblock \emph{\bibinfo{journal}{\apj}} \textbf{\bibinfo{volume}{972}}, \bibinfo{pages}{140} (\bibinfo{year}{2024}).

\bibitem{Kasen2010}
\bibinfo{author}{{Kasen}, D.} \& \bibinfo{author}{{Bildsten}, L.}
\newblock \bibinfo{title}{{Supernova Light Curves Powered by Young Magnetars}}.
\newblock \emph{\bibinfo{journal}{\apj}} \textbf{\bibinfo{volume}{717}}, \bibinfo{pages}{245--249} (\bibinfo{year}{2010}).

\bibitem{Woosley2010}
\bibinfo{author}{{Woosley}, S.~E.}
\newblock \bibinfo{title}{{Bright Supernovae from Magnetar Birth}}.
\newblock \emph{\bibinfo{journal}{\apjl}} \textbf{\bibinfo{volume}{719}}, \bibinfo{pages}{L204--L207} (\bibinfo{year}{2010}).

\bibitem{nicholl+17}
\bibinfo{author}{{Nicholl}, M.}, \bibinfo{author}{{Guillochon}, J.} \& \bibinfo{author}{{Berger}, E.}
\newblock \bibinfo{title}{{The Magnetar Model for Type I Superluminous Supernovae. I. Bayesian Analysis of the Full Multicolor Light-curve Sample with MOSFiT}}.
\newblock \emph{\bibinfo{journal}{\apj}} \textbf{\bibinfo{volume}{850}}, \bibinfo{pages}{55} (\bibinfo{year}{2017}).

\bibitem{nicholl+15}
\bibinfo{author}{{Nicholl}, M.} \emph{et~al.}
\newblock \bibinfo{title}{{On the diversity of superluminous supernovae: ejected mass as the dominant factor}}.
\newblock \emph{\bibinfo{journal}{\mnras}} \textbf{\bibinfo{volume}{452}}, \bibinfo{pages}{3869--3893} (\bibinfo{year}{2015}).

\bibitem{lunnan+18}
\bibinfo{author}{{Lunnan}, R.} \emph{et~al.}
\newblock \bibinfo{title}{{Hydrogen-poor Superluminous Supernovae from the Pan-STARRS1 Medium Deep Survey}}.
\newblock \emph{\bibinfo{journal}{\apj}} \textbf{\bibinfo{volume}{852}}, \bibinfo{pages}{81} (\bibinfo{year}{2018}).

\bibitem{prentice+18}
\bibinfo{author}{{Prentice}, S.~J.} \emph{et~al.}
\newblock \bibinfo{title}{{The Cow: Discovery of a Luminous, Hot, and Rapidly Evolving Transient}}.
\newblock \emph{\bibinfo{journal}{\apjl}} \textbf{\bibinfo{volume}{865}}, \bibinfo{pages}{L3} (\bibinfo{year}{2018}).

\bibitem{margutti+19}
\bibinfo{author}{{Margutti}, R.} \emph{et~al.}
\newblock \bibinfo{title}{{An Embedded X-Ray Source Shines through the Aspherical AT 2018cow: Revealing the Inner Workings of the Most Luminous Fast-evolving Optical Transients}}.
\newblock \emph{\bibinfo{journal}{\apj}} \textbf{\bibinfo{volume}{872}}, \bibinfo{pages}{18} (\bibinfo{year}{2019}).

\bibitem{Kuin2019}
\bibinfo{author}{{Kuin}, N. P.~M.} \emph{et~al.}
\newblock \bibinfo{title}{{Swift spectra of AT2018cow: a white dwarf tidal disruption event?}}
\newblock \emph{\bibinfo{journal}{\mnras}} \textbf{\bibinfo{volume}{487}}, \bibinfo{pages}{2505--2521} (\bibinfo{year}{2019}).

\bibitem{schulze+21}
\bibinfo{author}{{Schulze}, S.} \emph{et~al.}
\newblock \bibinfo{title}{{The Palomar Transient Factory Core-collapse Supernova Host-galaxy Sample. I. Host-galaxy Distribution Functions and Environment Dependence of Core-collapse Supernovae}}.
\newblock \emph{\bibinfo{journal}{\apjs}} \textbf{\bibinfo{volume}{255}}, \bibinfo{pages}{29} (\bibinfo{year}{2021}).

\bibitem{short+20}
\bibinfo{author}{{Short}, P.} \emph{et~al.}
\newblock \bibinfo{title}{{The tidal disruption event AT 2018hyz - I. Double-peaked emission lines and a flat Balmer decrement}}.
\newblock \emph{\bibinfo{journal}{\mnras}} \textbf{\bibinfo{volume}{498}}, \bibinfo{pages}{4119--4133} (\bibinfo{year}{2020}).

\bibitem{gomez+20}
\bibinfo{author}{{Gomez}, S.} \emph{et~al.}
\newblock \bibinfo{title}{{The Tidal Disruption Event AT 2018hyz II: Light-curve modelling of a partially disrupted star}}.
\newblock \emph{\bibinfo{journal}{\mnras}} \textbf{\bibinfo{volume}{497}}, \bibinfo{pages}{1925--1934} (\bibinfo{year}{2020}).

\bibitem{cendes+22}
\bibinfo{author}{{Cendes}, Y.} \emph{et~al.}
\newblock \bibinfo{title}{{A Mildly Relativistic Outflow Launched Two Years after Disruption in Tidal Disruption Event AT2018hyz}}.
\newblock \emph{\bibinfo{journal}{\apj}} \textbf{\bibinfo{volume}{938}}, \bibinfo{pages}{28} (\bibinfo{year}{2022}).

\bibitem{sfaradi+24a}
\bibinfo{author}{{Sfaradi}, I.} \emph{et~al.}
\newblock \bibinfo{title}{{An off-axis relativistic jet seen in the long lasting delayed radio flare of the TDE AT 2018hyz}}.
\newblock \emph{\bibinfo{journal}{\mnras}} \textbf{\bibinfo{volume}{527}}, \bibinfo{pages}{7672--7680} (\bibinfo{year}{2024}).

\bibitem{cendes+26}
\bibinfo{author}{{Cendes}, Y.} \emph{et~al.}
\newblock \bibinfo{title}{{Continued Rapid Radio Brightening of the Tidal Disruption Event AT2018hyz}}.
\newblock \emph{\bibinfo{journal}{\apj}} \textbf{\bibinfo{volume}{998}}, \bibinfo{pages}{111} (\bibinfo{year}{2026}).

\bibitem{Somalwar2023}
\bibinfo{author}{{Somalwar}, J.~J.} \emph{et~al.}
\newblock \bibinfo{title}{{A Candidate Relativistic Tidal Disruption Event at 340 Mpc}}.
\newblock \emph{\bibinfo{journal}{\apj}} \textbf{\bibinfo{volume}{945}}, \bibinfo{pages}{142} (\bibinfo{year}{2023}).

\bibitem{Dykaar2024}
\bibinfo{author}{{Dykaar}, H.} \emph{et~al.}
\newblock \bibinfo{title}{{An Untargeted Search for Radio-emitting Tidal Disruption Events in the VAST Pilot Survey}}.
\newblock \emph{\bibinfo{journal}{\apj}} \textbf{\bibinfo{volume}{973}}, \bibinfo{pages}{104} (\bibinfo{year}{2024}).

\bibitem{Hammerstein+23a}
\bibinfo{author}{{Hammerstein}, E.} \emph{et~al.}
\newblock \bibinfo{title}{{The Final Season Reimagined: 30 Tidal Disruption Events from the ZTF-I Survey}}.
\newblock \emph{\bibinfo{journal}{\apj}} \textbf{\bibinfo{volume}{942}}, \bibinfo{pages}{9} (\bibinfo{year}{2023}).

\bibitem{roth+kasen18}
\bibinfo{author}{{Roth}, N.} \& \bibinfo{author}{{Kasen}, D.}
\newblock \bibinfo{title}{{What Sets the Line Profiles in Tidal Disruption Events?}}
\newblock \emph{\bibinfo{journal}{\apj}} \textbf{\bibinfo{volume}{855}}, \bibinfo{pages}{54} (\bibinfo{year}{2018}).

\bibitem{Charalampopoulos+22}
\bibinfo{author}{{Charalampopoulos}, P.} \emph{et~al.}
\newblock \bibinfo{title}{{A detailed spectroscopic study of tidal disruption events}}.
\newblock \emph{\bibinfo{journal}{\aap}} \textbf{\bibinfo{volume}{659}}, \bibinfo{pages}{A34} (\bibinfo{year}{2022}).

\bibitem{decia+18}
\bibinfo{author}{{De Cia}, A.} \emph{et~al.}
\newblock \bibinfo{title}{{Light Curves of Hydrogen-poor Superluminous Supernovae from the Palomar Transient Factory}}.
\newblock \emph{\bibinfo{journal}{\apj}} \textbf{\bibinfo{volume}{860}}, \bibinfo{pages}{100} (\bibinfo{year}{2018}).

\bibitem{lunnan+20}
\bibinfo{author}{{Lunnan}, R.} \emph{et~al.}
\newblock \bibinfo{title}{{Four (Super)luminous Supernovae from the First Months of the ZTF Survey}}.
\newblock \emph{\bibinfo{journal}{\apj}} \textbf{\bibinfo{volume}{901}}, \bibinfo{pages}{61} (\bibinfo{year}{2020}).

\bibitem{greiner+15}
\bibinfo{author}{{Greiner}, J.} \emph{et~al.}
\newblock \bibinfo{title}{{A very luminous magnetar-powered supernova associated with an ultra-long {\ensuremath{\gamma}}-ray burst}}.
\newblock \emph{\bibinfo{journal}{\nat}} \textbf{\bibinfo{volume}{523}}, \bibinfo{pages}{189--192} (\bibinfo{year}{2015}).

\bibitem{Holoien2014}
\bibinfo{author}{{Holoien}, T.~W.-S.} \emph{et~al.}
\newblock \bibinfo{title}{{ASASSN-14ae: a tidal disruption event at 200 Mpc}}.
\newblock \emph{\bibinfo{journal}{\mnras}} \textbf{\bibinfo{volume}{445}}, \bibinfo{pages}{3263--3277} (\bibinfo{year}{2014}).

\bibitem{Cannizzaro2021}
\bibinfo{author}{{Cannizzaro}, G.} \emph{et~al.}
\newblock \bibinfo{title}{{Accretion disc cooling and narrow absorption lines in the tidal disruption event AT 2019dsg}}.
\newblock \emph{\bibinfo{journal}{\mnras}} \textbf{\bibinfo{volume}{504}}, \bibinfo{pages}{792--815} (\bibinfo{year}{2021}).

\bibitem{ho+20}
\bibinfo{author}{{Ho}, A. Y.~Q.} \emph{et~al.}
\newblock \bibinfo{title}{{The Koala: A Fast Blue Optical Transient with Luminous Radio Emission from a Starburst Dwarf Galaxy at z = 0.27}}.
\newblock \emph{\bibinfo{journal}{\apj}} \textbf{\bibinfo{volume}{895}}, \bibinfo{pages}{49} (\bibinfo{year}{2020}).

\bibitem{perley+21}
\bibinfo{author}{{Perley}, D.~A.} \emph{et~al.}
\newblock \bibinfo{title}{{Real-time discovery of AT2020xnd: a fast, luminous ultraviolet transient with minimal radioactive ejecta}}.
\newblock \emph{\bibinfo{journal}{\mnras}} \textbf{\bibinfo{volume}{508}}, \bibinfo{pages}{5138--5147} (\bibinfo{year}{2021}).

\bibitem{berger+12}
\bibinfo{author}{{Berger}, E.} \emph{et~al.}
\newblock \bibinfo{title}{{Radio Monitoring of the Tidal Disruption Event Swift J164449.3+573451. I. Jet Energetics and the Pristine Parsec-scale Environment of a Supermassive Black Hole}}.
\newblock \emph{\bibinfo{journal}{\apj}} \textbf{\bibinfo{volume}{748}}, \bibinfo{pages}{36} (\bibinfo{year}{2012}).

\bibitem{zauderer+13}
\bibinfo{author}{{Zauderer}, B.~A.} \emph{et~al.}
\newblock \bibinfo{title}{{Radio Monitoring of the Tidal Disruption Event Swift J164449.3+573451. II. The Relativistic Jet Shuts Off and a Transition to Forward Shock X-Ray/Radio Emission}}.
\newblock \emph{\bibinfo{journal}{\apj}} \textbf{\bibinfo{volume}{767}}, \bibinfo{pages}{152} (\bibinfo{year}{2013}).

\bibitem{Rhodes2023}
\bibinfo{author}{{Rhodes}, L.} \emph{et~al.}
\newblock \bibinfo{title}{{Day-time-scale variability in the radio light curve of the Tidal Disruption Event AT2022cmc: confirmation of a highly relativistic outflow}}.
\newblock \emph{\bibinfo{journal}{\mnras}} \textbf{\bibinfo{volume}{521}}, \bibinfo{pages}{389--395} (\bibinfo{year}{2023}).

\bibitem{alexander+16}
\bibinfo{author}{{Alexander}, K.~D.}, \bibinfo{author}{{Berger}, E.}, \bibinfo{author}{{Guillochon}, J.}, \bibinfo{author}{{Zauderer}, B.~A.} \& \bibinfo{author}{{Williams}, P.~K.~G.}
\newblock \bibinfo{title}{{Discovery of an Outflow from Radio Observations of the Tidal Disruption Event ASASSN-14li}}.
\newblock \emph{\bibinfo{journal}{\apjl}} \textbf{\bibinfo{volume}{819}}, \bibinfo{pages}{L25} (\bibinfo{year}{2016}).

\bibitem{stein+21}
\bibinfo{author}{{Stein}, R.} \emph{et~al.}
\newblock \bibinfo{title}{{A tidal disruption event coincident with a high-energy neutrino}}.
\newblock \emph{\bibinfo{journal}{Nature Astronomy}} \textbf{\bibinfo{volume}{5}}, \bibinfo{pages}{510--518} (\bibinfo{year}{2021}).

\bibitem{cendes+21}
\bibinfo{author}{{Cendes}, Y.} \emph{et~al.}
\newblock \bibinfo{title}{{Radio Observations of an Ordinary Outflow from the Tidal Disruption Event AT2019dsg}}.
\newblock \emph{\bibinfo{journal}{\apj}} \textbf{\bibinfo{volume}{919}}, \bibinfo{pages}{127} (\bibinfo{year}{2021}).

\bibitem{bright+22}
\bibinfo{author}{{Bright}, J.~S.} \emph{et~al.}
\newblock \bibinfo{title}{{Radio and X-Ray Observations of the Luminous Fast Blue Optical Transient AT 2020xnd}}.
\newblock \emph{\bibinfo{journal}{\apj}} \textbf{\bibinfo{volume}{926}}, \bibinfo{pages}{112} (\bibinfo{year}{2022}).

\bibitem{berger+03}
\bibinfo{author}{{Berger}, E.} \emph{et~al.}
\newblock \bibinfo{title}{{A common origin for cosmic explosions inferred from calorimetry of GRB030329}}.
\newblock \emph{\bibinfo{journal}{\nat}} \textbf{\bibinfo{volume}{426}}, \bibinfo{pages}{154--157} (\bibinfo{year}{2003}).

\bibitem{vanderhorst+07}
\bibinfo{author}{{van der Horst}, A.~J.} \emph{et~al.}
\newblock \bibinfo{title}{{GRB 030329: 3 years of radio afterglow monitoring}}.
\newblock \emph{\bibinfo{journal}{Philosophical Transactions of the Royal Society of London Series A}} \textbf{\bibinfo{volume}{365}}, \bibinfo{pages}{1241--1246} (\bibinfo{year}{2007}).

\bibitem{perley+14}
\bibinfo{author}{{Perley}, D.~A.} \emph{et~al.}
\newblock \bibinfo{title}{{The Afterglow of GRB 130427A from 1 to {}10$^{16}$ GHz}}.
\newblock \emph{\bibinfo{journal}{\apj}} \textbf{\bibinfo{volume}{781}}, \bibinfo{pages}{37} (\bibinfo{year}{2014}).

\bibitem{vanderHorst2014}
\bibinfo{author}{{van der Horst}, A.~J.} \emph{et~al.}
\newblock \bibinfo{title}{{A comprehensive radio view of the extremely bright gamma-ray burst 130427A}}.
\newblock \emph{\bibinfo{journal}{\mnras}} \textbf{\bibinfo{volume}{444}}, \bibinfo{pages}{3151--3163} (\bibinfo{year}{2014}).

\bibitem{laskar+23}
\bibinfo{author}{{Laskar}, T.} \emph{et~al.}
\newblock \bibinfo{title}{{The Radio to GeV Afterglow of GRB 221009A}}.
\newblock \emph{\bibinfo{journal}{\apjl}} \textbf{\bibinfo{volume}{946}}, \bibinfo{pages}{L23} (\bibinfo{year}{2023}).

\bibitem{kulkarni+98}
\bibinfo{author}{{Kulkarni}, S.~R.} \emph{et~al.}
\newblock \bibinfo{title}{{Radio emission from the unusual supernova 1998bw and its association with the {\ensuremath{\gamma}}-ray burst of 25 April 1998}}.
\newblock \emph{\bibinfo{journal}{\nat}} \textbf{\bibinfo{volume}{395}}, \bibinfo{pages}{663--669} (\bibinfo{year}{1998}).

\bibitem{ho+22a}
\bibinfo{author}{{Ho}, A. Y.~Q.} \emph{et~al.}
\newblock \bibinfo{title}{{Luminous Millimeter, Radio, and X-Ray Emission from ZTF 20acigmel (AT 2020xnd)}}.
\newblock \emph{\bibinfo{journal}{\apj}} \textbf{\bibinfo{volume}{932}}, \bibinfo{pages}{116} (\bibinfo{year}{2022}).

\bibitem{ho+23b}
\bibinfo{author}{{Ho}, A. Y.~Q.} \emph{et~al.}
\newblock \bibinfo{title}{{Minutes-duration optical flares with supernova luminosities}}.
\newblock \emph{\bibinfo{journal}{\nat}} \textbf{\bibinfo{volume}{623}}, \bibinfo{pages}{927--931} (\bibinfo{year}{2023}).

\bibitem{Pursiainen+25}
\bibinfo{author}{{Pursiainen}, M.} \emph{et~al.}
\newblock \bibinfo{title}{{Optical evolution of AT 2024wpp: the high-velocity outflows in Cow-like transients are consistent with high spherical symmetry}}.
\newblock \emph{\bibinfo{journal}{\mnras}} \textbf{\bibinfo{volume}{537}}, \bibinfo{pages}{3298--3309} (\bibinfo{year}{2025}).

\bibitem{chrimes+24}
\bibinfo{author}{{Chrimes}, A.~A.} \emph{et~al.}
\newblock \bibinfo{title}{{Multi-wavelength observations of the luminous fast blue optical transient AT 2023fhn: Up to {\ensuremath{\sim}}200 days post-explosion}}.
\newblock \emph{\bibinfo{journal}{\aap}} \textbf{\bibinfo{volume}{691}}, \bibinfo{pages}{A329} (\bibinfo{year}{2024}).

\bibitem{chen+18}
\bibinfo{author}{{Chen}, T.-W.} \emph{et~al.}
\newblock \bibinfo{title}{{SN 2017ens: The Metamorphosis of a Luminous Broadlined Type Ic Supernova into an SN IIn}}.
\newblock \emph{\bibinfo{journal}{\apjl}} \textbf{\bibinfo{volume}{867}}, \bibinfo{pages}{L31} (\bibinfo{year}{2018}).

\bibitem{vestrand+14}
\bibinfo{author}{{Vestrand}, W.~T.} \emph{et~al.}
\newblock \bibinfo{title}{{The Bright Optical Flash and Afterglow from the Gamma-Ray Burst GRB 130427A}}.
\newblock \emph{\bibinfo{journal}{Science}} \textbf{\bibinfo{volume}{343}}, \bibinfo{pages}{38--41} (\bibinfo{year}{2014}).

\bibitem{Chevalier2000}
\bibinfo{author}{{Chevalier}, R.~A.} \& \bibinfo{author}{{Li}, Z.-Y.}
\newblock \bibinfo{title}{{Wind Interaction Models for Gamma-Ray Burst Afterglows: The Case for Two Types of Progenitors}}.
\newblock \emph{\bibinfo{journal}{\apj}} \textbf{\bibinfo{volume}{536}}, \bibinfo{pages}{195--212} (\bibinfo{year}{2000}).

\end{thebibliography}

\newpage

\setcounter{figure}{0}
\begin{figure}[H]
\centering
\includegraphics{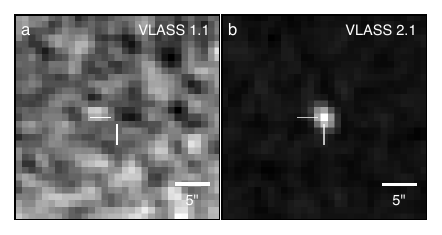}
\captionsetup{name=Extended Data Figure}
\caption{\small \textbf{VLASS images of location of AT 2019ijn. a,} VLASS Epoch 1.1 non-detection at 3 GHz at a $3\sigma$ threshold of 0.4 mJy on September 25, 2017. Background variation is due to calibration effects from a bright nearby quasar (see \ref{subsec:xray}). Crosshairs show the location of the transient from future epochs. \textbf{b,} VLASS Epoch 2.1 detection at 3 GHz on July 21, 2020. Radio source is not resolved.}\label{fig:discovery}
\end{figure}

\begin{figure}[H]
\centering
\includegraphics{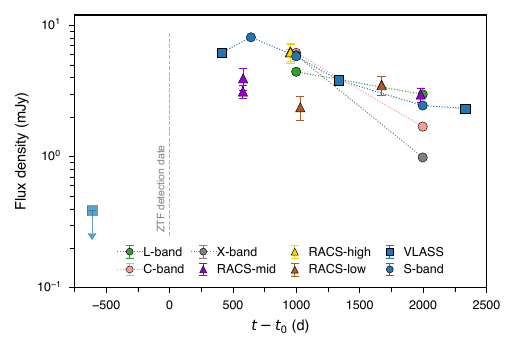}
\captionsetup{name=Extended Data Figure}
\caption{\small \textbf{Radio light curve of AT 2019ijn.} Radio light curve containing all archival data from VLASS (2-4 GHz; squares) and RACS (0.9-1.6 GHz; triangles), as well as all VLA follow-up data (1-12 GHz; circles). Dotted line corresponds to date of initial ZTF detection $t_0$ on May 31, 2019. Earliest point corresponds to the VLASS Epoch 1.1 non-detection - the exact onset of the radio emission is not constrained.}\label{fig:radioLCs}
\end{figure}

\begin{figure}[H]
\centering
\includegraphics[width=1\textwidth]{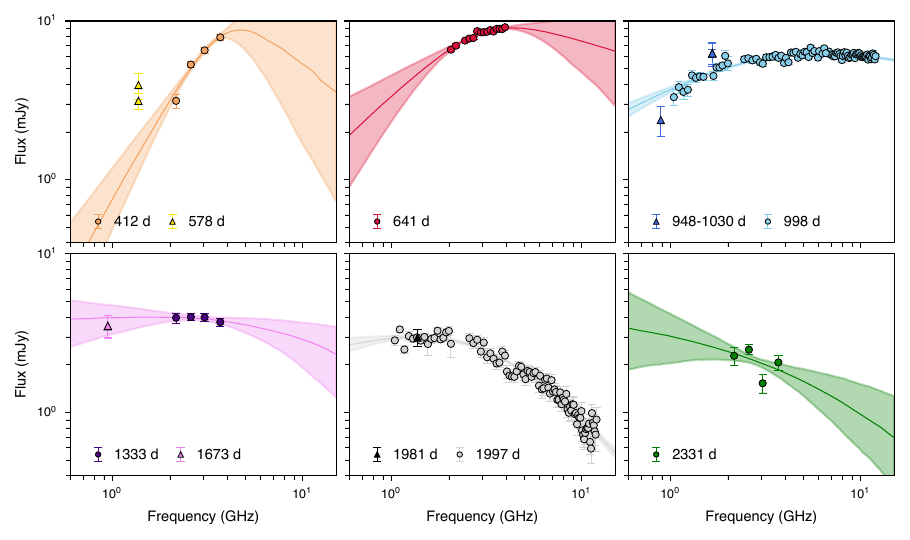}
\captionsetup{name=Extended Data Figure}
\caption{\small \textbf{Observed VLA spectra of radio transient,} including VLASS in-band (2-4 GHz) spectra (412 days post-$t_0$, gold; 1333 d, indigo; 2331 d, green), VLA single-band (2-4 GHz) follow-up (641 d, red) and VLA multi-band (1-12 GHz) spectra (998 d, blue and 1997 d, gray). Spectra are divided into sub-bands for greater spectral resolution and are shown with the best-fit to a synchrotron spectrum within the observed frequency regime. 
Data from RACS in the low, mid, and high bands are also shown (triangles), though not included in the spectral fitting.}\label{fig:radspec}
\end{figure}

\begin{figure}[H]
\centering
\begin{subfigure}[b]{1\textwidth}
    \includegraphics{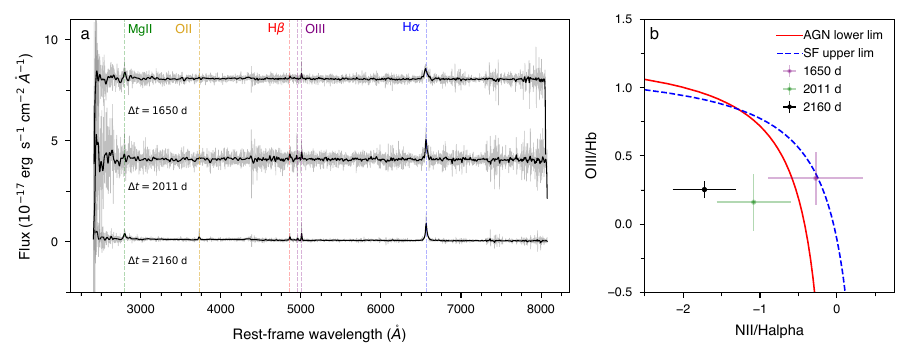}
\end{subfigure}
\hfill
\begin{subfigure}[b]{1\textwidth}
    \includegraphics{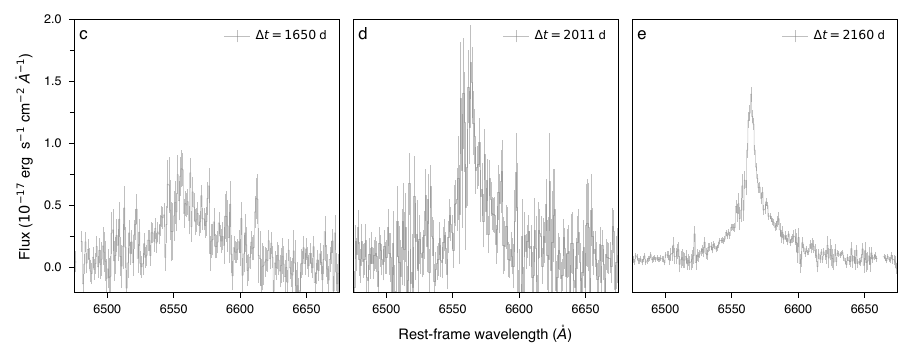}
\end{subfigure}
\captionsetup{name=Extended Data Figure}
\caption{\small \textbf{All LRIS spectra, H$\alpha$ lines, and BPT analysis of the host galaxy of AT 2019ijn. a,} LRIS spectra at 1650 d (shifted by $+ 8\times10^{-17}~\rm{erg~s^{-1}~cm^{-2}~\mathring{A}^{-1}}$), 2011 d (shifted by $+ 4\times10^{-17}~\rm{erg~s^{-1}~cm^{-2}~\mathring{A}^{-1}}$),  and 2160 d. Observed emission lines marked, with positions and redshift determined by the highest-quality spectrum at 2160 d. For ease of visualization, the raw spectrum is overlaid with a smoothed spectrum, created by convolving the raw spectrum with a 1D Gaussian Kernel with a standard deviation of 5 \r{A}. \textbf{b,} BPT diagram of the host galaxy properties of AT 2019ijn. Spectrum at 2160 d lies well within the star-forming regime. Results from earlier spectra are also shown, though less reliable due to worse data quality and higher transient contamination. \textbf{c, d, e,} window on the H$\alpha$ line only for spectra at 1650 d, 2011 d, and 2160 d, respectively.}\label{fig:alloptspec}
\end{figure}

\begin{figure}[H]
\centering
\includegraphics{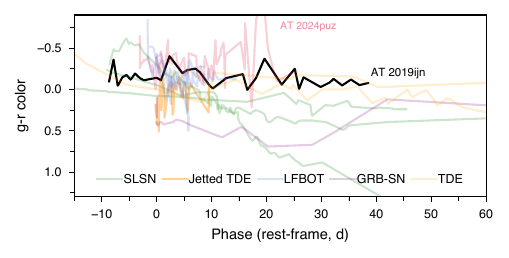}
\captionsetup{name=Extended Data Figure}
\caption{\small \textbf{g-r color evolution of AT 2019ijn compared to other energetic transients in the literature.} g-r colors are interpolated from available light curve data (see section~\ref{sec:figs}). SNe (SN 2018bgv, SN 2017ens, PTF 10hgi) are shown in green, LFBOTs (AT 2018cow, ZTFabvkwla, AT 2020xnd) are shown in blue, the jetted TDE AT 2022cmc in dark orange, non-jetted TDEs (AT 2018hyz, see Section \ref{sec:figs}; AT 2019dsg) in light orange, and GRB-SNe (SN 2011kl/GRB 111209A, GRB 130427A) in purple. AT 2024puz shown in crimson. Color evolution curves are shown for the duration of the AT 2019ijn optical flare for direct comparison.}\label{fig:colev}
\end{figure}

\begin{figure}[H]
\centering
\includegraphics[width=1\textwidth]{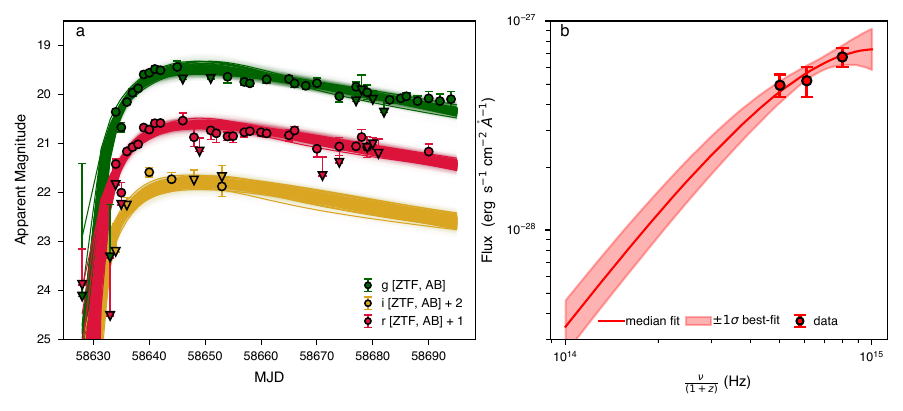}
\captionsetup{name=Extended Data Figure}
\caption{\small \textbf{Models of optical emission of AT 2019ijn. a,} \texttt{MOSFiT} fits, allowing super-Eddington accretion, to the optical light curve of AT 2019ijn, including measurements (circles) and upper limits (triangles) for g (green), r (red), and i (gold) bands. \textbf{b,} blackbody fits to the optical SED at peak light, with median fit and $1\sigma$ errors shown. }\label{fig:models}
\end{figure}

\newpage
\begin{longtable}[H]{lcccccc}
\captionsetup{name=Extended Data Table}
\caption{\small \textbf{Best-fit parameters for optical light curve fitting}}\label{tab:optLCfit} \\
\toprule
Band & $\nu$ & $t_p$ & $\sigma$ & $\tau$ & $R_{\rm{bb}}$ & $L_{\rm{bb}}$ \\
& (Hz) & (d) & (d) & (d) & ($10^{15}$ cm) & ($10^{44}$ erg/s) \\
\midrule
g & $6.3\times10^{14}$ & $8.9 \pm 0.5$ & $6.2 \pm 0.4$ & $ 73 \pm 3$  & $2.3 \pm 0.7$ & $3 \pm 4$ \\
r & $4.8\times10^{14}$ & $10.9 \pm 0.5$ & $7.4 \pm 0.5$ & $51 \pm 3$ & $3.0 \pm 0.8 $ & $6 \pm 7$ \\
i & $3.9 \times 10^{14}$ & $4 \pm 20$ & $50 \pm 30$ & $40 \pm 20$ & $3.8 \pm 0.9$ & $10 \pm 10$ \\
\bottomrule
\end{longtable}


\begin{longtable}[H]{lcccccccc}
\captionsetup{name=Extended Data Table}
\caption{\small \textbf{Best-fit parameters for radio SED fitting}}\label{tab:radfit} \\
\toprule
$\Delta t$ & $\nu_b$ & $s$ & $\beta_1$ & $\beta_2$ & $S_0$ & $\chi^2$ & BIC \\
(d) & (GHz) &  &  &  & (mJy) &  &  &  \\
\midrule
412 & $3.8 \pm 0.8$ & $1.7 \pm 0.8$ & $1 \pm 2$ & $1 \pm 2$ & $13 \pm 4$ & -16.1 & 22.3 \\
641 & $3 \pm 2$ & $1.8 \pm 0.8$ & $1 \pm 1$ & $-0.3 \pm 0.7$ & $13 \pm 3$ & 3.6 & 285.7 \\
998 & $3 \pm 1$ & $1.6 \pm 0.7$ & $0.7 \pm 0.2$ & $-0.2 \pm 0.1$ & $9 \pm 1$ & 1.8 & 51.4 \\
1333 & $11 \pm 10$ & $1.3 \pm 0.9$ & $-0.2 \pm 1$ & $-0.2 \pm 0.9$ & $5 \pm 3$ & -1.2 & 11.7 \\
1997 & $3 \pm 1$ & $0.9 \pm 0.4$ & $0.2\pm 0.9$ & $-1 \pm 1$ & $5 \pm 1$ & 1.6 & -34.0 \\
2331 & $9 \pm 10$ & $1.2 \pm 0.9$ & $-0.6 \pm 1$ & $-0.5 \pm 0.8$ & $3 \pm 3$ & -49.6 & 125.6 \\
\bottomrule
\end{longtable}

\newpage
\begin{longtable}[H]{lcccc}
\captionsetup{name=Extended Data Table}
\caption{\small \textbf{Best-fit parameters for optical spectral line fitting}}\label{tab:optfit} \\
\toprule
Line & Wavelength & Luminosity & Width & Velocity \\
 & $(\mathring{A})$ & ($10^{40}$ erg/s) & $(\mathring{A})$ & (km/s) \\
\midrule
\multicolumn{5}{c}{Spectrum at $\Delta t = 2160$ d} \\
H$\alpha$ narrow & 6564.6 & $1.6 \pm 0.2$ & $3.1 \pm 0.3$ & $330 \pm 40$ \\
H$\alpha$ broad & $6568 \pm 1$ & $4.6 \pm 0.2$ & $18 \pm 1$ & $1900 \pm 100$ \\
$\rm{[NII]}$ & 6549.9 & $< 0.03 \pm 0.03$ & $3.9 \pm 0.9$ & $400 \pm 100$ \\ 
$\rm{[NII]}$ & 6585.3 & $< 0.03 \pm 0.03$ & $3.9 \pm 0.9$ & $400 \pm 100$ \\
H$\beta$ & 4832.7 & $0.6 \pm 0.1$ & $3.6\pm 0.6$ & $520 \pm 90$ \\ 
$\rm{[OIII]}$ & 4960.3 & $0.4 \pm 0.1$ & $2.53\pm 0.03$ & $360 \pm 5$ \\
$\rm{[OIII]}$ & 5008.2 & $1.1 \pm 0.1$ & $2.53\pm 0.03$ & $360 \pm 5$ \\
$\rm{[OII]}$ & 3727.1 & $0.4 \pm 0.4$ & $2.6 \pm 0.1 $ & $500 \pm 20$ \\
$\rm{[OII]}$ & 3729.9 & $0.5 \pm 0.3$ & $2.6 \pm 0.1 $ & $500 \pm 20$ \\
$\rm{[MgII]}$ & 2796.0 & $0.8 \pm 0.3$ & $4.2 \pm 0.8$ & $1100 \pm 200$ \\
$\rm{[MgII]}$ & 2803.0 & $0.5 \pm 0.3$ & $4.2 \pm 0.8$ & $1100 \pm 200$ \\
\midrule 
\multicolumn{5}{c}{Spectrum at $\Delta t = 2011$ d} \\
H$\alpha$ narrow & 6564.6 & $3.0 \pm 0.9$ & $4.6 \pm 0.5$ & $490 \pm 60$ \\
H$\alpha$ broad & $6572 \pm 3$ & $6 \pm 3$ & $40 \pm 30$ & $5000 \pm 3000$ \\
$\rm{[NII]}$ & 6549.9 & $<0.2\pm 0.2$ & $3.9 \pm 0.8$ & $420 \pm 90$ \\ 
$\rm{[NII]}$ & 6585.3 & $<0.2 \pm 0.3$ & $3.9 \pm 0.8$ & $420 \pm 90$ \\ 
H$\beta$ & 4832.7 & $0.9 \pm 0.4$ & $4.0 \pm 0.8$ & $600\pm 100$ \\ 
$\rm{[OIII]}$ & 4960.3 & $0.3 \pm 0.2$ & $3.7\pm 0.8$ & $500\pm100$ \\
$\rm{[OIII]}$ & 5008.2 & $1.3 \pm 0.3$ & $3.7\pm 0.8$ & $500\pm100$ \\
$\rm{[OII]}$ & 3727.1 & $0.4 \pm 0.4$ & $3.5\pm 0.7$ & $700 \pm 100$ \\
$\rm{[OII]}$ & 3729.9 & $0.5 \pm 0.4$ & $3.5\pm 0.7$ & $700 \pm 100$ \\
$\rm{[MgII]}$ & 2796.0 & $1.1 \pm 0.8$ & $3.7 \pm 0.9$ & $900 \pm 200$ \\
$\rm{[MgII]}$ & 2803.0 & $0.5 \pm 0.5$ & $3.7 \pm 0.9$ & $900 \pm 200$ \\
\midrule
\multicolumn{5}{c}{Spectrum at $\Delta t = 1650$ d} \\
H$\alpha$ narrow & 6564.6 & $0.3 \pm 0.3$ & $4.3\pm 0.7$ & $470 \pm 80$ \\
H$\alpha$ broad & $6556 \pm 3$ & $5 \pm 1$ & $23 \pm 8$ & $2500 \pm 800$ \\
$\rm{[NII]}$ & 6549.9 & $<0.3 \pm 0.3$ & $3.9 \pm 0.9$ & $400 \pm 100$ \\ 
$\rm{[NII]}$ & 6585.3 & $<0.2 \pm 0.2$ & $3.9 \pm 0.9$ & $400 \pm 100$ \\ 
H$\beta$ & 4832.7 & $0.4 \pm 0.2$ & $3.7 \pm 0.9$ & $500 \pm 100$ \\ 
$\rm{[OIII]}$ & 4960.3 & $0.1 \pm 0.1$ & $2.8\pm 0.3$ & $400\pm 50$ \\
$\rm{[OIII]}$ & 5008.2 & $0.9 \pm 0.2$ & $2.8\pm 0.3$ & $400\pm 50$ \\
$\rm{[OII]}$ & 3727.1 & $0.2 \pm 0.2$ & $3.7\pm 0.8$ & $700\pm200$ \\
$\rm{[OII]}$ & 3729.9 & $0.2 \pm 0.2$ & $3.7\pm 0.8$ & $700\pm200$ \\
$\rm{[MgII]}$ & 2796.0 & $0.9 \pm 0.6$ & $4.1 \pm 0.8$ & $1000 \pm 200$ \\
$\rm{[MgII]}$ & 2803.0 & $0.9 \pm 0.6$ & $4.1 \pm 0.8$ & $1000 \pm 200$ \\
\bottomrule
\end{longtable}

\end{document}